\documentclass[reprint,
superscriptaddress,
 amsmath,amssymb,
 aps,
showkeys]{revtex4-1}

\usepackage{graphicx}% Include figure files
\usepackage{dcolumn}% Align table columns on decimal point
\usepackage{bm}% bold math
\usepackage{hyperref}% add hypertext capabilities
\usepackage{physics}
\usepackage{mathtools}
\usepackage{subcaption}
\usepackage{graphicx,amssymb,amsmath,amsthm,amsfonts}
\usepackage[usenames]{color}
\usepackage{mathtools}

\begin{document}

\title{Canonical ensemble of a $d$-dimensional  
Reissner-Nordstr\"om 
black hole  in a cavity}

\author{Tiago V. Fernandes}
\email{tiago.vasques.fernandes@tecnico.ulisboa.pt}
\affiliation{Centro de Astrof\'isica e Gravita\c c\~ao 
- CENTRA, Departamento de F\'isica, Instituto Superior T\'{e}cnico 
- IST, Universidade de Lisboa - UL,\\
 Avenida Rovisco Pais 1, 1049-001 Lisboa, Portugal}
\author{Jos\'{e} P. S. Lemos}
\email{joselemos@ist.utl.pt}
\affiliation{Centro de Astrof\'isica e Gravita\c c\~ao 
- CENTRA, Departamento de F\'isica, Instituto Superior T\'{e}cnico 
- IST, Universidade de Lisboa - UL,\\ 
Avenida Rovisco Pais 1, 1049-001 Lisboa, Portugal}


\begin{abstract}

We construct the canonical ensemble of a $d$-dimensional
Reissner-Nordstr\"om black hole spacetime in a cavity surrounded by a
heat reservoir through the Euclidean path integral formalism.  The
cavity radius $R$ is fixed, and the heat reservoir is at a fixed
temperature $T$ and fixed electric charge $Q$.  We use York's approach
to find the reduced action by imposing the Hamiltonian and Gauss
constraints and the appropriate conditions to the Euclideanized
Einstein-Maxwell action with boundary terms, and then perform a zero
loop approximation so that the paths that minimize the action
contribute to the partition function. We find that, for an electric
charge smaller or equal than a critical saddle electric charge $Q_s$,
there are three solutions $r_{+1}$, $r_{+2}$, and $r_{+3}$, such that
$r_{+1} < r_{+2} < r_{+3}$. The solutions $r_{+1}$ and $r_{+3}$ are
stable within the ensemble, while $r_{+2}$ is unstable.  For an
electric charge equal to $Q_s$, the solution $r_{+2}$ merges with
$r_{+1}$ and $r_{+3}$ at a given specific temperature.  For an
electric charge larger than $Q_s$, there is only one solution
$r_{+4}$, which can be seen as the merging of the $r_{+1}$ and
$r_{+3}$ solutions, with $r_{+4}$ being stable.  Since the partition
function is directly related to the free energy in the canonical
ensemble, we read off the free energy and calculate the thermodynamic
variables, namely the entropy, the thermodynamic electric potential,
the thermodynamic pressure, and the mean energy. We investigate
thermodynamic stability, which is controlled by the positivity of the
heat capacity at constant area and electric charge, and show that the
heat capacity is discontinuous at the electric charge $Q_s$, signaling
a turning point.  We analyze the favorable states, examining the free
energies of the stable black hole solutions and the free energy of
electrically charged hot flat space, in order to check for possible
first and second order phase transitions between the possible states.
For instance, the two stable black hole solutions $r_{+1}$ and
$r_{+3}$ are in competition between themselves, more specifically, for
certain ensemble parameters there exists a first order phase
transition from one solution to the other, and at the critical charge
$Q_s$ this transition turns into a second order phase transition.  We
also compare the thermodynamic radius of zero free energy with the
generalized Buchdahl bound radius, which do not match, and comment on
the physical implications, such as the possibility of total
gravitational collapse of the thermodynamic system.  We study the
limit of infinite cavity radius and find two possibilities, the Davies
and the Rindler solutions.  The Davies thermodynamic solution of
electrically charged black holes in $d=4$ dimensions is recovered from
the general $d$-dimensional canonical ensemble analysis. We obtain, in
particular, the heat capacity given by Davies and the Davies point.
The Rindler solution describes the black hole horizon as a Rindler
horizon, and the boundary, which is at fixed temperature $T$ provided
by the reservoir, must have the necessary acceleration to reproduce
the corresponding Unruh temperature.  Going back to a cavity with
finite radius we find that the three solutions mentioned above are
related to the original York two Schwarzschild black hole solutions
and to the two Davies solutions, with the middle unstable solution
$r_{+2}$ belonging simultaneously to the two sets of solutions. In
this sense, York's and Davies' formalisms have been unified in our
approach.  In all instances we mention carefully the four-dimensional
case, for which we accomplish new results, and study in detail all
aspects of the five dimensional case.

\end{abstract}

%Use showkeys class option if keyword
\maketitle

%\tableofcontents
%%%%%%%%%%%%%%%%%%%%%%%%%%%%%%%%%%%%%%%%%%%%%%%%%%%%%%%%%%%%%%%%%%%%%%
\section{\label{sec:Intro}Introduction}
%%%%%%%%%%%%%%%%%%%%%%%%%%%%%%%%%%%%%%%%%%%%%%%%%%%%%%%%%%%%%%%%%%%%%%

\subsection{Background}

The hypothesis that black holes have a thermodynamic character emerged
through a series of notable developments.
Bekenstein~\cite{Bekenstein:1972} introduced the idea that a black
hole has an entropy proportional to the surface area of its event
horizon and formulated a generalized second law of thermodynamics.
Smarr found a mass formula involving all the black hole parameters
\cite{Smarr:1973}, which was extended in a formal basis to the four
laws of black hole mechanics \cite{Bardeen:1973}.  These laws were
strikingly similar to the four laws of thermodynamics.
The complete description came with the discovery by
Hawking~\cite{Hawking:1975} that black holes radiate quanta 
with a thermal spectrum at temperature $T_{\rm H} =
\frac{\kappa}{2\pi}$, the Hawking temperature in Planck units, where
$\kappa$ is black hole's surface gravity, and for
instance, for the simplest nonrotating black hole one has 
$\kappa=\frac{1}{2\pi r_+}$, so that 
$T_{\rm H} =\frac{1}{4\pi r_+}$,
 $r_+$ being the event horizon radius.
Furthermore, the vacuum
state sitting at the horizon that enables the radiation to be produced
was shown to be described by the Hartle-Hawking vacuum
state~\cite{Hartle:1976}.
By assuming that the black hole is in thermal equilibrium with the
radiation emitted, it was argued that black holes must indeed be
thermodynamic objects, and it was found that the entropy $S$ of a
black hole has the expression $S=\frac{A_+}{4}$, the
Bekenstein-Hawking entropy, where $A_+$ is the surface area of the
event horizon. The thermodynamics of black holes was expanded for
black holes with rotation and electric charge by Davies
\cite{Davies:1977}, by assigning the first law of thermodynamics
together with the Hawking temperature and the Bekenstein-Hawking
entropy to those black holes. In this case, it was noticed 
an abrupt change of the
heat capacity of the system, which
was presumed to yield a phase transition.

The rationale for the black hole entropy and thermodynamics has also
been obtained through statistical methods, in juxtaposition to the
results obtained through the analysis of quantum fields around black
holes.
The focus 
on black hole statistical methods
implies one has to set up 
an ensemble of a physical spacetime. To build
the ensemble, one needs to know a priori the microscopic description
of the system.  Such an accepted description is still unknown for
gravity.  Nevertheless, the partition function of spacetime can be
computed through the Euclidean path integral approach to quantum
gravity. In this approach, the partition function is given by a path
integral of the exponential of the Euclidean action $I[g,\phi]$, over
Euclidean metrics $g$ and fields $\phi$ that permeate the space, where
the integral is restricted to metrics that are periodic in the
imaginary time length, i.e., $Z=\int Dg D\phi {\rm
e}^{-I[g,\phi]}$. Depending on the ensemble considered, there are
always quantities that are fixed at some boundary, e.g., in the
canonical ensemble the temperature given by the inverse of the
imaginary time length at the boundary is fixed, with this boundary
then being a heat reservoir. This method of computing the partition
function inherits the difficulties of the Euclidean path integral
approach to quantum gravity. For example, the map between the physical
Lorentzian spacetime and the Euclidean space, that is performed
through a Wick transformation on a time coordinate, is not in general
well-defined or unique, covering only some sections of the Lorentzian
spacetime.  Moreover, there are difficulties on the convergence of the
path integral. To put these difficulties aside, a zero loop
approximation of the path integral is considered, where only the paths
that minimize the Euclidean action are taken into account.  The
partition function is then given by $Z = {\rm e}^{-I_0}$, where now
$I_0$ is the classical Euclidean action evaluated at one of these
paths, yielding a partition function in the semiclassical
approximation.  One can then relate the partition function to a
thermodynamic potential depending on the ensemble chosen, and the
thermodynamics of the system can be worked out through the derivatives
of the thermodynamic potential.

The application of the Euclidean path integral approach to known
spaces, such as the Schwarzschild black hole in the canonical
ensemble, and the Reissner-Nordstr\"om black in the grand canonical
ensemble with fixed electric potential at the boundary, was done with
a heat reservoir at infinite proper distance \cite{Gibbons:1977},
recovering the Hawking temperature and the Bekenstein-Hawking entropy
of a black hole.  Yet, for those configurations in the respective
ensembles, the obtained heat capacity of the black hole is negative,
which means the configuration is thermodynamically unstable.  It was
later found that the configurations correspond to a saddle point of
the action \cite{Gross:1982}.  And so, the zero loop approximation is
not valid for these configurations, although one can still treat them
as instantons. A perturbation of the instanton yields a negative mode
that makes the one-loop contribution of the path integral formally
divergent, but the path integral can still be continued to the complex
numbers resulting in a nonzero imaginary part.  However, when applied
to a Schwarzschild-anti de Sitter black hole, which can be considered
a configuration of a black hole in a box, the formalism produced
consistent results and stable black hole solutions
\cite{HawkingPage1983}.  It was then found that the negative mode of
the pure Schwarzschild black hole ceases to exist if the heat
reservoir sits at a radius equal or smaller than the photon sphere
radius~\cite{Allen:1984}.

Soon after, York realized that the construction of canonical or grand
canonical ensembles should be performed by putting a black hole space
inside a cavity with a heat reservoir at finite
radius~\cite{York:1986}. Using the path integral approach and the zero
loop approximation for a Schwarzschild black hole inside a cavity, he
found that there are two stationary points for the Euclidean action
$I_0$. From these two, the one with the least mass is unstable and
corresponds to the Hawking-Gibbons black hole in the limit of infinite
radius of the heat reservoir. The other one has the largest mass and
is stable, therefore the zero loop approximation is valid for this
stationary point.  This motivated a series of developments, namely,
the study of the canonical ensemble of a Schwarzschild black hole
inside a cavity by considering a class of paths in the context of
York's formalism~\cite{Whiting:1988}, and the application of York's
formalism to a system including matter was sketched
in~\cite{Martinez:1989}.  Moreover, the grand canonical ensemble of a
Reissner-Nordstr\"om black hole inside a cavity was considered
within York's formalism in~\cite{Braden:1990}, by fixing the
temperature and the electric potential at the boundary of the cavity.
The canonical ensemble for arbitrary configurations of
self-gravitating systems was studied in~\cite{Zaslavskii:1991}.

There have been applications of the Euclidean path integral approach,
both in Gibbons-Hawking form and in York's extended formalism, to
asymptotically anti-de Sitter and de Sitter black hole spaces as well
as to higher dimensions, which we now mention. Black hole spaces with
negative cosmological constant within general relativity were
discussed in three and four dimensions~\cite{Brown:1994}.  The two
dimensional black hole space in Teitelboim-Jackiw theory which is
asymptotically de Sitter was studied in~\cite{Lemos:1996}.  The grand
canonical ensemble of the Reissner-Nordstr\"om-anti-de Sitter space in
four dimensions was constructed and analyzed in~\cite{Peca:1999},
and of the electric charged toroidal anti-de Sitter
black hole in~\cite{Peca:2000}.
The
stability and the negative mode for
Schwarzschild-\hskip-0.05cm{}Tangherlini vacuum spacetime was
described in~\cite{Gregory:2001}.  The construction of the canonical
ensemble of a four-dimensional Reissner-Nordstr\"om black hole inside
a cavity was obtained in~\cite{carlipvaidya2003,Lundgren:2006} by
adding a boundary term on the action and fixing the electric charge
instead of the electric potential at the boundary of the cavity.  A
study of the canonical ensemble of four and higher dimensional
Schwarzschild-anti de Sitter black holes was done
in~\cite{Akbar:2010}.
The canonical ensemble of black branes  in
arbitrary dimensions along with their
phase structure was developed in \cite{Lu:2011}.
The canonical ensemble has been applied to
Schwarzschild black holes in a cavity in $d$ dimensions
in~\cite{Andre:2020,Andre:2021}.
The canonical ensemble of gravastars was analyzed
in~\cite{Miyashita:2021}.
The first law of de Sitter spaces with
black holes in an ensemble context was identified in \cite{bjsv2022}.
A  detailed analysis of the
inclusion of matter is given in~\cite{Lemos:2023}, with surprising
results concerning the equilibrium and the equation of state of the
matter. 
The grand canonical ensemble of Reissner-Nordstr\"om black holes
inside cavity in dimensions $d\geq4$ was considered
in~\cite{Fernandes:2023}.  The canonical ensemble of black hole in a
de Sitter background has been constructed and explored in
\cite{lemoszaslavskii2024}.  Using the
Gibbons\hskip0.03cm-\hskip-0.03cm{}Hawking action formalism for
electrically charged black holes in the canonical ensemble, the
Davies' thermodynamic theory of black holes has been recovered
in~\cite{fernandeslemosdavies:2024}.

It should be mentioned that motivated by supergravity theories, the
analysis of ensembles and the Euclidean path integral approach were
extended to black brane solutions.  It was found that the mechanical
stability of black branes is related to their local thermodynamic
stability~\cite{Gubser:2001}.  This relation was further studied and
proven in some cases, see~\cite{Reall:2001, Miyamoto:2007,
Collingbourne:2021}.

We also note that York's  path integral formalism and the 
thermodynamics of a hot thin shell of matter
analyzed from a first law of thermodynamics basis
share some
similarities. This was found in~\cite{Martinez:1996}
for thin shells with an outer Schwarzschild spacetime and
in~\cite{Lemos:2015a} for thin shells with an outer
Reissner-Nordstr\"om spacetime. The analysis of hot thin shells have
been extended to higher dimensions in~\cite{Andre:2019} for
Schwarzschild spacetimes, and
in~\cite{Fernandes:2022} for Reissner-Nordstr\"om spacetimes.
In~\cite{Reyes:2022}, the Reissner-Nordstr\"om case was revisited.
A radius that will appear somewhat naturally
in the analysis is the 
generalized Buchdahl
bound radius, also called Buchdahl-Andr\'easson-Wright bound
radius~\cite{Wright:2015}. It is a dynamical radius rather than a
thermodynamic one.

\subsection{Motivation}

\subsubsection{Scales}

It is important to know in which physical settings and at which scales
the situations we are studying here prevail and are of interest.
Black holes can exist in all scales, from Planck scales, through micro
scales, up to astrophysical and cosmic scales.  Planck scale and micro
scale black holes with very small radii can appear through pair
creation in strong field settings, or produced by head-on collisions
of elementary type particles of enormous high energy. On the other hand,
astrophysical and cosmic black holes arise through the gravitational
collapse of huge quantities of matter.  Different physical effects
turn up for each range of scales and one should pick the appropriate
ones that have the most impact for the black holes under study
\cite{Lemosreview:1996,Nicolini2017}.

The scales of interest here are scales where quantum effects determined by
the Hawking radiation in a black hole environment become important.  The
quantity that can be taken to set the scales is then the Hawking
temperature $T_{\rm H}$.
In $d=4$, one has 
$T_{\rm H}=\frac{1}{4\pi r_+}$ for the simplest 
black hole, $r_+$ being the horizon radius of the 
black hole.
This is a temperature
measured at infinity.  Surrounding the black hole there is thus a
cloud of radiation created via quantum processes.
The Hawking temperature can be written as
$T_{\rm H}=\frac{l_{\rm pl}}{r_+}\,T_{\rm
pl}$, where the subscript {\scriptsize pl} means 
Planck quantities.
Thus, we can write $T_{\rm H}=10^{32}\frac1{r_+}\,{\rm K}$
with  $r_+$
given in Planck units.
We want to study quantum effects that are far from the full quantum
gravity regime. If we put $r_+=10^{20}l_{\rm pl}$, then the Hawking
temperature is $T_{\rm H}=10^{-20}T_{\rm pl}$.  In usual units one has
that in this case
the horizon radius has value $r_+=10^{-13}\,{\rm cm}$, the temperature
is $T_{\rm H}=10^{12}\,{\rm K}$, and the black hole mass $m$ is
$m=10^{15}\, {\rm g}$.  Such a black hole has the size of a neutron
or proton, and the mass of a large Earth mountain, a really interesting
microscopic black hole in our context, for which semiclassical effects
have to be taken into account.  This is the kind of system we are
interested, it is a microscopic system with high temperatures, where
quantum effects are important, but not full quantum gravity
\cite{nw2011}.

Now, the continual emitting of Hawking radiation depletes the black
hole of its energy, the horizon radius shrinks and eventually the
black hole disappears.  One can think of ways to stabilize the black
hole.  One way is to enclose the black hole in a cavity and surround
it by a heat reservoir. Another way is to give the black hole a
charge, e.g., an electric or magnetic charge. It is of interest
to implement 
both ways.

\subsubsection{Geometric and physical structure}

By enclosing the black hole in a cavity surrounded by a heat reservoir
with a definite radius and at a given temperature, one is able to
maintain a thermodynamic equilibrium between both temperatures, the
black hole and the reservoir temperatures, and treat the system in a
time independent way.

This system, black hole plus reservoir, is relevant when the
temperatures of its components are sufficiently high. This means that
the black hole and heat reservoir have to be of microscopic size,
where the curvature of space is sufficiently big to generate a
significant emission of radiation from the black holes permitting the
emergence of non-negligible quantum effects within the system.

A heat reservoir at constant temperature is a physical situation that
points to the building of a statistical mechanics canonical ensemble.
Since the scales of the regime one is interested in
are far from quantum
gravity scales, but nevertheless quantum effects are important, these
gravitational systems involving black holes at these micro scales can
be treated semiclassically.

Due to the time independence of the system, one can 
use an Euclidean path integral
approach to calculate the partition function of the canonical
ensemble \cite{hawking1979} by Euclideanizing the
chosen time coordinate of the solution. 
This allows one to analyze important properties of the
system, like its full thermodynamics, its thermodynamic phases, and
the possible first and second
order phase transitions between black holes and hot
spaces. It also
clarifies the reciprocal thermodynamic responses operating
between
event horizons and cavity walls kept at finite temperature.
One could think of producing these tiny black holes in
the laboratory and test important features
some of them found here, notably the stability
behavior and the role of the thermodynamic phases.

\subsubsection{Electric charge}

By giving some charge to the black hole, e.g., electric or magnetic
charge, it is possible to stabilize it thermodynamically.  In general
relativity, black holes can have mass, electric charge, and angular
momentum. Black holes with mass and angular momentum are used to
describe astrophysical phenomena, whereas black holes with electric
charge are dismissed for such phenomena since they are quickly
discharged by the plasma in the surrounding medium.  Notwithstanding,
electrically charged black holes can be of importance when one is
dealing with micro objects. In these black holes, quantum effects come
into play. Vacuum polarization at the black hole event horizon can
discharge the black hole, as particles with opposite charge in the
polarized domains are more probable to be absorbed
\cite{Gibbons:1975}. This happens when the temperature is sufficiently
high to allow particle production of massive particles, since
electric charge of contrary sign is
superradiantly emitted.  However, when
the temperature is sufficiently low, there is not enough energy to
produce charged massive particles and the black hole does not
discharge.  One can find different ways to stabilize the charge. For
instance, if the charge is purely topological, there are no particles to
radiate.  Another instance is when the only particles of the theory
are sufficiently massive that their creation is highly suppressed,
such as a very massive magnetic monopole in a magnetically charged
black hole background \cite{KPerry:1996,Genolini20231}.  Central
charges that appear in the algebra of supergravity theories also do
not suffer from pair creation instability.
One can also fix the electric
charge in the 
cavity with the black hole inside. 
This allows us to find stable
and unstable electrically charged black holes. Indeed, for small
enough black holes and relatively small temperatures all the
packets of energy with positive or negative electric charge
are trapped within the gravitational field of the black
hole which is then
electrically stable, i.e., it does not discharge. Thus
electrically charged black holes, in particular Reissner-Nordstr\"om
black holes, have interest in practice.

Now, Reissner-Nordstr\"om black hole spacetimes can be
asymptotically anti-de Sitter, asymptotically flat,
and asymptotically de Sitter \cite{he}.
Certain particle theories, notably,
supergravity theories, work with a negative cosmological
constant, and their black hole solutions have
anti-de Sitter asymptotics.
Pure general relativity
has black holes which are asymptotically flat,
with these spacetimes yielding the appropriate
environment in the study of a sufficiently large neighborhood
surrounding a black hole.
In a cosmological setting and in other settings, one
might want to use black holes in asymptotically
de Sitter spacetimes.

\subsubsection{Higher dimensions}

The world seems to have $d=4$ spacetime dimensions, but speculations
on higher dimensions has always been in the forefront of gravitational
theories.  For instance, Schwarzschild and Reissner-Nordstr\"om black
holes in higher dimensions, $d\geq 4$, were first conceived in a
discussion connected to the problem of the dimensionality of space
\cite{tangherlini1963}.  Properties of the spaces might differ, the
Hawking temperature is now $T_{\rm H}=\frac{d-3}{4\pi r_+}$ for the
Schwarzschild-Tangherlini $d$-dimensional black hole, $r_+$ being the
horizon radius of the $d$-dimensional black hole \cite{mhewitt2015}.

Moreover, certain theories are well formulated only in
higher-dimensional spacetimes, $d\geq 4$, such is the case of several
supergravity theories, and of string and superstring theories, which
makes the study of black hole solutions in $d$-dimensions important.
In connection to these theories, there is a correspondence between
black hole physics in anti-de Sitter backgrounds and a conformal field
theory physics in the boundary of those same backgrounds, the AdS/CFT
conjecture, which is formulated in different dimensions
\cite{maldacena1998,Genolini20232}.

The extra dimensions can be small, normal, or large when compared to
the usual $d=4$ ones.  If one conceives relatively large extra
dimensions, then one can in principle produce higher dimensional black
holes in future particle accelerator machines, see, e.g.,
\cite{landsberg2015}.

As well, in studying spacetimes with $d$ generic dimensions, $d\geq
4$, one has the possibility of understanding what is peculiar to $d=4$
and what is generic, with some results for the
particular case $d=4$ being recovered.

\subsection{Aim}

Our aim is to understand
more deeply the quantum and thermodynamic properties of microscopic
gravitational systems involving black holes, notably the
interaction of a black hole with a heat bath,
using an electrically charged black hole in general relativity.
For that,
in this work, we construct the canonical ensemble of a 
$d$-dimensional Reissner-Nordstr\"om-Tangherlini, or
simply Reissner-Nordstr\"om spacetime, inside a cavity. The
construction is made by obtaining
the partition function through the Euclidean path integral approach in
zero loop approximation. The canonical ensemble is
defined by adding a boundary term to the action, by fixing the inverse
temperature as the Euclidean time length at the boundary of the
cavity, and by fixing as well the electric flux, i.e., by fixing the
electric charge. We find three
black hole solutions for the ensemble for an electric charge
smaller or equal than a critical charge from which two are stable,
and one
black hole
solution for an electric charge larger than a critical charge which is
stable. We study the thermodynamics of the stable solutions and also
analyze the thermodynamic stability.  We perform an analysis of the
thermodynamic phases, namely the phases corresponding to the 
two stable black holes and the phase of hot flat charged space,
and discuss the possible first and second order phase transitions.
We
compare the radius of zero free energy with the generalized Buchdahl
bound radius, also called Buchdahl-Andr\'easson-Wright bound
radius~\cite{Wright:2015}.  We make the analysis of the system in the
limit of infinite radius of the cavity for generic $d$, and find two
possible limits, the Davies and the Rindler solutions.  Applying the
results to $d=4$, we recover Davies' thermodynamic theory for electrically
charged black holes from the canonical ensemble in the limit
of infinite radius, and we also retrieve the Davies point
showing that it signals
a turning point rather than a second order
phase transition as originally argued. The Rindler limit
reveals that the cavity boundary is accelerated at the corresponding
Unruh temperature.  We note that the $d=4$ canonical
ensemble was mentioned in \cite{Braden:1990} and
analyzed in \cite{carlipvaidya2003,Lundgren:2006}.  When
we specifically put $d=4$ in our analysis, we confirm the results
obtained in \cite{carlipvaidya2003,Lundgren:2006}, as well as find
other interesting new results, such as the recovering 
of the full thermodynamic analysis of Davies from the
canonical ensemble when the cavity
radius, i.e., the reservoir, is at infinity.

\subsection{Organization}

This paper is organized as follows. 
In Sec.~\ref{sec:Canonical1}, we construct the canonical 
ensemble through the Euclidean path integral approach. In 
Sec.~\ref{sec:ZeroLoopSolutions}, we apply the zero loop
approximation, find the solutions to the canonical ensemble, 
and analyze their stability and the dimension dependence. We
comment on the four-dimensional case and
cover in detail the five-dimensional case, a feature
that will be provided in all sections. In 
Sec.~\ref{sec:thermo}, we study the thermodynamics given by 
the canonical ensemble in the zero loop approximation. In 
Sec.~\ref{sec:favorablephases}, we study the favorable 
states, comparing the stable black hole solutions with a 
configuration of an electrically
charged shell with gravity turned off that emulates
hot flat space with electric charge.
We also find and comment
on the thermodynamic black hole configurations
that have horizon radii higher than the
Buchdahl radius.
In Sec.~\ref{sec:infiniteradiuscavity}, we study the 
canonical ensemble in the limit of infinite radius of 
the cavity, recovering the Davies thermodynamic
theory of black holes
and finding the Rindler solution at the Unruh temperature.
In Sec.~\ref{sec:Concl}, we conclude.
There are two important appendices. In
Appendix \ref{bcRicciandEuler},
the Euclidean action for the canonical ensemble,
the boundary conditions, the Ricci scalar, the Euler characteristic,
and the reduced action
are derived and explained in detail. 
In Appendix \ref{app:BHzerofreeenergy},
we perform the calculation of the radius
where the free energy of the electrically charged black
hole is zero and give the
results for different 
ensembles and the generalized Buchdahl radius.

%%%%%%%%%%%%%%%%%%%%%%%%%%%%%%%%%%%%%%%%%%%%%%%%%%%%%%%%%%%%%%%%%%%%%%
\section{\label{sec:Canonical1}
The canonical ensemble of a charged 
black hole in a cavity through the Euclidean path integral approach}
%%%%%%%%%%%%%%%%%%%%%%%%%%%%%%%%%%%%%%%%%%%%%%%%%%%%%%%%%%%%%%%%%%%%%%

%%%%%%%%%%%%%%%%%%%%%%%%%%%%%%%%%%%%%%%%%%%%%%%%%%%%%%%%%%%%%%%%%%%%%%
\subsection{\label{sec:partition}Partition function as a Euclidean 
path integral}
%%%%%%%%%%%%%%%%%%%%%%%%%%%%%%%%%%%%%%%%%%%%%%%%%%%%%%%%%%%%%%%%%%%%%%

In statistical mechanics, the canonical ensemble of a system is a
statistical ensemble of possible configurations of the system in
thermodynamic equilibrium with a reservoir of temperature $T$, with 
fixed particle number and unspecified energy. Through the canonical 
ensemble, it is possible to obtain the thermodynamic properties of 
the system in equilibrium with the heat reservoir. The quantity that 
holds all the thermodynamic information of the canonical ensemble 
is the partition function, $Z = \sum_i {\rm e}^{-\beta E_i}$, where the 
sum of all the possible states $i$ is done, $\beta$ is the
ensemble inverse 
temperature, $\beta=\frac{1}{T}$, and
$E_i$ is the energy of each state $i$.

When the canonical ensemble is 
applied to a quantum system, one can calculate the partition function 
as
$Z = {\rm Tr}\,({\rm e}^{-\beta H}) =
\sum_i \bra{\psi_i}{{\rm e}}^{- \beta H}\ket{\psi_i}$, 
where ${\rm Tr}(\mathrm{e}^{-\beta H})$ is the trace of the quantum operator
${\rm e}^{-\beta H}$, $H$ is the 
Hamiltonian of the system and the $\psi_i$ are a basis of a Hilbert
space, not necessarily the Hamiltonian eigenstates. 
Consider now a quantum system to be in a state $\psi$ 
at time $t_1$ and in a state $\hat\psi$ at time $t_2$. Then, the 
amplitude of a system to evolve from the state $\psi$ to $\hat\psi$ 
is $\bra*{\hat\psi,t_2}\ket{\psi,t_1} 
= \bra*{\hat\psi}{ {\rm e}}^{-i H(t_2 - t_1)}
\ket{\psi}$, 
which can be calculated by the Feynman path integral approach 
as $\bra*{\hat\psi,t_2}\ket{\psi,t_1} = \int d[\psi] {\rm e}^{i I[\psi]}$, 
where the functional integration on $\psi$ is done from 
$\psi(t_1) = \psi$ to $\psi(t_2) = \hat\psi$. If we now evaluate 
the amplitude of the system to evolve from a state $\psi$ to 
the same state $\psi$ in a time interval 
$(t_2 - t_1) = -i \beta$ and sum the amplitudes for all the 
basis states, we have that this sum is the partition function 
now written in the 
Euclidean path integral form 
$Z = {\rm Tr}({\rm e}^{-\beta H}) = \int d\psi {\rm e}^{-I[\psi]}$, 
where $I$ is now the Euclidean action of the system and the 
integration is done for every possible periodic function $\psi$ 
with period $\beta=\frac1T$.
This is the Euclidean path integral approach to 
construct the canonical ensemble.

It is reasonable to extend this approach to self-gravitating systems.
Moreover, such extension provides a way to describe quantum gravity, 
i.e., the Euclidean path integral approach to quantum gravity.
The partition function is then given by the Euclidean path integral
\begin{align}
Z = \int DgD\phi {\rm e}^{-I[g,\phi]}\,,
\label{eq:partigionfunctionactiongrav}
\end{align}
where
$g$ is the 
Euclidean metric obtained from the Lorentzian metric by making a 
Wick transformation $t= -i\tau$, i.e., time is Euclideanized,
$\phi$ represents all the matter fields that might be
present in the system, and $Dg$ and $D\phi$ mean
integration measures of the path integral whose structures
are not of concern here.
Here, we construct the canonical ensemble by the Euclidean path 
integral approach to a spherically symmetric
electrically charged 
black hole inside a cavity, in arbitrary $d$ dimensions. 
The system will be in equilibrium with
a heat reservoir at the boundary of the cavity 
with fixed inverse temperature $\beta$, which is given by the total 
Euclidean proper time of the boundary of the cavity, and with fixed 
electric flux, i.e., with the black hole electric charge $Q$ fixed. 
The thermodynamics of the system can
then be obtained by considering that 
the partition function of the canonical ensemble is tied to the 
Helmholtz free energy $F$ through $Z = {\rm e}^{-\beta F}$, i.e.,
$\beta F=-\ln Z$.
With the free energy determined, the other thermodynamic
quantities are obtained by the derivatives of the free energy,
noting that 
$F = E - TS$, where $E$ is the thermodynamic energy
of the system and $S$ its entropy.

%%%%%%%%%%%%%%%%%%%%%%%%%%%%%%%%%%%%%%%%%%%%%%%%%%%%%%%%%%%%%%%%%%%%%%
\subsection{The Euclidean action for the canonical ensemble}
%%%%%%%%%%%%%%%%%%%%%%%%%%%%%%%%%%%%%%%%%%%%%%%%%%%%%%%%%%%%%%%%%%%%%%

The Euclidean action of the system consisting of an
electrically charged black hole 
in a cavity in $d$ dimensions is
\begin{align}
    I =& - \frac{1}{16\pi}\int_M R \sqrt{g}d^d x 
- \frac{1}{8\pi} \int_{\partial M} (K-K_0)\sqrt{\gamma} d^{d-1}x\nonumber\\
&+ \frac{(d-3)}{4\Omega_{d-2}}\int_M F_{ab}F^{ab}\sqrt{g}d^dx\nonumber\\ 
    & + \frac{(d-3)}{\Omega_{d-2}}\int_{\partial M}F^{ab}A_{a}n_b 
    \sqrt{\gamma}d^{d-1}x\,,
    \label{eq:action1}
\end{align}
where $R$ is the Ricci scalar given by derivatives and second 
derivatives of the Euclidean metric $g_{ab}$, 
$g$ is the determinant of $g_{ab}$, 
$K$ is the trace of the 
extrinsic curvature of the boundary of the cavity 
defined as $K_{ab}$, 
$K_0$ is the trace of the 
extrinsic curvature of the boundary of the cavity 
embedded in flat Euclidean space, 
$\gamma$
is the determinant of
the induced metric $\gamma_{\alpha\beta}$
on the boundary of the cavity,
$\Omega_{d-2}$ is the surface area of a
$d-2$  unit sphere and appears here basically
for practical purposes,  
$F_{ab} = \partial_a A_b - \partial_b A_a$ is the Maxwell
tensor given by derivatives of the vector
potential $A_a$, 
$n_b$ is the outward unit normal vector to the boundary
of the cavity,
$a,b$ are spacetime indices that run from 0 to $d-1$ in the
usual manner, and $\alpha,\beta$ are indices on the
boundary that run from 0 to $d-2$.
The boundary term depending on the Maxwell tensor 
must be present so that the canonical 
ensemble may be prescribed, see \cite{Braden:1990}. 
This term allows us to fix the electric flux given by the 
integral of the Maxwell tensor on a $(d-2)$-surface, 
which has the meaning of electric charge. Otherwise, the potential vector 
$A_a$
must be fixed, which means the grand canonical
ensemble should be prescribed, see \cite{Fernandes:2023} for this case.

%%%%%%%%%%%%%%%%%%%%%%%%%%%%%%%%%%%%%%%%%%%%%%%%%%%%%%%%%%%%%%%%%%%%%%
\subsection{Geometry,  electromagnetic field, and boundary conditions}
%%%%%%%%%%%%%%%%%%%%%%%%%%%%%%%%%%%%%%%%%%%%%%%%%%%%%%%%%%%%%%%%%%%%%%

\subsubsection{Geometry and boundary conditions}

We assume that the Euclidean path integral is done along metrics 
which are spherically symmetric. Therefore, the Euclidean space is 
given by the warped space product $\mathbb{R}^2\times \mathbb{S}^{d-2}$
with the warping function $r^2$, where $\mathbb{S}^{d-2}$ is a 
$(d-2)$-sphere with radius $r$. The 
Euclidean metric of such space is given by
\begin{align}
ds^2 = b^2(y) d\tau^2 + \alpha^2(y) dy^2 + r^2(y)d\Omega_{d-2}^2\,,
\label{eq:Euclideanmetric}
\end{align} 
where 
$\tau$ is the periodic Euclidean time with 
range $0\leq\tau<2\pi$, $y$ is a
radial coordinate with range $0\leq y\leq 1$,
$d\Omega_{d-2}^2$ is the line element of the unit 
$(d-2)$-sphere with total area 
$\Omega_{d-2} = \frac{2 \pi^{\frac{d-1}{2}}}{\Gamma(\frac{d-1}{2})}$, 
$\Gamma$ being the gamma function,
$b(y)$ and $\alpha(y)$ are functions of
$y$, and $r(y)$ 
represents the radius that gives the area of the $(d-2)$-sphere.
The functions 
$b(y)$, $\alpha(y)$, and $r(y)$ are unspecified for now and
are to be integrated in the path integral.

The hypersurface $y=0$ is assumed to correspond 
to the bifurcation two-surface 
of the event horizon of the 
charged black hole, so we must impose the conditions
\begin{align}
&  b(0) = 0\,, \label{eq:condb0}\\
&r(0) = r_+\label{eq:condrp}\,,
\end{align}
where $r_+$ is the horizon radius. 
The conditions
given in Eqs.~\eqref{eq:condb0} and~\eqref{eq:condrp}  
impose that the $y=0$ hypersurface corresponds 
to $\{y=0\}\times\mathbb{S}^{d-2}$, 
i.e., a point times a $(d-2)$-sphere.
The $y=0$ point in the $(\tau,y)$ sector
coincides with the central point 
of the $\mathbb{R}^2$ plane in polar coordinates, 
where $\tau$ is in fact the angular coordinate
and $y$ is the radial coordinate of the plane. 
The $y=0$ hypersurface can be seen as 
the limit $y\rightarrow 0$ of $y$ constant hypersurfaces,
with these latter having  an
$\mathbb{S}^1 \times \mathbb{S}^{d-2}$ topology. 
For the metric to be smooth, as $y$ goes to zero, the 
constant $y$ hypersurfaces $\mathbb{S}^1 \times \mathbb{S}^{d-2}$
must go smoothly to $\{y=0\}\times\mathbb{S}^{d-2}$.
There are more conditions other than the two above,
see Appendix 
\ref{bcRicciandEuler}
for a detailed derivation of these
conditions. 
One of the conditions for smoothness is 
\begin{align}
\left(\frac{b'}{\alpha}\right)(0)= 1\,,
    \label{eq:condbprime0}
\end{align}
where
$\left(\frac{b'}{\alpha}\right)(0)
\equiv\left(\frac{b'}{\alpha}\right)_{\hskip-0.1cm y=0}$.
This is a third condition and means
there are no conical singularities in the Euclidean manifold. 
With Eq.~\eqref{eq:condbprime0} considered, one can compute 
the Ricci scalar $R$ of the metric of
Eq.~\eqref{eq:Euclideanmetric}
when the metric is expanded
 around $y=0$ with the
conditions given
in Eqs.~\eqref{eq:condb0} and \eqref{eq:condrp}.
One obtains that to have a well-defined Ricci scalar $R$
and for the space to be smooth, one must impose 
the fourth and fifth conditions
\begin{align}
   & \left(\frac{r'}{\alpha}\right)(0) = 0\,,
    \label{eq:condalphar0}\\
    & \left(\frac{1}{\alpha}\left(\frac{b'}{\alpha}\right)'\right)
    \hskip-0.1cm (0) = 0\,,
        \label{eq:condb2prime0}
\end{align}
with $\left(\frac{r'}{\alpha}\right)(0)
\equiv\left(\frac{r'}{\alpha}\right)_{\hskip-0.1cm y=0}$
and
$ \left(\frac{1}{\alpha}\left(\frac{b'}{\alpha}\right)'\right)
\hskip-0.1cm (0)
\equiv
\left(\frac{1}{\alpha}\left(\frac{b'}{\alpha}\right)'\right)
_{\hskip-0.1cm y=0}$,
see Appendix 
\ref{bcRicciandEuler}.
The condition given in
Eq.~\eqref{eq:condalphar0} 
is 
equivalent, in even dimensions,
to requiring that the Euclidean space considered
has an Euler characteristic $\chi = 2$ by the Chern-Gauss-Bonnet 
formula. On the other hand,
in odd dimensions, the Euler characteristic vanishes, and 
so this requirement is not satisfactory. Nevertheless, the 
requirement that the Ricci scalar is well-defined suffices. 
One can also see that the condition
given in Eq.~\eqref{eq:condalphar0} is equivalent to requiring 
that the event horizon
of the black hole is a null hypersurface, if one performs a
Wick transformation back to the Lorentzian signature.
The condition given in Eq.~\eqref{eq:condb2prime0}
means for some coordinate $y$,  
that if $(b'\alpha^{-1})'|_{y=0}$ is nonzero finite, then 
$\alpha|_{y=0}$
must diverge. Indeed, this is satisfied by the 
Reissner-Nordstr\"om line element with coordinate choice $y=r$ 
found by solving the Einstein
equation, as it is seen below. We note that the condition 
given in
Eq.~\eqref{eq:condb2prime0} is not referred to
elsewhere, in particular
it is not mentioned in \cite{Braden:1990,Fernandes:2023}.

The hypersurface $y=1$ corresponds to the boundary of the cavity,
where two conditions are imposed
\begin{align}
&  b(1)=\frac{\beta}{2\pi}\,, \label{eq:betacond}\\
&   r(1) = R\label{eq:riR}\,.
\end{align}
The condition given by
 Eq.~\eqref{eq:betacond}
fixes $\beta$ at the boundary, with
$\beta$ being
the inverse temperature of the cavity,
$\beta=\frac1T$.
This condition enforces
that the total Euclidean proper time of the boundary of the cavity is fixed 
to be equal to the inverse temperature of the cavity, and the
condition comes from the definition of the path integral 
as stated in Sec.~\ref{sec:partition}.
The condition given by
Eq.~\eqref{eq:riR}
states that the boundary is at radius $R$.

\subsubsection{Electromagnetic field and boundary conditions}

For the electromagnetic Maxwell field, 
due to spherical symmetry and admitting the nonexistence of 
magnetic monopoles, the only nonvanishing components of the 
Maxwell tensor $F_{ab}$ are $F_{y\tau} = -F_{\tau y}$. Moreover,
we choose a gauge where the only nonvanishing component of the 
vector potential is $A_{\tau}(y)$. Therefore, the Maxwell tensor 
$F_{ab}$ is described 
only by the term
\begin{align}
    F_{y\tau}(y) = \frac{d A_\tau(y)}{dy}\,.
    \label{eq:FytauAtau}
\end{align}
The boundary conditions can now be imposed.

At $y=0$, 
we require that 
\begin{align}
    A_{\tau}(0) = 0\,,\label{eq:Atau0}
\end{align}
to have regularity. This condition also fixes completely the gauge of 
the Maxwell field.

At $y=1$, we fix the electric charge by specifying the electric 
flux given by $\int_{\substack{y=1 \\ \tau = c}} F^{ab}dS_{ab} =
2 i\Omega_{d-2} Q$,
where $c$ is a constant, $Q$ is the electric charge in the cavity, 
$dS_{ab} = 2 u_{[a} n_{b]}dS$ is the surface 
element of the $y=1$ and $\tau = 0$ surface, 
$u_a dx^a = b d\tau$, $n_a dx^a = \alpha dy$, and $dS$ is 
the surface volume,
i.e.,
\begin{align}
    &  \Bigl(b \alpha 
    r^{d-2}\,F^{y\tau}\Bigr) (1) =
    - i Q\,.\label{eq:Fcond}
\end{align}

%%%%%%%%%%%%%%%%%%%%%%%%%%%%%%%%%%%%%%%%%%%%%%%%%%%%%%%%%%%%%%%%%%%%%%
\subsection{The action with boundary conditions}
%%%%%%%%%%%%%%%%%%%%%%%%%%%%%%%%%%%%%%%%%%%%%%%%%%%%%%%%%%%%%%%%%%%%%%

Putting together the conditions just found into the action 
Eq.~\eqref{eq:action1},
one finds that it
is a function of
the radius of the cavity $R$, the inverse temperature 
$\beta$, and the charge $Q$, which are
the fixed quantities of the system, and a
functional of $b$, $\alpha$, $r$ and $A_\tau$.
The partition function is
then given at this stage
by the action appropriately
integrated in all paths in the path integral.
We  now evaluate the action Eq.~\eqref{eq:action1} with the
considered boundary conditions. Here we sketch the calculation, see
Appendix \ref{bcRicciandEuler}
for full details.

We start by looking at the Ricci scalar $R$.
The Ricci scalar $R$ is the contraction of
the Ricci tensor $R_{ab}$ which itself is the
contraction of the Riemann tensor, and moreover,
one can form the Einstein tensor 
 $G_{ab}$ from $R_{ab}$ and $R$, 
$G_{ab}=R_{ab}-\frac12 g_{ab}R$.
The
Ricci scalar for the metric in Eq.~\eqref{eq:Euclideanmetric} is 
given by
$-\frac{1}{16\pi}R = 
    \frac{1}{8\pi \alpha b r^{d-2}}\left( \frac{r^{d-2}b'}{\alpha}\right)'
    + \frac{1}{8\pi}{G^{\tau}}_{\tau}$,
where ${G^{\tau}}_{\tau}$ is the time-time component of the 
Einstein tensor and is given by 
${G^{\tau}}_{\tau} = \frac{(d-2)}{2 r' r^{d-2}}\left[r^{d-3}
\left(\frac{r'^2}{\alpha^2} -1\right) \right]'$,
and the prime means derivative with respect to $y$.
By putting the expression of the Ricci scalar into 
 Eq.~\eqref{eq:action1},
we observe that the first term in the volume integration of the 
Ricci scalar
yields 
$\frac{\Omega_{d-2}}{4}\left(\frac{r^{d-2}b'}{\alpha}\right)_{y=1} 
- \frac{\Omega_{d-2}}{4}\left(\frac{r^{d-2}b'}{\alpha}\right)_{y=0}$, 
i.e., a boundary term at $y=1$ and a boundary term at $y=0$. 
The
term $-\int_{y=1} \frac{\sqrt{\gamma}}{8\pi}(K-K_0) d^{d-1}x$
is called the
Gibbons-Hawking-York 
boundary term, and is given by
$-\int_{y=1} \frac{\sqrt{\gamma}}{8\pi}(K-K_0) d^{d-1}x 
=  \left(\frac{2\pi b r^{d-3}}{\mu}\left(1 -
\frac{r'}{\alpha}\right)\right)_{y=1}
- \frac{\Omega_{d-2}}{4}\left(\frac{r^{d-2}b'}{\alpha}\right)_{y=1}$,
where it was used that the extrinsic curvature of a constant $y$ 
hypersurface is $\mathbf{K} = \frac{bb'}{\alpha}d\tau 
+ \frac{rr'}{\alpha}d\Omega_{d-2}^2$, and
$\mathbf{K}_0=rd\Omega_{d-2}^2$
is the extrinsic curvature of the hypersurface embedded in flat 
space and $\mu = \frac{8\pi}{(d-2)\Omega_{d-2}}$.
The last term 
of the Gibbons-Hawking-York 
boundary term
cancels 
with the 
boundary term at $y=1$ of the Ricci scalar.
Moreover, by using the boundary condition 
Eq.~\eqref{eq:condbprime0}, the remaining boundary term of 
the volume integral of the Ricci scalar 
becomes 
$- \frac{\Omega_{d-2}}{4}\left(\frac{r^{d-2}b'}{\alpha}\right)_{y=0} 
= -\frac{\Omega_{d-2}}{4}r^{d-2}_+$.
It is useful to rewrite the Maxwell boundary term in the action
Eq.~\eqref{eq:action1}. Using the 
divergence theorem and that $\nabla_b(F^{ab}A_a) = - \frac{1}{2}
F_{ab}F^{ab} + A_a\nabla_bF^{ab} $, one transforms the boundary 
Maxwell term into bulk terms, obtaining that the Maxwell
part of the action is
$- \frac{(d-3)}{4\Omega_{d-2}}\int_M F_{ab}F^{ab}\sqrt{g}d^dx
+ \frac{(d-3)}{\Omega_{d-2}}\int_M  A_a\nabla_b F^{ab} \sqrt{g}d^dx$.
Now, 
$-\frac{(d-3)}{4\Omega_{d-2}}F_{ab}F^{ab} = 
    -\frac{(d-3)}{2\Omega_{d-2}} 
    \frac{{A'_{\tau}}^2}{\alpha^2 b^2}$,
where $F_{y\tau} =$ $A'_\tau$ was used, and
$\frac{(d-3)}{\Omega_{d-2}}\nabla_b F^{ab}A_{a}
= -\frac{(d-3)}{\Omega_{d-2} \alpha b r^{d-2}} 
\left(\frac{r^{d-2}A'_{\tau}}{b\alpha}\right)'
A_{\tau}$,
where $\nabla_a F^{\tau a} = 
-\frac{1}{\alpha b r^{d-2}}\left(\frac{r^{d-2}A'_\tau}{\alpha b}\right)'$
was used.

One can proceed with the integrations 
at the cavity, since the integrands do not depend on time or on 
the angles, and one obtains
from Eq.~\eqref{eq:action1} 
the full action 
\begin{align}
    &I[\beta,Q,R;b,\alpha,r,A_\tau] = \frac{\beta R^{d-3}}{\mu}
    \left(1 - \left(\frac{r'}{\alpha}\right)(1)\right) \notag \\
    &- \frac{\Omega_{d-2}}{4}r_+^{d-2} -\frac{(d-3)}{\Omega_{d-2}}
    \int_M \left(\frac{r^{d-2}A'_\tau}{b\alpha}\right)'
    A_\tau d\tau dy d\Omega_{d-2} \notag\\
    &  + 
    \int_M \frac{\alpha b r^{d-2}}{8\pi}
    \left({G^\tau}_\tau 
    - \frac{4\pi(d-3)}{\Omega_{d-2}}\frac{{A'_\tau}^2}
    {\alpha^2 b^2}\right)d\tau dy d\Omega_{d-2}\,,
    \label{eq:actionwithcond}
\end{align}
where it was used that the time length at the cavity is given by 
Eq.~\eqref{eq:betacond}, i.e., $\beta = 2\pi b(1)$, and that
from Eq.~\eqref{eq:riR} one has
$r(1)=R$. We have then 
the action as a functional of $b$, $\alpha$, $r$ and $A_\tau$
to be integrated in all paths, in the path integral.

%%%%%%%%%%%%%%%%%%%%%%%%%%%%%%%%%%%%%%%%%%%%%%%%%%%%%%%%%%%%%%%%%%%%%%
\section{\label{sec:ZeroLoopSolutions}The zero loop approximation: 
reduced action, solutions, and stability}
%%%%%%%%%%%%%%%%%%%%%%%%%%%%%%%%%%%%%%%%%%%%%%%%%%%%%%%%%%%%%%%%%%%%%%

%%%%%%%%%%%%%%%%%%%%%%%%%%%%%%%%%%%%%%%%%%%%%%%%%%%%%%%%%%%%%%%%%%%%%%
\subsection{Constraints and the reduced action}
%%%%%%%%%%%%%%%%%%%%%%%%%%%%%%%%%%%%%%%%%%%%%%%%%%%%%%%%%%%%%%%%%%%%%%

Due to the aforementioned difficulties in dealing with the path integral, 
we perform the zero loop approximation. We do this in steps. First, we find
the reduced action by imposing  
the Hamiltonian and momentum constraints to the metric and the 
Gauss constraint to the Maxwell field.
Second,  we implement the zero loop 
approximation, i.e., we only consider the path that minimizes the 
reduced action.

The Hamiltonian constraint is 
${G^{\tau}}_{\tau} = 8\pi {T^{\tau}}_\tau$, with
${G^\tau}_\tau$ given by
${G^{\tau}}_{\tau} = \frac{d-2}{2 r' r^{d-2}}\left[r^{d-3}
\left(\frac{r'^2}{\alpha^2} -1\right) \right]'$,
and
${T^{\tau}}_{\tau} = \frac{(d-3)}{\Omega_{d-2}}
\frac{A'^{2}_\tau}{2\alpha^2 b^2}$,
where ${T^\tau}_\tau$ is the time-time
component of the stress-energy tensor ${T^a}_b$. 
Thus, the Hamiltonian constraint is
\begin{align}
\frac{d-2}{ 2r' r^{d-2}}\left[r^{d-3}
\left(\frac{r'^2}{\alpha^2} -1\right) \right]' = 
\frac{4\pi (d-3) A'^{2}_\tau}{\Omega_{d-2}\alpha^2 b^2}\,.
\label{eq:Hamconsteq}
\end{align}
The momentum constraint is trivially satisfied since 
the metric Eq.~\eqref{eq:Euclideanmetric} is diagonal 
and does not depend on 
the Euclidean time.
The Gauss constraint is $\nabla_y F^{\tau y} = 0$,
which explicitly is 
\begin{align}
\left(\frac{r^{d-2}A'_\tau}{b\alpha}\right)' =0\,,
\label{eq:Gaussconstr0}
\end{align}
The two constraint equations, Eqs.~\eqref{eq:Hamconsteq}
and \eqref{eq:Gaussconstr0},
are coupled, nevertheless they can be integrated.
It is better to 
start first by integrating Eq.~\eqref{eq:Gaussconstr0}.
Its integration yields 
\begin{align}
    A'_\tau = -i\frac{ q\, }{r^{d-2}}b\alpha\,,
    \label{eq:Gaussconstr}
\end{align}
where $q$ is an integration constant. If one evaluates 
Eq.~\eqref{eq:Gaussconstr} at $y=1$ and uses the boundary 
condition Eq.~\eqref{eq:Fcond}, then one obtains that 
\begin{align}
    q=Q\,\,,\label{eq:qQ}
\end{align}
and so the integration constant $q$ of the Gauss constraint is 
precisely the fixed electric charge $Q$ of the ensemble. 
From this point onward we work with $Q$. By using 
Eq.~\eqref{eq:Gaussconstr} and Eq.~\eqref{eq:qQ}, the Hamiltonian 
constraint becomes $\frac{d-2}{2 r' r^{d-2}}\left[r^{d-3}
\left(\frac{r'^2}{\alpha^2} -1\right) \right]' = 
- \frac{4\pi (d-3) Q^2}{\Omega_{d-2}r^{2d-4}}$, which can be 
integrated to obtain
\begin{align}
\frac{r'^2}{\alpha^2} \equiv f(r,Q,r_+) 
\,,\label{eq:rprimealpha0}
\end{align} 
where
\begin{align}
f(r,Q,r_+) \equiv 1 - \frac{r_+^{d-3} 
+ \frac{\mu Q^2}{r_+^{d-3}}}{r^{d-3}} + \frac{\mu Q^2}{r^{2d-6}}
\,,
\label{eq:rprimealpha}
\end{align} 
with
\begin{align}
\mu = \frac{8\pi}{(d-2)\Omega_{d-2}}\,.
\label{eq:mu}
\end{align}
The function 
$f$ in Eq.~\eqref{eq:rprimealpha}
is defined for convenience, and the regularity 
conditions Eqs.~\eqref{eq:condrp} and~\eqref{eq:condalphar0} 
were used to determine the integration constant $r_+$.
Although 
the condition Eq.~\eqref{eq:condb2prime0} is not used anywhere, 
notice for bookkeeping that, 
if $y=r$ is chosen, $r' = 1$ and $\alpha$ diverges at 
$r=r_+$, therefore the condition Eq.~\eqref{eq:condb2prime0}
should be satisfied if
$\left(\frac{b'}{\alpha}\right)'_{\hskip-0.1cm y=0}$
is finite. 
The function 
$A'_\tau$ in Eq.~\eqref{eq:Gaussconstr} is related to the 
Coulomb electric field in Lorentzian curved spacetime as
$n_a E^a = \frac{i A'_\tau}{b\alpha} = \frac{Q}{r^{d-2}}$, 
where $E^a$ is the electric field.
It is important to write explicitly the extremal case,
i.e., when $
r_{+}^{2d-6}=
\mu Q^2 $,
and we write this special radius as ${r_{+}}_e$,
which is thus
given by
\begin{align}
{r_{+}}_e=\left(\mu Q^2\right)^{\frac{1}{2d-6}} 
\,.
\label{eq:rprimealphaextremal}
\end{align}
The function $f(r,Q,r_+)$ in Eq.~\eqref{eq:rprimealpha}
in the extremal case is then
$f(r,Q,{r_{+}}_e) = \left( 1 
- \frac{\sqrt{\mu}Q}{r^{d-3}}\right)^2$.

The Hamiltonian, momentum, and Gauss
constraints
simplify
the action of Eq.~\eqref{eq:actionwithcond} considerably. 
One can see that the third term in Eq.~\eqref{eq:actionwithcond}
has an integrand proportional to ${G^\tau}_\tau - 8\pi{T^\tau}_\tau$
and so, applying the Hamiltonian constraint
given in Eq.~\eqref{eq:Hamconsteq}, this term vanishes.
Moreover, the last term in Eq.~\eqref{eq:actionwithcond} is 
proportional to $\left(\frac{r^{d-2}A'_\tau}{b\alpha}\right)'$
which vanishes also if the Gauss constraint
given in Eq.~\eqref{eq:Gaussconstr0}
is applied.
Therefore, the action Eq.~\eqref{eq:actionwithcond} becomes
\begin{align}
    I_*[\beta,Q,R;r_+] =\!\frac{\beta R^{d-3}}{\mu}(1 - \sqrt{f(R,Q,r_+)}) 
    \!-\! \frac{\Omega_{d-2} r_+^{d-2}}{4}\,,
    \label{eq:actioninrp}
\end{align}
where $I_*$ is  the reduced action, which is 
the Euclidean action 
evaluated on the paths that obey the Hamiltonian and Gauss 
constraints, and $(r'\alpha^{-1})_{y=1}$ was substituted by the 
solution to the Hamiltonian constraint 
given in Eq.~\eqref{eq:rprimealpha0}.
From Eq.~\eqref{eq:rprimealpha}
we have that $f(r,Q,r_+)$ appearing in 
 Eq.~\eqref{eq:actioninrp} evaluated at the cavity
 radius $R$ is given by
\begin{align}
f(R,Q,r_+) \equiv 1 - \frac{r_+^{d-3} 
+ \frac{\mu Q^2}{r_+^{d-3}}}{R^{d-3}} + \frac{\mu Q^2}{R^{2d-6}}
\,.
\label{eq:Rprimealpha}
\end{align} 
The extremal case characterized by Eq.~\eqref{eq:rprimealphaextremal}
has this function at $R$ given by 
$f(R,Q,{r_{+}}_e)  = \left( 1 
- \frac{\sqrt{\mu}Q}{R^{d-3}}\right)^2$.

The Hamiltonian, momentum, and Gauss
constraints, 
together with the boundary conditions and the 
requirement of spherical symmetry, restrict the path integral 
substantially. The Euclidean space is determined by the functional 
$r_+$ and so the path integral is the sum of spaces with all 
possible $r_+$.
Indeed, the partition function is given by the path integral,
i.e.,
\begin{align}
Z = \int Dr_+ {\rm e}^{-I_*[\beta,Q,R;r_+]}\,,
\label{eq:partitionf1}
\end{align}
with $I_*[\beta,Q,R;r_+]$ 
being the reduced action described in Eq.~\eqref{eq:actioninrp}. 
There is formally another functional which is 
$A_\tau$, i.e., the Maxwell field, but the action does not depend
explicitly on $A_\tau$, it only depends on the electric charge 
which is fixed at the cavity. This means the integration over paths 
of $A_\tau$ can be absorbed by a normalization and does not yield 
any contribution to the constrained path integral.

%%%%%%%%%%%%%%%%%%%%%%%%%%%%%%%%%%%%%%%%%%%%%%%%%%%%%%%%%%%%%%%%%%%%%%
\subsection{Reduced action evaluated at stationary points, $I_0$:
Analytic investigation of the existence of stationary points in
generic $d$ dimensions\label{subsec:Zero}}
%%%%%%%%%%%%%%%%%%%%%%%%%%%%%%%%%%%%%%%%%%%%%%%%%%%%%%%%%%%%%%%%%%%%%%

\subsubsection{Reduced action evaluated at stationary points, $I_0$}

To further simplify the path integral in the partition function
of Eq.~\eqref{eq:partitionf1}, we perform the zero loop 
approximation, i.e., we only consider the path that minimizes the 
action given in Eq.~\eqref{eq:actioninrp}. The partition function
in this approximation is given by
\begin{align}
Z[\beta, R, Q] = {\rm e}^{-I_{0}[\beta,R,Q]}\,,
\label{eq:partitionf2}
\end{align}
where
\begin{align}
I_{0}[\beta,R,Q] = I_*[\beta,R,Q; r_+[\beta,R,Q]]\,,
\label{eq:I0}
\end{align}
is the 
action in Eq.~\eqref{eq:actioninrp} evaluated at its minimum with 
respect to $r_+$. The function $r_+[\beta,R,Q]$ corresponds
to a black hole solution that is in equilibrium with the cavity and 
it is determined by
a stationary point of the action, i.e.,
$\left(\frac{\partial I_*}{\partial r_+}\right)_{r_+=r_+[\beta,R,Q]} = 0$.

\subsubsection{Equations for the $d$-dimensional stationary points}

Thus, the stationary points of the reduced action $I_*$,
given by
$\left(\frac{\partial I_*}{\partial r_+}\right)_{r_+=r_+[\beta,R,Q]} = 0$,
can be found through Eq.~\eqref{eq:actioninrp}
to give
\begin{align}
\beta = \iota(r_+)\,,\,
\iota(r_+)\equiv  \frac{4\pi}{(d-3)}\frac{r_+^{d-2}}{r_+^{d-3}
- \frac{\mu Q^2}{r_+^{d-3}}}\sqrt{f(R,Q,r_+)}\,,
\label{eq:beta1}
\end{align}
where $\iota(r_+)$
is the inverse temperature function, defined here for convenience.
Note that for fixed $R$ and $Q$, 
$\iota$ is indeed a function of $r_+$ alone.
The solutions $r_+[\beta,R,Q]$ of Eq.~\eqref{eq:beta1}
are stationary points
of the action in Eq.~\eqref{eq:actioninrp}
and are obtained from inverting 
Eq.~\eqref{eq:beta1}. To help in the analysis,
we define in this section
a horizon radius parameter $x$ and an electric
charge parameter $y$ as
\begin{align}
x = \frac{r_+}{R}\,,\quad\quad\quad\quad
y= \frac{\mu Q^2}{R^{2d-6}}
\,.\label{eq:newv}
\end{align}
Rearranging
Eq.~\eqref{eq:beta1} we obtain
\begin{align}
\hskip -0.20cm
(x^{2d-6}
\hskip -0.15cm
-
\hskip -0.08cm
y)^2
\hskip -0.08cm
\left(
\hskip -0.08cm
\frac{(d-3)\beta}{4\pi R}
\hskip -0.08cm\right)^2
\hskip -0.25cm
-
\hskip -0.05cm
x^{3d-7} (1
\hskip -0.08cm
-
\hskip -0.08cm
x^{d-3})(x^{d-3}
\hskip -0.15cm
-
\hskip -0.08cm
y) 
\hskip -0.1cm
=
\hskip -0.1cm
0.\hskip -0.1cm
\label{eq:sols1}
\end{align}
The equation above, Eq.~\eqref{eq:sols1},
can be reduced at most to sixth polynomial order 
for $d=5$, while for other dimensions 
the polynomial order is higher. Therefore, we did not find any 
analytical solution to this equation for
any specific value of $d$. 

It should be noted that
the nonextremal condition for the black hole
is best seen by putting in the form
\begin{align}
x_e\leq x \leq 1\,, 
\label{eq:extremal1}
\end{align}
where $x_e$ is the extremal $x$ related to the extremal $y$,
denoted as $y_e$,
by 
\begin{align}
y_e= x_e^{2d-6}\,,
\label{eq:extremal2}
\end{align}
see Eq.~\eqref{eq:rprimealphaextremal}.

%%%%%%%%%%%%%%%%%%%%%%%%%%%%%%%%%%%%%%%%%%%%%%%%%%%%%%%%%%%%%%%%%%%%%%
\subsubsection{Saddle points of the action in $d$ dimensions}
%%%%%%%%%%%%%%%%%%%%%%%%%%%%%%%%%%%%%%%%%%%%%%%%%%%%%%%%%%%%%%%%%%%%%%

Although it is not
possible to find exact solutions for $x$ and $y$,
nevertheless, we are able to obtain analytically
the limiting values for
the solutions. These are determined by
the saddle points of the action $I_*$ described as 
$\left(\frac{\partial^2 I_*}{\partial r_+^2}\right)_0 = 0$,
where the subscript  $0$ 
means that the quantity inside parenthesis is evaluated at the 
stationary point.
Now, 
$\left(\frac{\partial^2 I_*}{\partial r_+^2}\right)_0 = 
-\frac{\Omega_{d-2} (d-2) r_+^{d-3}}{4} \beta^{-1} 
\frac{\partial \iota}{\partial r_+}$,
so
the saddle points of the action are given by the
equation 
$\frac{\partial \iota}{\partial r_+}=0$
together with Eq.~\eqref{eq:beta1}.
This condition can
 be put as a  
function of the variables $x$ and $y$ and it 
simplifies to 
\begin{align}
    &\frac{d-1}{2}x^{4d-12} - (1+y)x^{3d - 9} - 3(d-3)y x^{2d-6} 
    \nonumber \\
    &+ (2d-5)y(1+y)x^{d-3} -\frac{3d-7}{2}y^2 = 0\,\,.
    \label{eq:critbeta}
\end{align}
Equation~\eqref{eq:critbeta} is a polynomial equation of order four 
in $x^{d-3}$ and it can be solved analytically. 
We enumerate the solutions.

For $y=0$, one has the electrically uncharged case and it was
discussed in \cite{York:1986} for $d=4$,
 \cite{Andre:2020} for $d=5$,
and \cite{Andre:2021} for generic $d$.

For $0<y<y_s$, there are four real roots of Eq.~\eqref{eq:critbeta}, 
from which only two obey the condition $0<y<x^{2d-6}$
which is the nonextremal condition,
see Eq.~\eqref{eq:extremal1}, and where
$y_s$ is a saddle or critical electric charge parameter to be
determined.
We designate the
two saddle points of the action, i.e.,
the
two solutions
of interest of Eq.~\eqref{eq:critbeta}, as $x_{s1} = x_{s1}(y)$ 
and $x_{s2} = x_{s2}(y)$, where $x_{s1}\leq x_{s2}$.
Now, we
find explicitly the saddle points of the action.
They are
\begin{align}
&x_{s1}^{d-3} = \frac{1+y}{2(d-1)} + \xi- 
\frac{1}{2}\sqrt{2\eta + \frac{\zeta}{\xi} 
- 4 \xi^2}\,, \\
&x_{s2}^{d-3} = \frac{1+y}{2(d-1)} + \xi+ 
\frac{1}{2}\sqrt{2\eta + \frac{\zeta}{\xi} 
- 4 \xi^2}\,, 
\label{xs1xs1}
\end{align}
where
\begin{equation}
\begin{aligned}
& \eta = \frac{3(1+y)^2 + 12(d-1)(d-3)y}{2(d-1)^2}\,,\\
& \zeta
\hskip-0.05cm
= 
\hskip-0.05cm
\frac{(1+y)}{(d-1)^3}
\left(y^2 - (4d^3 -24d^2 +48d - 30)y +1 \right)\,,\\
& \xi= \frac{1}{2}\sqrt{\frac{2}{3}\eta 
+ \frac{2}{3(d-1)}\frac{\sigma^2 + \sigma_0}{\sigma}}\,,\\
& \sigma = \left(\frac{\sigma_1 
+ \sqrt{\sigma_1^2 - 4\sigma_0^3}}{2}\right)^{\frac13}\,,\\
& \sigma_0 = 3(2d-5)y(1-y)^2\,,\\
& \sigma_1 = 54 (d-3)(d-2)^2(1-y)^2y^2\,\,.
\end{aligned}
\label{defs}
\end{equation}

\begin{figure}[h]
\centering
\begin{subfigure}{0.5\linewidth}
\centering
\includegraphics[width=\linewidth]{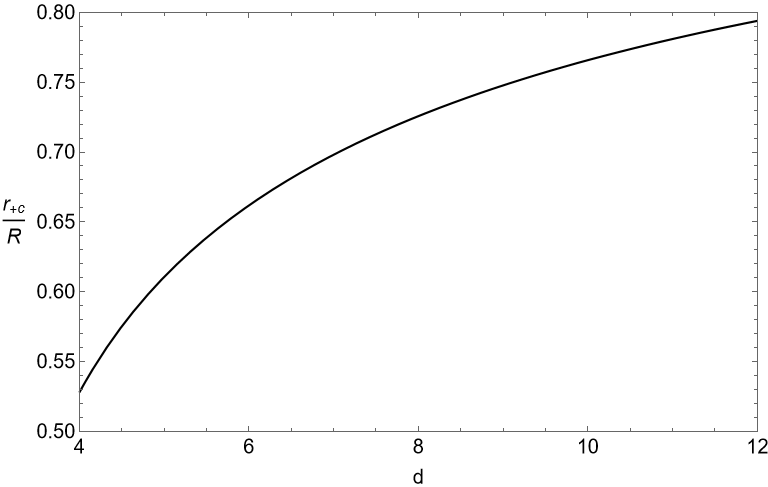}
\caption{\label{fig:rcd}}
\end{subfigure}%
\begin{subfigure}{0.5\linewidth}
\centering
\includegraphics[width=\linewidth]{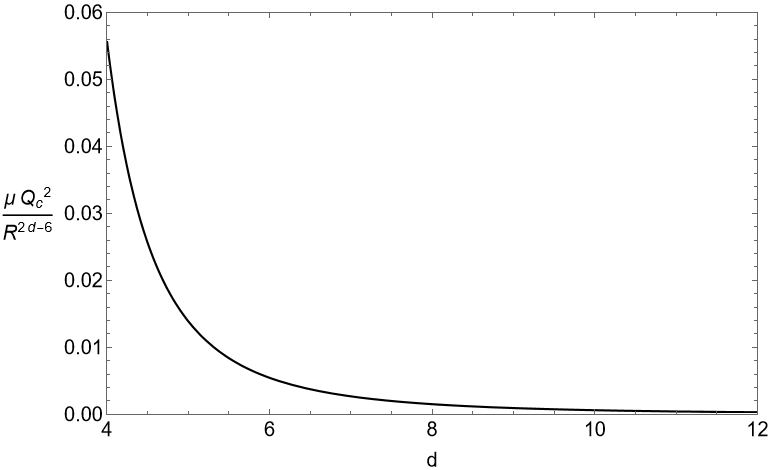}
\caption{\label{fig:qcd}}
\end{subfigure}
\begin{subfigure}{0.5\linewidth}
\centering
\includegraphics[width=\linewidth]{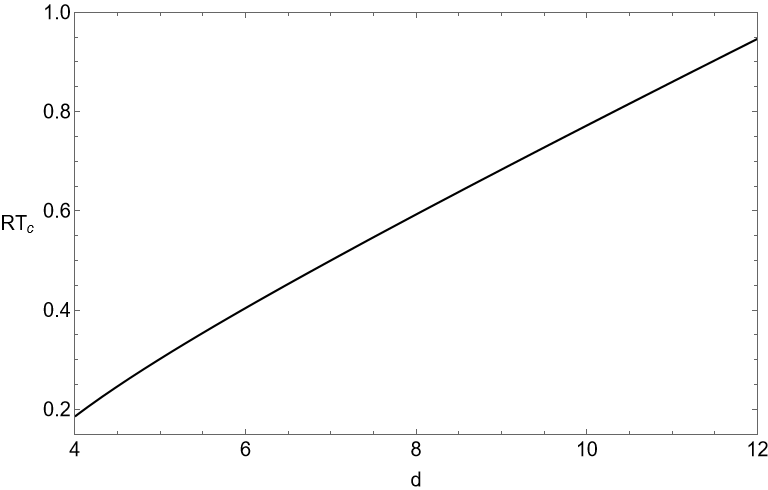}
\caption{\label{fig:Tcd}}
\end{subfigure}
\caption{
Plots of the saddle point
$(x_s,y_s,T_s)$ of the action, where the
third derivative of the action also vanishes, as functions of the
number of dimensions $d$. (a) Plot of $x_s = \frac{r_{+s}}{R}$ as a
function of $d$; (b) plot of $y_s = \frac{\mu Q_s^2}{R^{2d-6}}$ as a
function of $d$; (c) plot of $R T_s$ as a function of $d$. 
\label{fig:saddlej}}
\end{figure}

For $y=y_{s}$, both saddle points merge into a single one.
We write the value of the saddle point 
 $x_{s}\equiv x_{s1}= x_{s2}$ at $y=y_s$, 
 which is a saddle point with the feature
that the third derivative of the action 
also vanishes.
The saddle point $x_{s}\equiv x_{s1}= x_{s2}$ is given by 
\begin{align}
&x_s^{d-3} = \frac{1}{2(d-1)(2d-5)}\nonumber\\
&\times\bigg[(d-1)(3d-7)(3d^2-16d +22)\nonumber\\
& - 3\sqrt{3}(d-2)^2(d-3)\sqrt{(d-1)(3d-7)}\bigg]\,\,
\label{eq:xcd},
\end{align}
which occurs at $y=y_s$ given by
\begin{align}
& y_s = \frac{1}{4(d-1)(2d-5)^3(3d-7)}\nonumber\\
&\times\bigg[(d-1)(3d-7)(3d^2-16d + 22) \nonumber\\
& - 3\sqrt{3}(d-3)(d-2)^2\sqrt{(d-1)(3d-7)}\bigg]^2\,\,.
\label{eq:ycd}
\end{align}
Of course, to $x_s$ corresponds an $r_{+s}$ through
$
r_{+s}=x_sR$, 
 and to 
 $y_s$ corresponds a $Q_{s}$ through
$
Q_s= \frac{y_sR^{2d-6}}{\mu}$, where
we have not put the subscript
$\scriptsize s$ in $R$ in these formulas
because, for finite $R$, one can always assume
$R$ fixed.
Putting the values given in Eqs.~\eqref{eq:xcd}
and \eqref{eq:ycd}
into Eq.~\eqref{eq:sols1},
one finds $RT_s$, 
\begin{align}
RT_s=RT_s(x_s,y_s)\,
\label{eq:RTs}
\end{align}
the temperature parameter at which 
$x_s$ is a solution of the black hole for $y=y_s$.
The values of $x_s$, $y_s$, and $R T_s$
are displayed for different values of $d$ 
in Fig.~\ref{fig:saddlej}.
It can be seen that both $x_s$ and $RT_s$ increase as
$d$ increases, and $y_s$ decreases as $d$ increases.

For $ y_s<y<1$, there are no roots of 
Eq.~\eqref{eq:critbeta} that obey the condition $0<y<x^{2d-6}$
which is the nonextremal condition,
see Eq.~\eqref{eq:extremal1}, i.e., there are no
saddle points of the action.

%%%%%%%%%%%%%%%%%%%%%%%%%%%%%%%%%%%%%%%%%%%%%%%%%%%%%%%%%%%%%%%%%%%%%%
\subsubsection{The solutions in $d$ dimensions: Qualitative analysis}
%%%%%%%%%%%%%%%%%%%%%%%%%%%%%%%%%%%%%%%%%%%%%%%%%%%%%%%%%%%%%%%%%%%%%%

For $y=0$, one has the electrically uncharged case and it has been
discussed in \cite{York:1986} for $d=4$, \cite{Andre:2020} for $d=5$,
and \cite{Andre:2021} for generic $d$.

For $0<y<y_s$,
from the results
for the saddle points of the action,
one can make a qualitative analysis and find that
there are three 
solutions $x(\beta,y)$, or if one prefers $x(T,y)$,
of Eq.~\eqref{eq:sols1}. These three
solutions we designate by $x_1$, $x_2$,
and $x_3$. The solution $x_1$ exists in the interval of 
temperatures $0<T<T_1$
and is bounded by $x_e<x_1(T,y)< x_{s1}(y)$, where 
the values of the solution at the bounds are 
$x_1(0,y)= x_e$,
with $x_e$ defined in Eq.~\eqref{eq:extremal2},
and $x_1(T_1,y) = x_{s1}(y)$, 
with $T_1$ being defined by the latter relation. 
The solution $x_2$ exists in the interval of temperatures
$T_1>  T > T_2$ and is bounded by 
$x_{s1}(y) < x_{2}(T,y)< x_{s_2}(y)$, where the values of the 
solution at the bounds are
$x_{2}(T_1,y) = x_{s1}(y)$ 
and $x_{2}(T_2,y) = x_{s2}(y)$, with $T_2$ being defined 
by the former relation. The solution $x_{3}$ exists in the 
interval of temperatures $T_2 < T<\infty$, and 
is bounded by $x_{s2}(y)<x_3 (T,y)< 1$, where the values of the 
solution at the bounds are 
$x_3 (T_2,y) = x_{s2}(y)$ and $x_3 (T\rightarrow \infty,y) = 1$.
As $y_s$ decreases with the increase of $d$, this means that the 
area of the region of existence of these solutions decreases with 
the increase of $d$, as it is 
squeezed toward lower values of the electric charge.

For $y=y_{s}$, there are still three solutions $x_1$, $x_2$, and $x_3$, 
with the solution $x_2$ having been
reduced to a point, more precisely to the 
saddle point of $\iota(r_+)$ given as $x_2 (T_s,y_s) = x_{s}$, with $T_s$ 
being defined by the latter relation. 
The bounds of $x_1$ and $x_3$ are the same as the case 
$0 < y < y_s$, except that $x_{s1}(y_s) = x_{s2}(y_s) = x_{s}$ and 
$T_s = T_1 = T_2$.

For $y_s < y<1$, there is only one solution $x_4$ that 
exists for all $T$ and is bounded by 
$x_e< x_4 (T,y) < 1$, where $x_4(0,y)
= x_e$ and $x_4 (T\rightarrow \infty,y) = 1$.

\subsection{Stability condition}

To determine if the solutions are minima of the action and thus
stable, we must go beyond the zero loop approximation and expand
the action and 
the path integral around the stationary point.
Thus, we write
$I_*= I_0 +
\left(\frac{\partial I_*}{\partial r_+}\right)_0 \delta r_+
+\left(\frac{\partial^2 I_*}{\partial r_+^2}\right)_0 
\delta r_+^2$,
where the subscript $0$ 
means that the quantity inside parenthesis is evaluated at the 
stationary point, $I_0= I_*(\beta,Q,R;(r_+)_0)$,
and 
$\delta r_+ = r_+ - (r_+)_0$.
Then, the partition function can be expanded as 
\begin{align}
Z = {\rm e}^{-I_0}\int D\delta r_+ 
{\rm e}^{- \left(\frac{\partial^2 I_*}{\partial r_+^2}\right)_0 
\delta r_+^2}\,.
\label{eq:oneloop}
\end{align}
The partition function in Eq.~\eqref{eq:oneloop} 
contains one loop contributions that obey the spherical symmetry 
of the geometry, the boundary conditions, and the Hamiltonian and 
Gauss constraints. For the path integral to be well defined, 
we must have 
\begin{align}
\left(\frac{\partial^2 I_*}{\partial r_+^2}\right)_{0} > 0\,,
\label{eq:minimum}
\end{align}
so
that 
the stationary point is a minimum and stable, otherwise the integral 
may blow up or be continued to a complex function, 
indicating that the stationary point is not a minimum 
and is therefore an instanton. The second derivative of the action
Eq.~\eqref{eq:actioninrp}
can be simplified into 
$\left(\frac{\partial^2 I_*}{\partial r_+^2}\right)_{0} = 
-\frac{\Omega_{d-2} (d-2) r_+^{d-3}}{4\beta}  
\frac{\partial \iota}{\partial r_+}$.
Thus, the stability condition reduces to 
$\frac{\partial \iota}{\partial r_+}<0$, meaning
that the solution is stable
when $\frac{r_+}{R}$ increases with a decrease in the inverse
temperature, and so with an increase in the temperature.
Now, 
$\frac{\partial \iota}{\partial r_+} = \frac{\beta}{r_+}[(d-2) 
- (d-3)\frac{r_+^{2d-6} + \mu Q^2}{r_+^{2d-6}
- \mu Q^2} 
 + \frac{d-3}{2f}
(
\frac{\mu Q^2}{r_+^{d-3}R^{d-3}}
-\frac{r_+^{d-3}}{R^{d-3}}
) ]$,
where the 
expression of $\iota$ in Eq.~\eqref{eq:beta1} was used, and
so stability means
$
(d-2) 
- (d-3)\frac{r_+^{2d-6} + \mu Q^2}{r_+^{2d-6}
- \mu Q^2} 
 + \frac{d-3}{2f}
(\frac{\mu Q^2}{r_+^{d-3}R^{d-3}}
-\frac{r_+^{d-3}}{R^{d-3}}
)
<0$.
In terms of the variables $x$ and $y$
defined as $x = \frac{r_+}{R}$ and $y= \frac{\mu Q^2}{R^{2d-6}}$,
see Eq.~\eqref{eq:newv},
 the stability condition is
\begin{align}
&\frac{d-1}{2}x^{4d-12} - (1+y)x^{3d - 9} - 3(d-3)y x^{2d-6} 
\nonumber \\
&+ (2d-5)y(1+y)x^{d-3}
-\frac{3d-7}{2}y^2 > 0\,.
\label{eq:stabcond1}
\end{align}
The range of $x$ is $x_e<x<1$, where $x_e$ is a function of $y_e$, see
Eq.~\eqref{eq:extremal2}.
In the case of 
$0\leq y < y_s$, the condition of stability reduces to
two intervals in $x$, one is
$0<x<x_{s1}(y)$ and
the other is
$x_{s2}(y)<x<1$. Therefore, the solutions $x_{1}$ and $x_{3}$ 
are stable, while the solution $x_{2}$ is unstable.
Moreover, 
the points $x=x_{s1}$ and $x=x_{s2}$ are saddle points of the action
as previously stated, and so they are neutrally stable.
In the case of 
$y = y_s$, the same applies as the previous case. 
In the case 
of $y_s<y<1$, the stability condition is satisfied in the interval 
$x_e<x<1$ and so the solution $x_{4}$ is stable.

It is of interest to pick specific dimensions $d$.
Due to its real importance we study $d=4$, and
as a typical case of higher dimension we analyze
carefully $d=5$.

\subsection{$d=4$: Stationary points and stability in four
dimensions}

We now comment on the particular case of four dimensions, $d=4$.  The
original results were presented in
\cite{carlipvaidya2003,Lundgren:2006}, we show here that they are in
agreement with ours and we also display new and interesting features
for this case.

First, we want to understand qualitatively $x\equiv\frac{r_+}{R}$ as a
function of the temperature parameter $RT$, i.e., $x(RT)$, for the
several distinct electric charge parameter $y$ regions.  Recall that the
value of $y_s$ is important since it separates the behavior of the
solutions. From Eq.~\eqref{eq:ycd}, in $d=4$ it is $y_s=(\sqrt{5}
-2)^2=0.056$, the latter equality being approximate.  The solutions
can then be divided using the electric charge parameter $y$ in the
solution for the no charge case $y=0$, solutions for the charge
parameter in the region $0<y<(\sqrt{5} -2)^2$, the solution for
$y=y_s=(\sqrt{5} -2)^2$, and solutions for the charge parameter in the
region $(\sqrt{5} -2)^2<y<1$.
We can comment now on $x(RT)$ within each $y$ division.
For $y=0$, it describes the uncharged case and the solution
is known, it is the original York
solution \cite{York:1986}, and consists of two
solutions, here represented as $x_2$ and $x_3$.
The solution $x_{s2}$ happens
when $x_2$ and $x_3$ meet at temperature
$RT = \frac{3\sqrt{3}}{8\pi}=0.207$,
the latter equality being approximate.
For the electric charge in the range $0<y<(\sqrt{5} -2)^2$,
there are three solutions 
$x_1$, $x_2$ and $x_3$, where $x_1$ is stable,
 $x_2$ is unstable,
and $x_3$
is stable.
For very small charges, the temperature $T_1$, which is the 
temperature at which $x_{s1}$ is a solution
for the black hole at the given charge, is very high, tending to
infinite when the charge tends to zero.
For very small charges, the temperature $T_2$, which is the 
temperature at which $x_{s2}$ is a solution
for the black hole at the given charge,
is very near
the minimum temperature of the solutions of the canonical 
ensemble of the Schwarzschild black hole in four dimensions,
i.e., 
$RT = \frac{3\sqrt{3}}{8\pi}$, mentioned above.
Increasing the electric charge from small values, one has that the saddle
points $x_{s1}$ and $x_{s2}$ approach each other.
For the electric charge parameter given by
$y = (\sqrt{5} -2)^2=y_s$, the saddle points
$x_{s1}$ and $x_{s2}$ meet, and at this
electric charge, the solution $x_1$ is 
described by a curve, the solution $x_2$ is now reduced to a
point that coincides with
$x_s = x_{s1} = x_{s2}$, and the solution $x_3$ is described by another
curve.  All
solutions are stable,
more precisely, $x_1$ is stable, $x_2$ is neutrally stable, and
$x_3$ is stable.
For electric charge in
the range $ (\sqrt{5} -2)^2 < y
<1$, there is only one solution $x_4$ which
represents the union of $x_1$ and $x_3$, with 
$x_2$ having disappeared. Also, the solution $x_4$ is stable.

Second, we want to understand qualitatively $x\equiv\frac{r_+}{R}$ as
a function of the electric charge parameter $y\equiv\frac{\mu
Q^2}{R^2}$, with $\mu=1$ here, i.e., $x(y)$,  for the several distinct
temperature parameter $RT$ regions.  Recall that the value of $RT_s$
and the value of minimum temperature in the uncharged case 
$RT = \frac{3\sqrt{3}}{8\pi}$
are important since they separate the behavior of the solutions. In
$d=4$, the value of the temperature corresponding to $y_s$ and $x_s$
is $RT_s= 0.185$, this equality being approximate.  Thus,
the temperature parameter regions are $0<RT< 0.185$, $RT_s= 0.185$,
$0.185< RT \leq \frac{3\sqrt{3}}{8\pi}=0.207$, and
$\frac{3\sqrt{3}}{8\pi}<RT<\infty$.
We can comment now on $x(y)$ within each $RT$ division.
For $0<RT< 0.185$, 
there are only two solutions, which are $x_1$ in the 
interval $0 < y < y_s$ and $x_4$ in the interval $y_s \leq y < 1$,
with  $y_s= (\sqrt{5} -2)^2$. 
For $RT_s= 0.185$ corresponding
to $y_s$ and $x_s$, with this equality being approximate, 
there are four solutions,
but two of them are degenerate. Indeed, there is the
$x_1$ solution, there are the $x_2$ and $x_3$ solutions
that degenerate
into a point $x_2=x_3$ at $y=y_s$, and there
is the $x_4$ solution.
For $0.185< RT \leq \frac{3\sqrt{3}}{8\pi}=0.207$,
the latter equality being approximate,
there are the four solutions 
$x_1$, $x_2$, $x_3$ and $x_4$.
The solutions $x_1$, $x_2$, and $x_3$
lie in the range $0 < y < y_s$, and
the solution $x_4$ exists only for $y_s < y <1$.
We can note again that 
the solution $x_4$ is a continuation in $y$,
i.e., in $Q$, of the solutions $x_1$ and 
$x_3$, and so in a sense $x_4$
is the union of $x_1$ and $x_3$.
For $\frac{3\sqrt{3}}{8\pi}<RT<\infty$, 
there are also the four solutions but $x_2$ and 
$x_3$ are discontinuous.

%\vfill

\subsection{$d=5$: Stationary points
and stability in five dimensions}

\subsubsection{Behavior of the solutions and plots}

We now present in some detail the particular case of five dimensions,
$d=5$, as a typical higher dimensional case. The behavior of the
solutions will be developed for this case with explanations
and plots.

\begin{figure}[h]
\centering
\includegraphics[width=\linewidth]{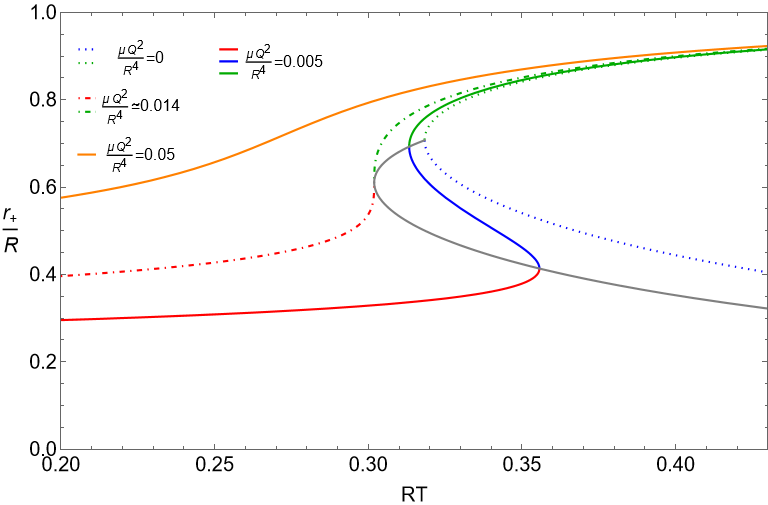}
\caption{Plots of the solutions $x\equiv\frac{r_+}{R}$ as a function
of $RT$ of the canonical ensemble in five dimensions, $d=5$, for four
values of the electric charge parameter $y\equiv\frac{\mu Q^2}{R^4}$,
with $\mu=\frac{4}{3\pi}$ here.  The four values of the electric
charge parameter $y$ are $y=0$ in dotted lines, $y=0.005$ in full
lines, $y= \frac{(68 - 27\sqrt{6})^2}{250}=0.014$ in dot dashed lines,
the latter equality being approximate, and $y=0.05$ in an orange full
line. The solution $x_1=\frac{r_{+1}}{R}$ is represented in red,
$x_2=\frac{r_{+2}}{R}$ is represented in blue, $x_3=\frac{r_{+3}}{R}$
is represented in green, and $x_4=\frac{r_{+4}}{R}$ is represented in
orange.  The gray curve describes the trajectory of the saddle points
of the action $x_{s1}=\frac{r_{+s1}}{R}$ and
$x_{s2}=\frac{r_{+s2}}{R}$ by changing the electric charge parameter,
and it separates the regions of existence of the solutions
$x_1=\frac{r_{+1}}{R}$, $x_2=\frac{r_{+2}}{R}$, and
$x_3=\frac{r_{+3}}{R}$.
\label{fig:rt5}
}
\end{figure}

First, we analyze $x\equiv\frac{r_+}{R}$ as a function
of the temperature parameter $RT$, for the several regions of the
electric charge parameter $y$.  Once more, the value
of $y_s$ is important for the analysis 
since it separates the regions of different behavior for the
solutions. From Eq.~\eqref{eq:ycd}, in $d=5$ it is $y_s=\frac{(68 -
27\sqrt{6})^2}{250}=0.014$, the latter equality being approximate.
We can divide the analysis into the following 
regions of the electric charge parameter
$y$: the no charge case $y=0$, the electric
charge parameter in the region $0<y<\frac{(68 - 27\sqrt{6})^2}{250}$,
the specific case of the critical 
charge $y=y_s=\frac{(68 - 27\sqrt{6})^2}{250}$, and 
the electric charge parameter in the region $\frac{(68 -
27\sqrt{6})^2}{250}<y<1$.
We now describe the solutions 
$x(RT)$ for each region of $y$, and for that we
display in Fig.~\ref{fig:rt5} the plots of the solutions
$x\equiv\frac{r_+}{R}$ as a function of $RT$ of the canonical ensemble
in five dimensions, $d=5$.
%
%%%%%%%%%%%%%%%%%%%%%%%%%%%%%%%%%%%%%%%%%%%%%%%%%%%%%%%%%%%%%%%%%%%%%%
%%%%%%%%%%%%%%%%%%%%%%%%%%%%%%%%%%%%%%%%%%%%%%%%%%%%%%%%%%%%%%%%%%%%%%
%
An important line in such plots is the line in gray in the figure,
that represents the trajectory of the saddle points $x_{s1}$ and
$x_{s2}$ of the action by varying the electric charge. This 
gray line separates
the regions where the solutions $x_1$, $x_2$, and $x_3$ can be
found. More precisely, the two saddle points $x_{s1}$ and $x_{s2}$ are
the bounds of the solution $x_2$.
For $y=0$, we have the uncharged case. 
The solution has been analyzed in
\cite{Andre:2020}, and consists of two solutions, here represented as
$x_2$ and $x_3$. At the saddle point $x_{s2}$, the solutions 
$x_2$ and $x_3$ meet at temperature $RT = \frac{1}{\pi}$.
For the electric charge parameter $y$ in the region $0<y<\frac{(68 -
27\sqrt{6})^2}{250}$, which can be visualized by the $y=0.005$ case 
in the plot, there are
three solutions $x_1$, $x_2$, and $x_3$, where
again $x_1$ is stable, $x_2$
is unstable, and $x_3$ is stable, 
see below for the discussion of thermodynamic
stability.  This case is representative of small electric charges.
For very small charges, the temperature $T_1$ 
corresponding to the saddle point $x_{s1}$ 
assumes very large values and tends to infinity when the
charge tends to zero. Moreover, the temperature $T_2$,
corresponding to the saddle point $x_{s2}$ 
is close to the minimum temperature of the
solutions of the canonical ensemble of the Schwarzschild black hole in
five dimensions $RT = \frac{1}{\pi}$. Note that the
figure with the plots 
for small electric charge parameter yields a unification of York
and Davies, as the two solutions are here represented. 
More precisely, the blue and green lines
correspond to the unstable and stable black holes of York
\cite{York:1986}, respectively, and the red and blue lines correspond
to the stable and unstable black holes of Davies \cite{Davies:1977},
respectively, see below for these latter black holes. Increasing the
electric charge from small values, one sees that the saddle points
$x_{s1}$ and $x_{s2}$ approach each other along the gray curve.
For the saddle electric charge $y = y_s\frac{(68 -
27\sqrt{6})^2}{250}=0.014$, with the latter equality being approximate, 
the saddle points $x_{s1}$ and $x_{s2}$ are equal as 
$x_{s1} = x_{s2} = x_{s}$. While $x_1$ and 
$x_3$ are described by a curve, the solution $x_2$ reduces to a 
point $x_2= x_s$ that connects both solutions $x_1$ and $x_3$.  
Regarding stability, $x_1$ is
stable, $x_2$ is neutrally stable, and $x_3$ is stable.
For the electric charge parameter $y$ in the region $\frac{(68 -
27\sqrt{6})^2}{250} < y <1$, which is represented in the plot 
by the case $y=0.05$, 
there is only one solution $x_4$, that is in a sense
the continuation of $x_1$ and $x_3$, with $x_2$ having disappeared. 
We note that $x_4$ is a stable solution.

\begin{figure}[b]
\centering
\includegraphics[width=\linewidth]{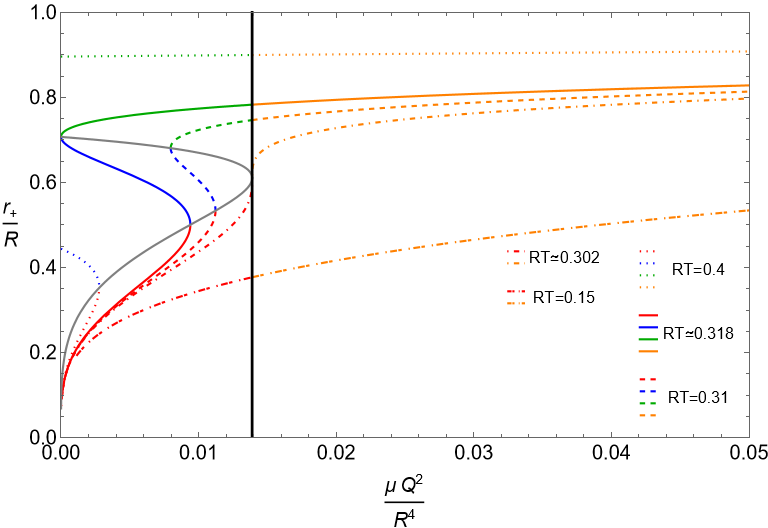}
\caption{
Plots of the solutions $x\equiv\frac{r_+}{R}$ as a function of
$y\equiv\frac{\mu Q^2}{R^4}$ of the canonical ensemble in five
dimensions, $d=5$, for five values of the temperature parameter $RT$,
with $\mu=\frac{4}{3\pi}$.  The five values of $RT$ are $RT=0.15$ in
double dashed lines, $RT=RT_s= 0.302$ in dot dashed lines, $RT=0.31$
in dashed lines, $RT=\frac{1}{\pi}=0.318$, in full lines, the latter
equality being approximate, and $RT = 0.4$ in dotted lines.  The
solution $x_1=\frac{r_{+1}}{R}$ is represented in red,
$x_2=\frac{r_{+2}}{R}$ is represented in blue, $x_3=\frac{r_{+3}}{R}$
is represented in green, and $x_4=\frac{r_{+4}}{R}$ is represented in
orange. The black line, corresponding to $y= y_s = \frac{(68 -
27\sqrt{6})^2}{250}$, separates the solution $x_4=\frac{r_{+4}}{R}$
from the remaining solutions.  The gray line corresponds to the
trajectory of the saddle points of the action
$x_{s1}=\frac{r_{+s1}}{R}$ and $x_{s2}=\frac{r_{+s2}}{R}$, which
bounds the region where $x_2=\frac{r_{+2}}{R}$ exists.
\label{fig:rq5}
}
\end{figure}

Second, we want to describe $x\equiv\frac{r_+}{R}$ as
a function of the electric charge parameter $y\equiv\frac{\mu
Q^2}{R^4}$, with $\mu=\frac{4}{3\pi}$ here,
for the several regions of the
temperature parameter $RT$. Here, the value of $RT_s$ and 
the value of the minimum temperature of the uncharged case 
$RT = \frac{1}{\pi}$
are important since they separate the regions of 
different behavior for the solutions. In
$d=5$, the temperature corresponding to $x_s(y_s)$ is
$RT_s= 0.302$, with this equality being approximate. We then 
consider the temperature parameter regions
$0<RT< 0.302$, $RT_s= 0.302$,
 $0.302< RT \leq \frac1\pi=0.318$, the latter equality
 being approximate, 
and $\frac1\pi<RT<\infty$.
We now describe $x(y)$ within each $RT$ region, and for that we
display in Fig.~\ref{fig:rq5} plots of the solutions
$x\equiv\frac{r_+}{R}$ as a function of $y\equiv\frac{\mu
Q^2}{R^4}$, $\mu=\frac{4}{3\pi}$,  of the canonical ensemble
in five dimensions, $d=5$.
%
%%%%%%%%%%%%%%%%%%%%%%%%%%%%%%%%%%%%%%%%%%%%%%%%%%%%%%%%%%%%%%%%%%%%%%
%%%%%%%%%%%%%%%%%%%%%%%%%%%%%%%%%%%%%%%%%%%%%%%%%%%%%%%%%%%%%%%%%%%%%%
%
For the temperature parameter $RT$
in the range 
$0<RT< 0.302$, of which $RT=0.15$
is represented in the figure, 
there are only two solutions to display, which are $x_1$ in the 
interval $0 < y < y_s$, and $x_4$ in the interval $y_s \leq y < 1$,
with  $y_s= \frac{(68 - 27\sqrt{6})^2}{250}$. 
For the temperature parameter $RT$
given by $RT= RT_s= 0.302$, this equality being approximate,
one has the curves of the
$x_1$ solution and the $x_4$ solution, 
while the $x_2$ and $x_3$ solutions degenerate 
into a point $x_2=x_3$ at $y=y_s$.
For the temperatures $0.302< RT \leq \frac1\pi=0.318$,
of which  $RT = 0.31$ and $RT = \frac1\pi$ 
are represented in the figure, 
one has the solutions 
$x_1$, $x_2$ and $x_3$ lying in the range 
$0 < y < y_s$, while the solution $x_4$ 
lies in the range $y_s < y < 1$.
The figure shows explicitly that the solution $x_4$ 
is a continuation in
the electric charge parameter $y$
of the solutions $x_1$ and 
$x_3$. Note also that the gray curve in the figure
bounds the solution $x_2$.
For $\frac1\pi<RT<\infty$, which is represented by $RT=0.4$ 
in the figure,
one has also the four solutions but the segments of $x_2$ and 
$x_3$ are discontinuous.

\subsubsection{Interpretation of the results}

These results merit some underlying understanding of the physics,
which we now give in terms of the thermal wavelength $\lambda$ 
which is proportional to the inverse of the temperature,
$\lambda=\frac1T$.  
We give the reasoning for the plots of the
solutions $x\equiv\frac{r_+}{R}$ as a function of $RT$ of the
canonical ensemble shown in Fig.~\ref{fig:rt5}. We 
analyze the solutions from small electric charge to 
large electric charge, and from low to high temperature $T$
with $R$ fixed. We note that small $RT$ corresponds to low $T$ here.  

To start with, we analyze the case for a given small electric charge.
For small $T$, the associated thermal wavelength $\lambda$ is large
and is stuck to the cavity walls, which means that if there were no
electric charge, there would be no black hole.  But since there is a
fixed electric charge, there is a small black hole with radius $r_+$
of the order of the length scale set by the charge itself. This black hole
does not form by collapse, its presence comes from
topological constraints.  The black hole is stable, small
perturbations cannot evaporate it.  For the smallest possible $T$,
$T=0$, the black hole is an extremal black hole.
For small temperature, there
is only one black hole solution which is this one.
For an intermediate $T$, as the temperature increases, one has that
the associated thermal wavelength $\lambda$ decreases. The black hole
with small $r_+$ is still there, but there is now the possibility of
forming black holes via collapse, indeed the thermal
wavelengths are no more stuck to the cavity walls and the existent
thermal energy can collapse.  One black hole that can form in this way
has radius $r_+$ of the order of $\lambda$ and is thermodynamically 
unstable since
clearly it can evaporate. The other black hole that can form in this way
has radius $r_+$ large such that $R-r_+$ is of the order of $\lambda$,
and is thermodynamically stable, the reservoir and the black hole
exchange quanta of $\lambda$ in a stabilizing way.  For intermediate
temperatures, there are thus three black hole solutions for each
temperature.
For high $T$, as the temperature increases and the associated
wavelength $\lambda$ gets even smaller.  The smallest black hole $r_+$
ceases to exist because, due to the turbulence created by the high
temperature, there is no way to maintain the electric charge
coherently at the center of the cavity.  The intermediate black hole
$r_+$ ceases to exist because the electric charge repulsion is
sufficient to halt gravitational collapse
of this black hole with
intermediate $r_+$.  The large black hole $r_+$
still exists, as it has sufficient mass to overcome the electric
repulsion and still collapses. 
For high $T$, therefore only the large black hole exists.
This is for a typical reasonably low electric charge $Q$, and we see
there is an interplay between the two quantities that characterize the
ensemble, namely, the temperature $T$ and the electric charge $Q$.

We now analyze the case of high electric charge. 
Again here, for small $T$, the associated thermal wavelength $\lambda$
is large and is stuck and cannot collapse.  But since there is a fixed
electric charge, there is a small black hole with radius $r_+$ of the
order of the length set by the charge itself, its presence comes from
topological constraints, is stable, i.e., small perturbations cannot
evaporate it. $T=0$ yields an extremal black hole.
At intermediate $T$, there is turbulence to disperse the black hole
with topological features but it is possible
to have sufficient mass to collapse the existent
thermal energy into the large black hole, with $R-r_+$ starting to be 
comparable to $\lambda$. Note that the intermediate black hole 
does not exist because the electric charge is large enough to counter 
its collapse.
For high $T$, as the temperature increases and the associated
wavelength $\lambda$ gets smaller, the large black hole $r_+$ has
sufficient mass to overcome the electric repulsion and the thermal
energy collapses, being stable.
For all temperatures, there is thus one black hole solution
only for each
temperature. It is in a sense the union of the
topological black hole with the large  collapsed black
hole as the temperature $T$
increases, the intermediate one having disappeared.
Following this
reasoning, one can also extend this interpretation to
the plots of the solutions
$x\equiv \frac{r_+}{R}$ as a function of $\frac{\mu Q^2}{R^{4}}$ in
Fig.~\ref{fig:rq5}.

%\newpage

%\centerline{}
%\vfill

%\newpage

%%%%%%%%%%%%%%%%%%%%%%%%%%%%%%%%%%%%%%%%%%%%%%%%%%%%%%%%%%%%%%%%%%%%%%
\section{Thermodynamics from the canonical ensemble of a charged black 
hole inside a cavity in $d$ dimensions 
\label{sec:thermo}}
%%%%%%%%%%%%%%%%%%%%%%%%%%%%%%%%%%%%%%%%%%%%%%%%%%%%%%%%%%%%%%%%%%%%%%

%%%%%%%%%%%%%%%%%%%%%%%%%%%%%%%%%%%%%%%%%%%%%%%%%%%%%%%%%%%%%%%%%%%%%%
\subsection{Free energy, entropy,
pressure, electric potential, and  thermodynamic energy
\label{sec:thermoquantities}}
%%%%%%%%%%%%%%%%%%%%%%%%%%%%%%%%%%%%%%%%%%%%%%%%%%%%%%%%%%%%%%%%%%%%%%

With the zero loop approximation performed, the partition function
is $Z={\rm e}^{-I_0[\beta,R,Q]}$, and simultaneously, in the canonical 
ensemble, it is also given by $Z = {\rm e}^{-\beta F}$, where $F$ 
is the Helmholtz
free energy. Therefore, we have 
the relation $I_0[\beta,R,Q] = \beta F$, i.e., the action is related 
to the free energy of the charged black hole in the cavity by
\begin{align}
F = T\,I_0[\beta,R,Q]\,.
\label{eq:freeennergygeneric}
\end{align}
The free energy of the system containing the charged black hole is 
then
\begin{align}
F = \frac{R^{d-3}}{\mu}\left(
1 - \sqrt{f \left(R,Q,r_+\right)}
\right) 
- T\frac{\Omega_{d-2} r_+^{d-2}}{4}\,,
\label{eq:freeennergy}
\end{align}
with $f(R,Q,r_+) \equiv 1 - \frac{r_+^{d-3} 
+ \frac{\mu Q^2}{r_+^{d-3}}}{R^{d-3}} + \frac{\mu Q^2}{R^{2d-6}}$,
see Eq.~\eqref{eq:Rprimealpha}.

The Helmholtz
free energy
is given in terms of the internal energy
$E$, the temperature $T$, and the entropy
$S$ by the relation
\begin{align}
F = E - TS\,.
\label{eq:free}
\end{align}
By definition $F$ has the differential
\begin{align}
dF = - S dT - p dA + \phi dQ\,,
\label{eq:difffree}
\end{align}
where, in addition to the entropy 
$S$, the area 
$A$, and the electric charge
$Q$, 
there is the thermodynamic 
pressure $p$,
and the thermodynamic electric potential $\phi$. 
And so, we can obtain the thermodynamic quantities from the 
derivatives of the free energy $F$, more precisely, the entropy is 
$S = - \left(\frac{\partial F}{\partial T}\right)_{A,Q}$, the 
pressure is $p = - \left(\frac{\partial F}{\partial A}\right)_{T,Q}$,
and the electric potential is 
$\phi = \left(\frac{\partial F}{\partial Q}\right)_{T,A}$, where here
the subscript indicates the quantities that are fixed while 
performing the derivative. In Eq.~\eqref{eq:difffree}, some of the
dependence on $T$, $A$, and 
$Q$ is implicit on the solution for the 
horizon radius $r_+ = r_+(T,A,Q)$, as it is evaluated at the 
minima of the action. To simplify the calculation of the derivatives,
we can perform the chain rule and the fact that, since
$r_+ = r_+(T,A,Q)$, the derivative of the reduced action obeys
$\left(\frac{\partial I_*}{\partial r_+}\right)_{T,R,Q} = 
 \left(\frac{\partial F}{\partial r_+}\right)_{T,R,Q}= 0$, to get 
for example
$S = - \left(\frac{\partial F}{\partial T}\right)_{A,Q} = 
-\left(\frac{\partial F}{\partial T}\right)_{R,Q,r_+} 
- \left(\frac{\partial F}{\partial r_+}\right)_{T,R,Q}
 \frac{\partial r_+}{\partial T}
= -\left(\frac{\partial F}{\partial T}\right)_{R,Q,r_+}$,
and this also holds similarly for the computation of the 
pressure and the electric potential. Therefore, the thermodynamic 
quantities can be computed to be
$S = -\left(\frac{\partial F}{\partial T}\right)_{R,Q,r_+}$,
$p = -\frac{1}{(d-2)\Omega_{d-2} R^{d-3}}
\left(\frac{\partial F}{\partial R} \right)_{T,Q,r_+}$,
$\phi = \left(\frac{\partial F}{\partial Q}\right)_{T,R,r_+}$
and $E = F - TS$.
The entropy is then given as
\begin{align}
S = \frac{A_+}4\,,
\label{eq:entropy}
\end{align}
where $A_+\equiv \Omega_{d-2} r_+^{d-2}$ is
the area of the event horizon, 
and so this is the usual Hawking-Bekenstein expression
for the entropy of a black hole. The thermodynamic pressure is
\begin{align}
p = \frac{d-3}{16\pi R \sqrt{f}}\left((1 - \sqrt{f})^2 
- \frac{\mu Q^2}{R^{2d-6}} \right)\,,
\label{eq:pressure}
\end{align}
the thermodynamic electric potential is
\begin{align}
\phi = \frac{Q}{\sqrt{f}}\left(\frac{1}{r_+^{d-3}} 
- \frac{1}{R^{d-3}} \right)\,,
\label{eq:phi}
\end{align}
and finally, from Eq.~\eqref{eq:free}, the
thermodynamic energy is given by
\begin{align}
E = \frac{R^{d-3}}{\mu}\left[1 - \sqrt{\left(1 - \frac{r_+^{d-3}}
{R^{d-3}} \right)
\left(1 - \frac{\mu Q^2}{r_+^{d-3}R^{d-3}}\right)} \right]\,,
\label{eq:energy}
\end{align}
Collecting Eqs.~\eqref{eq:entropy}-\eqref{eq:energy},
one finds that the first law of thermodynamics in the form
\begin{align}
dE = TdS-pdA+\phi dQ\,,
\label{eq:firstlawoft}
\end{align}
holds.
It is interesting to note,
and surely not a coincidence,
that these thermodynamic quantities are identical to the ones calculated 
for a self-gravitating charged shell, where the first law of 
thermodynamics is imposed, and, the charged shell assumes the
temperature equation 
of state of a black hole and the thermodynamic pressure equation of state 
of the cavity, see \cite{Fernandes:2022}.

%%%%%%%%%%%%%%%%%%%%%%%%%%%%%%%%%%%%%%%%%%%%%%%%%%%%%%%%%%%%%%%%%%%%%%
\subsection{Euler relation equation and Gibbs-Duhem relation
\label{sec:thermoEuler}}
%%%%%%%%%%%%%%%%%%%%%%%%%%%%%%%%%%%%%%%%%%%%%%%%%%%%%%%%%%%%%%%%%%%%%%

With the thermodynamic quantities obtained in 
Eqs.~\eqref{eq:entropy}-\eqref{eq:energy}, one can get an integrated 
first law of thermodynamics known as Euler equation.
For that, one rewrites
the energy in Eq.~\eqref{eq:energy} in terms of the entropy in 
Eq.~\eqref{eq:entropy}, the area $A = \Omega_{d-2} R^{d-2}$, and the
electric charge
$Q$
as
\begin{align}
E =& \frac{(d-2)A^{\frac{d-3}{d-2}}\Omega_{d-2}^{\frac{1}{d-2}}}{8\pi}\times
\nonumber\\&\left(1-\sqrt{\left(1-\left(\frac{4 S}{A}
\right)^{\frac{d-3}{d-2}}\right)\left(1-\frac{\mu Q^2 
\Omega_{d-2}^{2\frac{d-3}{d-2}}}{(4SA)^{\frac{d-3}{d-2}}}\right)} \right)\,
\,.\label{eq:energyintermsof}
\end{align}
We have that the energy is a function $E = E(S,A,Q)$, and if a 
scaling is performed on the thermodynamic quantities 
$S \rightarrow \nu S$, $A \rightarrow \nu A$ and 
$Q \rightarrow \nu^{\frac{d-3}{d-2}} Q$, then it can be verified that 
$E(\nu S, \nu A, \nu^{\frac{d-3}{d-2}}Q) = \nu^{\frac{d-3}{d-2}} 
E(S,A,Q)$. According to the Euler relation theorem, and considering 
that the differential of the energy is given by the first law of 
thermodynamics Eq.~\eqref{eq:firstlawoft}, we have the Euler equation
\begin{align}
E = \frac{d-2}{d-3} (T S - p A) + \phi Q\,.
\label{eq:EulerEquation}
\end{align} 
One can furthermore differentiate Eq.~\eqref{eq:EulerEquation} and 
use the first law of thermodynamics to obtain
\begin{align}
\hskip -0.25cm
\frac1{d-3}\left(
{TdS - pdA}\right) + \frac{d-2}{d-3}(SdT-Adp) + Q d\phi
\hskip -0.06cm
=
\hskip -0.06cm
0\,,
\label{eq:gibbsduhem}
\end{align}
which is the Gibbs-Duhem relation.

%%%%%%%%%%%%%%%%%%%%%%%%%%%%%%%%%%%%%%%%%%%%%%%%%%%%%%%%%%%%%%%%%%%%%%
\subsection{Heat capacity}
%%%%%%%%%%%%%%%%%%%%%%%%%%%%%%%%%%%%%%%%%%%%%%%%%%%%%%%%%%%%%%%%%%%%%%

A system to be thermodynamically stable must have positive
heat capacity  at constant 
area and constant electric charge $C_{A,Q}$, i.e., 
\begin{align}
C_{A,Q}\geq0\,,
\label{eq:caq>0}
\end{align}
where
$
C_{A,Q}\equiv T\left(\frac{\partial S}{\partial T}\right)$.
In section~\ref{subsec:Zero}, we have shown that the stability 
condition in the ensemble formalism was reduced to the condition 
$\frac{\partial \iota}{\partial r_+} < 0$.
The derivative above can be put in terms of thermodynamic variables,
and then in terms of the heat capacity.
The inverse temperature function $\iota(r_+)$
is a function of $r_+$, $R$ and $Q$. 
The variables $Q$ and $R$ are already thermodynamic variables. 
The quantity $r_+$ is also a thermodynamic variable since we have obtained
that $S = \frac{\Omega_{d-2} r_+^{d-2}}{4}$.
Therefore, since $\beta=\iota(r_+)$, we have that
$\frac{\partial \iota}{\partial r_+} = 
- \frac1T \left(\frac{\partial S}{\partial r_+}\right)
\frac{1}{C_{A,Q}}$,
where we have used the definition of the heat capacity at constant 
area and constant electric charge.

The heat capacity is then
\begin{align}
C_{A,Q}
\hskip -0.1cm
=
\hskip -0.1cm
\frac{  (d-2)  R^{d-2}f
\left(\frac{r_+^{d-3}}{R^{d-3}} 
- \frac{\mu Q^2}{R^{d-3}r_+^{d-3}}\right)
\frac{\Omega_{d-2}r_+^{d-2}}{4R^{d-2}}
}
{
\hskip -0.1cm
\frac{d-3}{2}
\hskip -0.1cm
\left(
\hskip -0.1cm
\frac{r_+^{d-3}}{R^{d-3}}
\hskip -0.07cm
-
\hskip -0.07cm
\frac{\mu Q^2}{r_+^{d-3}R^{d-3}}
\hskip -0.12cm
\right)^{\hskip -0.1cm 2}
\hskip -0.15cm
-
\hskip -0.1cm
f
\hskip -0.1cm
\left(
\hskip -0.1cm
\frac{r_+^{d-3}}{R^{d-3}}
\hskip -0.07cm
-
\hskip -0.07cm
(2d
\hskip -0.1cm
-
\hskip -0.1cm
5)\frac{\mu Q^2}{r_+^{d-3}R^{d-3}}
\hskip -0.1cm
\right)}.
\label{eq:Caq}
\end{align}
Since to be thermodynamically stable one has that
$C_{A,Q}\geq0$,
thermodynamic stability reduces to 
Eq.~\eqref{eq:stabcond1} after rearrangements and
definitions.
Thus,
the physical interpretation is that the stability of the solutions 
is controlled by the heat capacity at constant area and charge, as 
it should be in the canonical ensemble. This quantity is tied to 
the derivative of the inverse temperature given by 
Eq.~\eqref{eq:beta1} and so the condition 
reduces to the intervals given by the stationary points of 
$\iota(r_+, R,Q)$, or the saddle points of the action. Moreover, 
solutions where $r_+$ increases as $T$ increases are stable and
solutions where $r_+$ decreases as $T$ increases are unstable.

It is interesting to see what happens when one fixes
$\frac{r_+}{R}$  and change
the electric charge parameter $\frac{\mu Q^2}{R^{2d-6}}$.
For $\frac{r_+}{R}> \left(\frac{2}{d-1}\right)^{\frac1{d-3}}$, the 
heat capacity is always positive.
The limit of the bound happens for the uncharged black hole,
black holes
that obey this inequality 
and have any finite electric charge have positive
heat capacity.
For
$0\leq
\frac{r_+}{R} \leq \left(\frac{2}{d-1}\right)^{\frac1{d-3}}$, 
the sign of the heat capacity
$C_{A,Q}$ changes according to the electric charge. $C_{A,Q}$ is
positive for sufficiently high electric charge parameter
$\frac{\mu Q^2}{R^{2d-6}}$,
and is negative for sufficiently low electric charge parameter
$\frac{\mu Q^2}{R^{2d-6}}$, the change in sign happening
at  the definite value of the charge satisfying
Eq.~\eqref{eq:critbeta} with
fixed $\frac{r_+}{R}$.
We note that this does not indicate a phase transition since 
$\frac{r_+}{R}$ is not a thermodynamic variable controlled in the 
ensemble. At that definite value of the charge parameter,
there is rather a turning point describing the 
ratio of scales at which there is stability.

The thermodynamic variables are the temperature and
the electric charge, and therefore
the heat capacity must be
analyzed in terms of these quantities,
instead of $\frac{r_+}{R}$ and the electric charge.
For the
range of electric charges $0 < \frac{\mu Q^2}{R^{2d-6}}<
\frac{\mu Q_s}{R^{2d-6}}$, one has
three curves for the heat capacity
as a function of the temperature, one
for each solution. The heat capacity is positive for the solutions
$r_{+1}$ and $r_{+3}$, while it is negative for $r_{+2}$. The heat
capacity diverges when the solutions reach the temperatures of the
saddle points of the action, which are the turning points.  For the
critical charge parameter $\frac{\mu Q_s^2}{R^{2d-6}}$, one has two
curves for the heat capacity as a function of the temperature.  In
this
particular case, the two curves are described by the solutions
$r_{+1}$ and $r_{+3}$ and it is positive for both. Moreover, there is
a discontinuity between the two curves at $RT_s$, where the heat
capacity diverges. This point indeed does mark a second order phase
transition between $r_{+1}$ and $r_{+3}$, as both solutions are stable
and it can be seen that the free energy is continuous at $RT_s$ for
$\frac{\mu Q_s^2}{R^{2d-6}}$.  For the range $\frac{\mu
Q^2}{R^{2d-6}}>\frac{\mu Q_s^2}{R^{2d-6}}$,
there is only one curve for the heat capacity as a 
function of the temperature, corresponding to the solution $r_{+4}$
and it is always positive.

We can now give the thermodynamic expressions
with commentaries for the
particular dimensions $d=4$ and $d=5$.

\subsection{$d=4$: Thermodynamics
in four  dimensions}

For $d=4$, we write the results
explicitly. The entropy is given as
$S = \pi r_+^2$,
which is the usual Hawking-Bekenstein formula $S = \frac{A_+}4$, with 
$A_+=4\pi r_+^2$ being the area of the event horizon. 
The pressure is
$ p = \frac{1}{16\pi R \sqrt{f}}\left((1 - \sqrt{f})^2 
- \frac{ Q^2}{R^{2}} \right)$ 
where we have used $\mu=1$
and
$f= 1 - \frac{r_+
+ \frac{Q^2}{r_+}}{R} + \frac{Q^2}{R^2}$.
The electric potential is
$\phi = \frac{Q}{\sqrt{f}}\left(\frac{1}{r_+} 
- \frac{1}{R} \right)$.
Finally, the mean energy is given by
$E = R\left[1 - \sqrt{
\left(1 - \frac{r_+}{R} \right)
\left(1 - \frac{Q^2}{r_+R}\right)} \right]$.
One can then write
the energy in terms of $S$, $A = 4\pi R^2$, and 
$Q$, i.e., $E=E(S,A,Q)$ to obtain
the Euler relation 
$E = 2(T S - p A) + \phi Q$.
The  Gibbs-Duhem 
relation is
$TdS - pdA + 2(SdT-Adp) + Qd\phi = 0$.

The heat capacity, the quantity that controls
thermodynamic stability, is
\begin{align}
C_{A,Q} = \frac{
2R^2f
\left(\frac{r_+}{R} 
- \frac{Q^2}{R^2}\frac{R}{r_+} \right)
\frac{\pi r_+^2}{R^2}
}{
\frac12\left(\frac{r_+}{R} -\frac{Q^2}{R^2}\frac{R}{r_+} \right)^2 
- f\left(\frac{r_+}{R} 
- \frac{3Q^2}{R^2}\frac{R}{r_+}\right)}\,.
\label{eq:CAq4d}
\end{align}
One could fix $\frac{r_+}{R}$ and change
the electric charge parameter $\frac{Q^2}{R^2}$ in 
Eq.~\eqref{eq:CAq4d}.
As seen in the general $d$ case, we find that for 
$\frac{r_+}{R} > \frac23$, the heat capacity is always 
positive, 
and for
$0\leq
\frac{r_+}{R}\leq \frac23$, the sign of the heat capacity
$C_{A,Q}$ changes depending on the electric charge, being
positive for a region of high electric charge parameter
$\frac{Q^2}{R^2}$,
and being negative for a region of low electric charge parameter
$\frac{Q^2}{R^2}$.
This does not indicate a phase transition but rather a turning
point. To see this fact and verify the true phase transitions,
one must analyze the heat capacity in terms of the fixed
quantities of the ensemble, i.e., the temperature and the electric
charge.  For the range
of charge parameters 
$0 <
\frac{\mu Q^2}{R^{2}}<(\sqrt{5} - 2)^2$,
where in $d=4$ one has $\frac{\mu Q_s^2}{R^{2}}=(\sqrt{5} - 2)^2$,
the heat capacity has a curve
for each solution $r_{+1}$, $r_{+2}$, and $r_{+3}$, being positive for
$r_{+1}$ and $r_{+3}$, and being negative for $r_{+2}$.  When the
solutions reach the temperatures of the saddle points of the action,
i.e., 
the turning points, the heat capacity diverges but this only indicates
conditions for stability of the ensemble,
there are no phase transitions at these points.
For the critical charge
$\frac{\mu Q_s^2}{R^{2}}= (\sqrt{5} - 2)^2$, the heat capacity has two
curves as a function of the temperature, $r_{+1}$ and $r_{+3}$,
being positive for both solutions.  For this case, there is a
discontinuity between the two curves at $RT_s = 0.185$, where the heat
capacity diverges.  This point indeed signals a second order phase
transition between $r_{+1}$ and $r_{+3}$, as both solutions are stable
and it can be seen that the free energy is continuous at $RT_s= 0.185$
for $\frac{\mu Q^2}{R^{2}}= (\sqrt{5} - 2)^2$.  For the range
of charge parameters
$\frac{\mu Q^2}{R^{2}}>(\sqrt{5} - 2)^2$, one only has that the heat
capacity of $r_{+4}$ as a function of the temperature is always
positive.
In
\cite{carlipvaidya2003,Lundgren:2006}
some of these results for $d=4$
are presented.

\subsection{$d=5$: Thermodynamics
in five dimensions}

Here, we make the results explicit for the case $d=5$.
The entropy is given as
$
S = \frac{\pi^2 r_+^3}{2}
$, matching the usual Hawking-Bekenstein formula $S = \frac{A_+}4$, with 
$A_+=2\pi^2 r_+^3$ being the area of the event horizon. 
The pressure yields
$p = \frac{2}{16\pi R \sqrt{f}}\left((1 - \sqrt{f})^2 
- \frac{4Q^2}{3\pi R^4} \right)
$,
where we have used $\mu=\frac{4}{3\pi}$
and
$f= 1 - \frac{r_+^2
+ \frac{4 Q^2}{3\pi r_+^2}}{R^2} + \frac{4 Q^2}{3\pi R^4}$.
The electric potential yields
$
\phi = \frac{Q}{\sqrt{f}}\left(\frac{1}{r_+^2} 
- \frac{1}{R^2} \right)
$. And the energy has the expression
$
E = \frac{3\pi R^2}{4}\left[1 - \sqrt{\left(1 - \frac{r_+^2}
{R^2} \right)
\left(1 - \frac{4 Q^2}{3\pi r_+^2R^2}\right)} \right]
$.
These thermodynamic quantities are identical to the ones calculated 
for a self-gravitating charged shell, where the first law of 
thermodynamics is imposed, and the charged shell assumes the equation 
of state of the black hole, see \cite{Fernandes:2022}.
The energy can be written 
in terms of $S$, $A = 2\pi^2 R^3$, and the electric charge
$Q$, as $E=E(S,A,Q)$ to obtain
the Euler relation 
$E = \frac32(T S - p A) + \phi Q$.
The  Gibbs-Duhem 
relation yields
$\frac12\left(TdS - pdA\right) +
\frac32\left(SdT-Adp\right) + Qd\phi = 0$.

\begin{figure}[h]
\centering
\includegraphics[width=\linewidth]{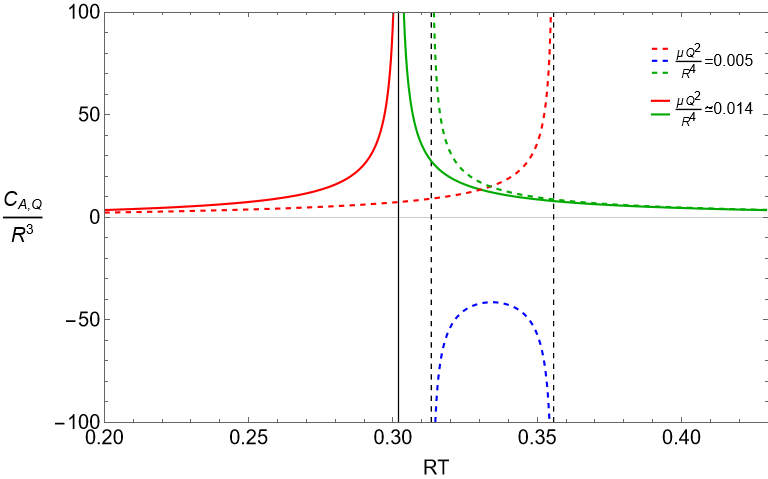}
\caption{\label{fig:heatcapacity5d}
The heat capacity $C_{A,Q}$ in $R^3$ units, $\frac{C_{A,Q}}{R^3}$,
in $d=5$,
as a function of the temperature for two values of the electric charge, 
$\frac{\mu Q^2}{R^4}=0.005$ and $\frac{\mu Q^2}{R^4}= 
\frac{\mu Q^2_s}{R^4}=0.014$ approximately, for solutions $r_{+1}$ in 
red, $r_{+2}$ in blue, and $r_{+3}$ in green. The dashed black lines 
mark the turning points of the solutions and the solid black line marks 
the second order phase transition between the stable
solutions $r_{+1}$ and $r_{+3}$.
}
\end{figure}

The heat capacity is
\begin{align}
C_{A,Q} = \frac{
3R^3 f
\left(\frac{r_+^2}{R^2} 
- \frac{4 Q^2}{3\pi R^2r_+^2} \right)\frac{\pi^2r_+^3}{2R^3}
}
{
\left(
\frac{r_+^2}{R^2} -\frac{4 Q^2}{3\pi R^4}\frac{R^2}{r_+^2}\right)^2 
- f\left(\frac{r_+^2}{R^2} 
- \frac{20 Q^2}{3\pi R^4}\frac{R^2}{r_+^2}
\right)
}
\,.
\label{eq:CAq5d}
\end{align}
Regarding the behavior 
of the heat capacity with fixed $\frac{r_+}{R}$ as a function of
the electric charge parameter $\frac{Q^2}{R^4}$, 
one has that the heat capacity is always positive for 
$\frac{r_+}{R} > \frac{\sqrt{2}}{2}$, and the 
heat capacity changes signs for 
$0\leq \frac{r_+}{R}\leq \frac{\sqrt2}{2}$,
being positive for 
high electric charge parameter $\frac{Q^2}{R^4}$, and being
negative for low electric charge parameter $\frac{Q^2}{R^4}$.
As already noted, to understand the turning points
and the possible phase transitions of the
solutions, one must 
analyze the behavior of the 
heat capacity through its dependence in the temperature 
and the electric charge, see Fig.~\ref{fig:heatcapacity5d}.
%%%%%%%%%%%%%%%%%%%%%%%%%%%%%%%%%%%%%%%%%%%%%%%%%%%%%%%%%%%%%%%%%%%%%%%%%%
%%%%%%%%%%%%%%%%%%%%%%%%%%%%%%%%%%%%%%%%%%%%%%%%%%%%%%%%%%%%%%%%%%%%%%%%%%%%
For a fixed electric charge
parameter in the range $0 < \frac{\mu Q^2}{R^4}
< \frac{(68-27\sqrt{6})^2}{250}$,
where in $d=5$ one
has $\frac{\mu Q_s^2}{R^{4}}=\frac{(68-27\sqrt{6})^2}{250}$,
the heat capacity is described by three 
curves, one for each solution $r_{+1}$, $r_{+2}$, and $r_{+3}$, being 
positive for $r_{+1}$ and $r_{+3}$, and being negative for 
$r_{+2}$, see Fig.~\ref{fig:heatcapacity5d} for the case 
$\frac{\mu Q^2}{R^4} = 0.005$. The heat capacity in this range of 
charges diverges at the turning points of the solutions, as seen 
by the dashed black lines, indicating the conditions for stability of 
the solutions and not
signaling any
phase transition. For the electric charge 
$\frac{\mu Q^2_s}{R^4}= \frac{(68-27\sqrt{6})^2}{250}$, the heat capacity 
is positive, as it is described by the curves of the solution $r_{+1}$ 
and $r_{+3}$. The heat capacity diverges at $RT_{s} =  0.302$, the 
solid black line, and here 
one in fact has a second order transition, from $r_{+1}$ to $r_{+3}$, 
as these are both stable solutions, and the free energy is continuous there.
For $\frac{\mu Q^2}{R^4} > \frac{(68-27\sqrt{6})^2}{250}$, the heat capacity is 
always positive, as it is described only by the solution $r_{+4}$.

%\vfill

%%%%%%%%%%%%%%%%%%%%%%%%%%%%%%%%%%%%%%%%%%%%%%%%%%%%%%%%%%%%%%%%%%%%%%
\section{\label{sec:favorablephases}
Favorable phases of the
$d$ dimensional canonical ensemble of an electrically charged
black hole in a cavity and phase transitions}
%%%%%%%%%%%%%%%%%%%%%%%%%%%%%%%%%%%%%%%%%%%%%%%%%%%%%%%%%%%%%%%%%%%%%%

%%%%%%%%%%%%%%%%%%%%%%%%%%%%%%%%%%%%%%%%%%%%%%%%%%%%%%%%%%%%%%%%%%%%%%
\subsection{Black hole sector of the canonical ensemble and favorable
phases}
%%%%%%%%%%%%%%%%%%%%%%%%%%%%%%%%%%%%%%%%%%%%%%%%%%%%%%%%%%%%%%%%%%%%%%

Consider the black hole sector of the canonical ensemble and the
corresponding free energy.  Since free energy $F$ and action $I_0$ are
related by $F=\frac{I}{\beta}=T\,I_0$, the black hole free energy
$F_{\rm bh} $ can be taken directly from Eq.~\eqref{eq:freeennergy} to
be rewritten as
\begin{align}
F_{\rm bh} = \frac{ R^{d-3}}{\mu}\left(1 - \sqrt{f(R,Q,r_+)}\right) -
T\frac{A_+}{4}
\,,
\label{eq:freeennergybh}
\end{align}
where in this section we put a bh subscript in $F$
to denote that it is a black hole free energy to distinguish from other
possible free energies.
Since $A_+\equiv \Omega_{d-2} r_+^{d-2}$ and
$r_+=r_+(T,R,Q)$,
the black hole
solutions have their free energies of the
form $F_{\rm bh}(T,R,Q)$. For
a system characterized by the free energy,
the one that
has the lower free energy $F_{\rm bh}$,  for given $R$, $T$, and $Q$,
is the one that is
thermodynamically favored. Thus, we can find the
black hole that is favored.

We have shown that in the zero loop approximation, there are different
black hole solutions depending on the electric charge and temperature
of the reservoir, see Sec.~\ref{sec:Canonical1}. 
For sufficiently low electric charge parameter, i.e.,
for $0\leq\frac{\mu Q^2}{R^{2d-6}}
<\frac{\mu Q_s^2}{R^{2d-6}}$,
where $Q_s$ is the saddle electric charge value,
corresponding to the saddle electric charge parameter
$y_s=\frac{\mu Q_s^2}{R^{2d-6}}$, we have seen that
there can be up to
three solutions $\frac{r_{+1}}{R}$ , $\frac{r_{+2}}{R}$, and
$\frac{r_{+3}}{R}$.
Now we comment on the free energies $F_{\rm bh}$ of these
three solutions.
The solution $\frac{r_{+1}}{R}$ has
positive free energy for all the temperatures in which
the solution exists.  The solution $\frac{r_{+2}}{R}$
has also 
positive free energy always,
but it is unstable, so we are not interested in it here. The solution
$\frac{r_{+3}}{R}$, has a temperature for each electric charge 
parameter
$\frac{\mu Q^2}{R^{2d-6}}$ at which the free energy becomes zero,
which we define as
$T_{F_{\rm bh}=0}(Q)$
or $T_{F_{\rm bh}=0}(\frac{\mu Q^2}{R^{2d-6}})$,
thus $\frac{r_{+3}}{R}$
can have
positive or negative free energy.
For the saddle charge parameter $\frac{\mu Q^2}{R^{2d-6}}
=\frac{\mu Q_s^2}{R^{2d-6}}$, the solution
$\frac{r_{+1}}{R}$ has 
positive free energy, there is a solution where
$\frac{r_{+1}}{R} =
\frac{r_{+2}}{R}=
\frac{r_{+3}}{R}$ 
 which has 
positive free energy, and the solution $\frac{r_{+3}}{R}$
has again a temperature
$T_{F_{\rm bh}=0}(Q_s)$
or $T_{F_{\rm bh}=0}(\frac{\mu Q_s^2}{R^{2d-6}})$, 
at which the free energy becomes zero,
thus $\frac{r_{+3}}{R}$ can have
positive or negative free energy.
For higher values of the
electric charge parameter, i.e.,
for
$\frac{\mu Q_s^2}{R^{2d-6}}<
\frac{\mu Q^2}{R^{2d-6}}<1$,
the
solution $\frac{r_{+4}}{R}$ has also a
temperature $T_{F_{\rm bh}=0}(Q)$, or
$T_{F_{\rm bh}=0}(\frac{\mu Q_s^2}{R^{2d-6}})$,
at which the free energy becomes zero,
thus $\frac{r_{+4}}{R}$  can have
positive or negative free energy.
The temperature $T_{F_{\rm bh}=0}(Q)$
can be calculated by solving $F_{\rm
bh}=0$, with $F_{\rm bh}$ given in Eq.~\eqref{eq:freeennergybh} for
either the solution $\frac{r_{+3}}{R}$
or $\frac{r_{+4}}{R}$.  One can instead put the free energy 
in terms of the mass $m$ and electric charge $Q$ through
Eq.~\eqref{eq:beta1} and through the relation $2\mu m = r_+^{d-3} +
\frac{\mu Q^2}{r_{+}^{d-3}}$, so that $F_{\rm bh}=0$ reduces to a
quartic equation for the mass $m$ as a function of the electric
charge, see Appendix~\ref{app:BHzerofreeenergy}. After solving it, one
can then recover the value of $r_+$ and consequently the value
$T_{F_{\rm bh}=0}(\frac{\mu Q^2}{R^{2d-6}})$.  For temperatures lower
than $T_{F_{\rm bh}=0}(\frac{\mu Q^2}{R^{2d-6}})$, the solutions have
positive free energy and for temperatures higher than $T_{F_{\rm
bh}=0}(\frac{\mu Q^2}{R^{2d-6}})$, the solutions have negative free
energy.

There is another important temperature, $T_f$, which depends on the
electric charge $Q$, i.e., on the electric charge
parameter $\frac{\mu Q^2}{R^{2d-6}}$, and
at which the favorability of one phase over the other changes.
For
the electric charge parameter within the region
 $0\leq\frac{\mu Q^2}{R^{2d-6}}
<\frac{\mu Q_s^2}{R^{2d-6}}$, 
there is a phase favorability
temperature $T_f$ at
which the solutions $\frac{r_{+1}}{R}$ and $\frac{r_{+3}}{R}$ have the
same free energy.  In other words, the solutions $\frac{r_{+1}}{R}$
and $\frac{r_{+3}}{R}$ are stable, and thus within the black hole
sector they compete between themselves to be the most favored phase.
Specifically, for temperatures lower than $T_f$, the solution
$\frac{r_{+1}}{R}$ is either more favorable than $\frac{r_{+3}}{R}$,
or is the only existing solution if the temperature is low enough.
For a temperature equal to $T_f$, the solutions $\frac{r_{+1}}{R}$ and
$\frac{r_{+3}}{R}$ are equally favorable, i.e., they coexist equally.
For temperatures higher than $T_f$, either the solution
$\frac{r_{+3}}{R}$ is more favorable than $\frac{r_{+1}}{R}$, or is
the only existing solution if the temperature is high enough.
For the electric charge parameter given by 
$\frac{\mu Q^2}{R^{2d-6}}
=\frac{\mu Q_s^2}{R^{2d-6}}$,
the temperature $T_f$
is the temperature at which
$\frac{r_{+1}}{R} =
\frac{r_{+2}}{R}=
\frac{r_{+3}}{R}$ 
and all have the same free energy, i.e.,
$\frac{r_{+1}}{R}$ and
$\frac{r_{+3}}{R}$ 
coexist. For temperatures lower than $T_f$, the solution
$\frac{r_{+1}}{R}$   
is
the only existing solution.  For temperatures higher than $T_f$, the
solution 
$\frac{r_{+3}}{R}$   
 is the only existing solution.
For the electric charge parameter within the region
$\frac{\mu Q_s^2}{R^{2d-6}}<
\frac{\mu Q^2}{R^{2d-6}}<1
$,
there is only one black hole solution, it is
$\frac{r_{+4}}{R}$. Within the black hole sector
it is surely the most favored state since
it is stable and there is no other solution.
It can 
have
positive or negative free energy.

%%%%%%%%%%%%%%%%%%%%%%%%%%%%%%%%%%%%%%%%%%%%%%%%%%%%%%%%%%%%%%%%%%%%%%
\subsection{Hot flat space sector of the electrically
charged canonical ensemble}
%%%%%%%%%%%%%%%%%%%%%%%%%%%%%%%%%%%%%%%%%%%%%%%%%%%%%%%%%%%%%%%%%%%%%%

Let us
consider a possible electrically charged hot flat space sector, i.e.,
a cavity with nothing in it with its boundaries defined by $R$, $T$,
and $Q$, the settings of the canonical ensemble.

To have such a solution one can think in trying to decrease $r_+$ up
to zero, to a point where there is no more a black hole and
thus obtain flat
space. However, this is not possible, since there is a minimum limit
for $r_+$ given by $r_+ = {r_{+}}_e$ corresponding to the extremal black
hole. At ${r_{+}}_e$, the free energy tends to 
$F_{\rm bh} = \frac{Q}{\sqrt{\mu}}$, 
and it is then
impossible to decrease $r_+$ further. Regarding extremal
black holes, the only temperature that
such solutions exist is at $T=0$ and we do not
consider them here as it is only one point of the ensemble, although
it is a very interesting one.
We simply note, that there is no
other immediate solution of
the action that can be a candidate for a stationary point of the
reduced action.
Thus, to emulate electrically charged hot flat space one has to go
beyond the black hole sector.
One can consider, for example, a shell with radius $r_{\rm shell}$,
coated with the required electric charge $Q$, and with gravity
turned off, i.e., the constant of gravitation is set to zero.  The
action of the system if one considers terms depending only on the
Maxwell field can be calculated to give the free energy as
$F_{\rm shell} = \frac{Q^2}{2}
\left(\frac{1}{r_{\rm shell}^{d-3}} - \frac{1}{R^{d-3}}\right)$,
i.e.,
\begin{align}
F_{\rm shell} = \frac{ Q^2}{2r_{\rm shell}^{d-3}}
\left(1 -
\frac{r_{\rm shell}^{d-3}}{R^{d-3}}\right)
\,.
\label{eq:freeennergyshell}
\end{align}
Thus, for a
given $r_{\rm shell}$, one has
that $F_{\rm shell}$ has a given constant fixed value.  There are two
limits that one can mention.  One limit is when $r_{\rm shell}$ is
very small. One could see this limit as an electrically
charged central point surrounded by hot
flat space, where quantum fluctuations of the hot flat space generate
electric charge.
But this seems to lead to a
divergent free energy.
Note that the behavior mentioned for $r_{\rm shell}$ very small
contrasts with the grand canonical ensemble
case~\cite{Fernandes:2023}, where $r_{\rm shell}=0$ corresponds to a
zero grand potential.
The other limit is when $r_{\rm shell}=R$ and
so the free energy is zero. This means that all the charge is
infinitesimally near the boundary of the cavity, i.e., it is at the
boundary of the cavity itself and there is hot flat space inside the
cavity.  Thus, the more interesting limit is the latter one, when
$r_{\rm shell}=R$, and the charge is gathered near the boundary of the
cavity giving $F_{\rm shell} = 0$.
Since in this case the shell emulates hot flat
space with electric charge
at the boundary, one has $F_{\rm shell} = F_{\rm hfs} =0$.
Nevertheless, it is interesting to
compare the toy model of a shell with free energy
$F_{\rm shell}$ given in Eq.~\eqref{eq:freeennergyshell}
for several $\frac{r_{\rm shell}}{R}$, and in
particular for $\frac{r_{\rm shell}}{R}=1$, with the black hole free energy
$F_{\rm bh}$  given in Eq.~\eqref{eq:freeennergybh}.

One could further
think in building an equivalent system with the constant of
gravitation turned on, such as an electrically charged
self-gravitating shell close to the boundary of the cavity. Still, it
is unclear if there is a possible conversion of this system to
a charged black hole, and vice versa,
since the two systems correspond to different
topologies and also to a different action, as here we do not consider
the matter sector. So we stick to the electric shell with
gravitation turned off.

\subsection{Favorable phases:
First and second order phase transitions}

It is thus of interest to understand what are the favorable states of
the ensemble, i.e., of an ensemble
of a cavity with fixed radius $R$, fixed
temperature $T$, and fixed electric charge $Q$, all values of these
quantities set by the reservoir.

A thermodynamic system tends to be in a state in which its
thermodynamic potential, associated to the ensemble considered, has the
lowest value. In our case, the thermodynamic potential is the
Helmholtz free energy $F$, and so a state is favored relatively to
another if it has lower $F$ for given $R$, $T$, and $Q$. If a system
is in a stable state but with a higher free energy $F$ than another
stable state, it is probable that the system undergoes
a conversion, i.e., a phase transition, to the stable state with
the lowest free
energy. Indeed, in the calculation of the partition function by the
path integral approach, if there are two stable configurations, i.e.,
two states that minimize the action, then the largest contribution to
the partition function is given by the configuration with the lowest
action or, in thermodynamic language, with the lowest free energy.
This type of phase transitions are first order since the
free energy is continuous, but the first derivatives
are discontinuous.

In the case of the canonical ensemble of an electrically charged black
hole inside a cavity in $d$ dimensions, we must compare the free
energy between all the stable black hole solutions of the ensemble,
i.e., one has to compute $F_{\rm bh}$
given in 
Eq.~\eqref{eq:freeennergybh}, for the possible solution $r_+(R,T,Q)$.
For any $d$ 
we note that in this ensemble one can have three
solutions for the same temperature, two of them are stable.
The stable black hole with lowest $F_{\rm bh}$ is the one that is
favored.
This means that considering
only the two stable black hole solutions,
one 
would then have a first order phase transition from 
$r_{+1}$ to $r_{+3}$, for the electric charge parameter
in the range 
$0 < \frac{\mu Q^2}{R^{2d-6}} <\frac{\mu Q_s^2}{R^{2d-6}}$,
and in the limit of the charge parameter with value
$\frac{\mu Q^2}{R^{2d-6}}=\frac{\mu Q^2_s}{R^{2d-6}}$,
this first order phase transition 
becomes a second order phase transition.
It is also interesting to compare the black hole solutions with 
the nongravitating electrically
charged shell case for the same boundary data, which has
free energy given in Eq.~\eqref{eq:freeennergyshell}.
As we argued above, this shell is useful in 
mimicking charged
hot flat space inside the cavity. 
Depending on the value of the radius of the shell
$\frac{r_{\rm shell}}{R}$, this free energy can go from infinity, when
$\frac{r_{\rm shell}}{R}=0$,
to zero, when
$\frac{r_{\rm shell}}{R}=1$. 
In the case of $\frac{r_{\rm shell}}{R}=0$, the shell is never 
favored, while for $\frac{r_{\rm shell}}{R}=1$,
i.e., the case of hot flat space with
the electric charge at the boundary, there is a region 
in which it is favored. 
We proceed,
by essentially assuming a shell with 
$\frac{r_{\rm shell}}{R}=1$, so that 
 $F_{\rm shell}=F_{\rm hfs}=0
$.

Another issue that should be raised in the connection to favorable
states, although it does not come directly from the ensemble formalism
and its thermodynamics, is that there is a black hole radius $r_+$,
more precisely, there is
a ratio $\frac{r_+}{R}$, for which the thermodynamic
energy contained within $R$ is higher than the Buchdahl bound
or, in our context, the 
generalized Buchdahl bound~\cite{Wright:2015}.  When this
happens, that energy content should collapse into a black hole. In
this situation there is no more favorable phase considerations, the
unique phase is a black hole.  Indeed, the generalized
Buchdahl bound yields the maximum mass, or maximum
energy, that can be enclosed in a $d$-dimensional cavity with electric
charge $Q$, before the system shows up
some kind of singularity. At the bound or above, the system
most likely tends to gravitational collapse.  Since the mass of a
system is related to the gravitational radius, it also sets a bound on
the ratio $\frac{r_+}{R}$.  In our context, one should consider this
bound as yielding, for a fixed $R$, the mass $m$, or the gravitational
radius $r_+$, above which the energy within the system is sufficiently
large that the system cannot support itself gravitationally and
collapses.  We can now apply this concept to the case that interest
us here.

In the Schwarzschild black hole case in $d$ dimensions it was found in
\cite{Andre:2021}, that the canonical ensemble yields $F_{\rm bh}=0$
when $\frac{r_+}{R}$ has the Buchdahl bound
value, $\left(\frac{r_+}{R}\right)_{\rm Buch}$.
Since we are envisaging $R$ as fixed, we write 
$\left(\frac{r_+}{R}\right)_{\rm Buch}\equiv
\frac{r_{+\rm Buch}}{R}$ to simplify the notation.
In a $d$-dimensional Schwarzschild spacetime
one has $\frac{r_{+\rm Buch}}{R}=
\left( \frac{ 4(d-2) }
{ (d-1)^2 }
 \right)^{\frac{1}{d-3}}$.
One can infer that black hole
solutions with higher $\frac{r_+}{R}$, i.e., higher
temperatures $RT$, yield gravitational collapse. Since zero free
energy in this electrically
uncharged case, is also the free energy of hot flat space, $F_{\rm
hfs}=0$, one sees that in the uncharged case
one passes
directly from a situation where a hot flat
space phase is favored relatively to a black hole phase, to a situation
where the phase is a
phase where surely there is a black hole,
not merely a phase in which the black hole is favored.

Now, in our setting, i.e.,
in the canonical ensemble
for a black hole with electric charge,
one finds
that for $F_{\rm bh}=0$ only the
bigger black hole exists, and it gives a value
for $\frac{r_{+}}{R}$ that is higher
than the 
Buchdahl bound value.
Thus, there is a definite $F_{\rm bh}$ value
greater than zero where 
the 
Buchdahl value
$\frac{r_{+\rm Buch}}{R}$ 
is met, as we have
found by numerical means up to
$d=16$, but did not prove for all $d$.
For this definite value of $F_{\rm bh}$
or lower values of it, the system
has high enough temperature and
high enough self-thermodynamic
energy to undergo gravitational collapse.
When this happens there is no more
coexistence of phases, there is
only the black hole phase.
Below the saddle, or critical, charge, i.e.,
below the
electric charge parameter given
by $\frac{\mu Q_s^2}{R^{2d-6}}$,
it is the black hole solution
$\frac{r_{+3}}{R}$ that achieves 
$\frac{r_{+\rm Buch}}{R}$.
Above the saddle charge, i.e.,
above $\frac{\mu Q_s^2}{R^{2d-6}}$,
it is the black hole solution
$\frac{r_{+4}}{R}$ that achieves 
$\frac{r_{+\rm Buch}}{R}$.
In contrast,
if we consider the grand canonical ensemble with electric charge,
rather than the canonical ensemble we
are studying here, it was found \cite{Fernandes:2023}
that 
$W_{\rm bh}=0$, where $W_{\rm bh}$ is
the grand potential free energy related to
the grand canonical ensemble, gives a value
for $\frac{r_+}{R}$ which is lower than the 
Buchdahl bound value.
In the grand canonical ensemble,
there is only one stable black hole. So, this means
that for
$W_{\rm bh}=0$, the two phases black hole
and hot flat space coexist equally. For 
$W_{\rm bh}<0$ up to some definite
negative value, then the two phases, black hole
and hot flat space, coexist but
the black hole dominates.
For the definite negative value of $W_{\rm bh}$,
the radius $\frac{r_+}{R}$ is 
the
Buchdahl 
bound value $\frac{r_{+\rm Buch}}{R}$.
For even lower $W_{\rm bh}$, i.e., for
higher temperature
parameter $RT$, one has $\frac{r_+}{R}$ larger than 
$\frac{r_{+\rm Buch}}{R}$ 
and 
the system collapses, or is
collapsed, there is thus no coexistence, only the black
hole phase remains.
Although numerically all three radii $\frac{r_+}{R}$, namely, the
canonical zero free
energy, the Buchdahl, and the grand canonical zero
grand potential, are very close, see
Appendix~\ref{app:BHzerofreeenergy}, it seems that a connection
between the ensemble stability and the mechanical stability of matter
is elusive here.
A comment is in order. The Buchdahl bound applies to a
self-gravitating mechanical system consisting of a ball of matter of
radius $R$.  Our system is a thermodynamic system, with boundary data,
namely $R$, $T$, and $Q$, and contains no matter. One can argue that
in higher orders of approximation, the system contains packets of
energy and one can plausibly deduce that the system must collapse once
the Buchdahl bound is surpassed.  Be as it may, the inference we have
made comes from dynamics, not thermodynamics, and therefore is
strictly outside our approach.

To better understand the
issues and 
make progress one has to pick up  definite
dimensions. We now 
specify our generic $d$-dimensional results
to the dimensions $d=4$ and $d=5$.
We comment on the dimension
$d=4$, and will do a thorough analysis
for the dimension $d=5$.

\subsection{$d=4$: Analysis}

For $d=4$, as for any $d$, 
this ensemble can have either one or three
black hole solutions for a given temperature.
When there are three, two of them are stable
and are of
interest in the consideration
of the most favorable phase, while the remaining
solution is unstable and is of no interest in 
the consideration of the most favorable phase.
The two that are stable
have to be compared against one another
to see which is the most favorable phase. 

We start by comparing the free energy of 
the several black hole solutions that exist in
this ensemble between themselves. From
Eq.~\eqref{eq:freeennergybh}, 
in $d=4$, the black hole free energy
is
\begin{align}
F_{\rm bh} = R\left(1 - \sqrt{f(R,Q,r_+)}\right) 
- T\frac{A_+}{4}
\,,
\label{eq:freeennergybh4d}
\end{align}
where here $\frac{A_+}{4}=\pi r_+^2$,
$f(R,Q,r_+) \equiv 1 - \frac{r_+
+ \frac{Q^2}{r_+}}{R} + \frac{ Q^2}{R^2}$,
we have used $\mu=1$, and $r_+=r_+(T,R,Q)$.
In $d=4$, 
the saddle electric charge parameter value
$\frac{Q_s^2}{R^2} = (\sqrt{5} -2)^2
= 0.056$, 
the last equality being approximate,
separates the region with only one solution 
from the region with three solutions.

A first set
of general and specific comments can be made, namely about the positivity 
of the free energy for each solution. 
For
$0\leq \frac{Q^2}{R^2}<\frac{Q_s^2}{R^2}$, the stable black
hole solution $\frac{r_{+1}}{R}$
has positive $F_{\rm bh}$
for all the temperatures in which the solution exists.
The same happens for the solution $\frac{r_{+2}}{R}$,
but this solution is of not interest here
since it is unstable.
The other stable black
hole solution $\frac{r_{+3}}{R}$
has a temperature $T_{F_{\rm bh}=0}$ depending on the
electric charge,
at which the free energy becomes zero, and
so the black
hole solution
$\frac{r_{+3}}{R}$ can have $F_{\rm bh}$ positive or negative.
For the critical charge
$\frac{Q^2}{R^2}=\frac{Q_s^2}{R^2}$,
with
$\frac{Q_s^2}{R^2} = 0.056$ approximately, the stable black hole solution
$\frac{r_{+1}}{R}$
has positive free energy, the point
$\frac{r_{+1}}{R}=
\frac{r_{+2}}{R}=
\frac{r_{+3}}{R}
$ has positive free energy,
and the stable black hole solution $\frac{r_{+3}}{R}$
has a temperature
$T_{F_{\rm bh}=0}$ at which the free
energy becomes zero.
For
$\frac{Q_s^2}{R^2}< \frac{Q^2}{R^2}<1$,
the only black hole solution is $\frac{r_{+4}}{R}$, which is stable, 
and it 
has a temperature $T_{F_{\rm bh}=0}$ depending on the
electric charge,
at which the free energy becomes zero.
So, the free energy
of $\frac{r_{+4}}{R}$ can be positive or negative.
Quite generally one can calculate $T_{F_{\rm bh}=0}$
by solving $F_{\rm bh}=0$, with 
$F_{\rm bh}$ given in Eq.~\eqref{eq:freeennergybh4d},
for either the solution 
$\frac{r_{+3}}{R}$ or $\frac{r_{+4}}{R}$.
The free energy can be written in terms of 
$m$ and $Q$ through Eq.~\eqref{eq:beta1} in $d=4$ and 
through $2 m = r_+ + \frac{Q^2}{r_{+}}$, 
allowing us to reduce $F_{\rm bh}=0$ into a quartic
equation for the mass, see
Appendix~\ref{app:BHzerofreeenergy}.
The solutions have positive free energy for temperatures lower than 
$T_{F_{\rm bh}=0}$,
and the solutions have negative free energy
for 
temperatures higher
than $T_{F_{\rm bh}=0}$.

A second set
of general and specific comments can be made, namely about the favorability 
between black hole solutions.
For  $0\leq\frac{Q^2}{R^2}
<\frac{Q_s^2}{R^2}$,
there is a favorability
temperature $T_f$ which depends on the
electric charge,  and at which the solutions $\frac{r_{+1}}{R}$
and
$\frac{r_{+3}}{R}$
have the same free energy.  For temperatures lower than
$T_f$, the solution $r_{+1}$ is more favorable than $\frac{r_{+3}}{R}$,
or it is the
only existing solution. For temperatures higher than $T_f$, the
solution $\frac{r_{+3}}{R}$
is more favorable than $\frac{r_{+1}}{R}$, or it is the only
existing solution.
For the critical charge
 $\frac{Q^2}{R^2}=\frac{Q_s^2}{R^2}$,
the temperature $T_f$ is the temperature at which
$\frac{r_{+1}}{R}=
\frac{r_{+2}}{R}=
\frac{r_{+3}}{R}
$
and all have the same free energy, i.e., the
stable solutions $\frac{r_{+1}}{R}$
and
$\frac{r_{+3}}{R}$ coexist.
For
$\frac{Q_s^2}{R^2}<\frac{Q^2}{R^2}<1$,
there is only one black hole solution, it is
$\frac{r_{+4}}{R}$, and, since it is stable, it is favored.
We can now consider phase transitions between the two
stable black hole solutions.
There is 
a first order phase transition from 
$r_{+1}$ to $r_{+3}$, for the electric charge
parameter in the range 
$0 < \frac{\mu Q^2}{R^{2}} <\frac{\mu Q^2_s}{R^{2}}$ 
and, additionally, in the limit of
the electric charge parameter with value
$\frac{\mu Q^2}{R^{2}}=\frac{\mu Q^2_s}{R^{2}}$, 
this first order phase transition 
turns into a second order phase transition.

We  now comment on the comparison in $d=4$ between
the black hole phases just discussed above with
hot flat space phase, which we have emulated by a
nonself-gravitating shell.
In $d=4$,
the free energy of the shell is
\begin{align}
F_{\rm shell} = \frac{Q^2}{2r_{\rm shell}}
\left(1- \frac{r_{\rm shell}}{R}\right)
\,,
\label{eq:freeennergyshell4d}
\end{align}
where $r_{\rm shell}$ is
the radius of the shell, see Eq.~\eqref{eq:freeennergyshell}.
So $F_{\rm shell}$ depends on the
electric charge $Q$, on $r_{\rm shell}$,
and on $R$, but
is a constant as a function of the
temperature $T$.
The case of
a very small shell will lead to a very high free energy due 
to the dependence on $\frac{Q}{r_{\rm shell}}$,
and therefore, for this case the region of 
favorability for the shell
lies in very small values of the charge.
There are also
the cases of intermediate shell radius which
would have to be analyzed specifically.
The other limiting  
case is when the charge is near the boundary of the cavity,
with the 
free energy of this case tending to zero.
Ultimately, the black hole is favored when 
$F_{\rm bh} <F_{\rm shell}$, both coexist equally
when 
$F_{\rm bh} =F_{\rm shell}$, and
the black hole is not favored when 
$F_{\rm bh} > F_{\rm shell}$.
When the radius of the shell is at the cavity radius,
$\frac{r_{\rm shell}}{R}=1$,
then the shell has zero free energy
and emulates hot flat space with electric charge
at the boundary.
Then, the free energy of hot flat space is
 $F_{\rm shell}=F_{\rm hfs}=0$. 
 The black hole is not favored when 
$F_{\rm bh} >0$, both the black hole and hot flat space 
coexist equally
when 
$F_{\rm bh} =0$, and
the black hole is favored when 
$F_{\rm bh} <0$.
When the system finds itself in a phase that is not
favored, it will make a first order phase transition
to the favored phase.

The problem of the thermodynamic phases is even more complicated as we
have mentioned already.  When there is no electric charge, i.e., for the
Schwarzschild space in $d=4$, it was found in \cite{Andre:2021} that,
in the canonical ensemble, the condition $F_{\rm bh}=0$ yields a value
for $\frac{r_+}{R}$ that is equal to the generalized Buchdahl
bound~\cite{Wright:2015}, i.e., the limiting value
$\left(\frac{r_+}{R}\right)_{\rm Buch}$ for gravitational collapse of
a self-gravitating system of energy $E$ and radius $R$.  Since we are
envisaging $R$ as fixed, we write $\left(\frac{r_+}{R}\right)_{\rm
Buch}\equiv \frac{r_{+\rm Buch}}{R}$ to simplify the notation, and in
$d=4$ one has $ \frac{r_{+\rm Buch}}{R}=\frac89=0.89$, the latter
equality being approximate.  This result means that,
in the uncharged case, 
as soon as the black hole phase is favored, there is no further
coexistence with hot flat space, and the system collapses.
For nonzero electric charge there is
no more coincidence. 
Here,  to discuss this issue of favorability 
between black hole and hot flat space,
we are going to consider the case for which the free energy of the
shell is zero, $F_{\rm shell}=0$,
i.e., the case of hot flat space with electric charge
at the boundary, $\frac{r_{\rm shell}}{R}=1$.  In this case,
the shell is situated at the cavity, and so $F_{\rm shell}$ is the
free energy of hot flat space, $F_{\rm hfs}$, which is zero.
For
nonzero electric charge $Q$, i.e., nonzero charge parameter
$\frac{Q^2}{R^2}$, we
find that in the canonical ensemble, the condition $F_{\rm bh}=0$
yields a $\frac{r_+}{R}$ value, both for $\frac{r_{+3}}{R}$ and
$\frac{r_{+4}}{R}$, that is higher than the generalized Buchdahl bound. 
Notice that the generalized Buchdahl bound here is the limiting 
value of $\frac{r_+}{R}$ for gravitational collapse of a
self-gravitating system of energy $E$, electric charge $Q$, and radius $R$.
For an electric charge parameter
lower or equal than the saddle
value $\frac{Q_s^2}{R^2}$,
only the solution $\frac{r_{+3}}{R}$ can take the value of 
the Buchdahl bound, corresponding to a positive free energy and 
some temperature value $RT$. For a system with
this $RT$ or higher, then the system collapses gravitationally into a
black hole with the corresponding $\frac{r_{+3}}{R}$.  For an electric
charge higher or equal than the saddle value
$\frac{Q_s^2}{R^2}$, the solution
$\frac{r_{+4}}{R}$ can take the value of the 
Buchdahl bound, having a definite positive value of
$F_{\rm bh}$, at some temperature parameter $RT$.
For a system with this $RT$ or higher, the system again collapses
gravitationally into a black hole with the corresponding
$\frac{r_{+4}}{R}$.
Interesting to note that in the grand canonical ensemble, where there
is only one stable black hole solution, it was found
\cite{Fernandes:2023} that the equation $W_{\rm bh}=0$, $W_{\rm bh}$
denoting the grand potential, yields a $\frac{r_+}{R}$ value that is
lower than the Buchdahl bound.  Thus, in this case,
when $W_{\rm bh}=0$ for the system, the two phases coexist, black
hole and hot flat space. For $W_{\rm bh}<0$, the black hole phase
dominates in relation to hot flat space. And 
for a certain definite negative value of $W_{\rm bh}$, the
value of $\frac{r_+}{R}$ of the system is the same as the value of the
Buchdahl bound. From then on the system collapses, the only phase
being the black hole phase, and there is no coexistence of phases, see also
Appendix~\ref{app:BHzerofreeenergy}.
Here we have given plausible arguments
for the gravitational collapse of the system
when there is too much energy
inside the cavity, 
although we have
not performed a thermodynamic treatment of the collapsed phase.

\subsection{$d=5$: Analysis}

For $d=5$, as for any $d$, this ensemble has between one and three
black hole solutions for a given temperature.  When there are three
solutions, two of them are stable and are going to be considered here, 
while the remaining is unstable and is
of no interest in this analysis. 
The two that are stable have to be compared against
one another to see which is the most favorable phase.

\begin{figure}[h]
\centering
\includegraphics[width=\linewidth]{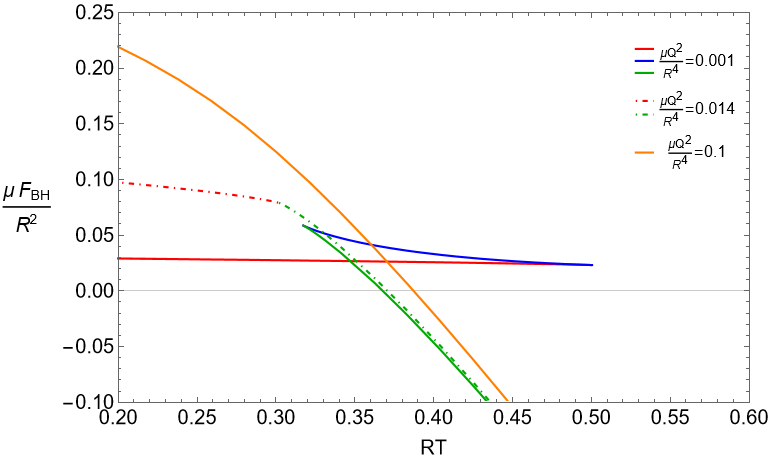}
\caption{\label{fig:freeenergyBH}
Free energy $F_{\rm bh}$ of the charged 
black hole solutions of the canonical ensemble in $d=5$,
given as a quantity with no units $\frac{\mu F_{\rm bh}}{R^2}$,
as a function of the temperature parameter $RT$ for
several electric charge parameters $\frac{\mu Q^2}{R^4}$,
where $\mu=\frac{4}{3\pi}$. 
For $\frac{\mu Q^2}{R^4} = 0.001$, the solution $r_{+1}$ is in red, 
the solution $r_{+2}$ is in blue, and the solution $r_{+3}$ is in 
green, all of them in solid lines. 
For $\frac{\mu Q^2}{R^4} = \frac{(68 - 27 \sqrt{6})^2}{250}
= 0.014$, the latter equality being approximate,
the solution $r_{+1}$ is in red 
and the solution $r_{+3}$ is in green, all of them in dashed lines.
For $\frac{\mu Q^2}{R^4} = 0.1$, the solution $r_{+4}$ is in orange,
in solid line. See text for all the details. 
}
\end{figure}

We start by comparing the free energy of 
the several black hole solutions that exist in
this ensemble between themselves.
In $d=5$, the black hole free energy
is
\begin{align}
F_{\rm bh} = \frac{ R^2}{\mu}\left(1 - \sqrt{f(R,Q,r_+)}\right) 
- T\frac{A_+}{4}
\,,
\label{eq:freeennergybh5d}
\end{align}
where here $\frac{A_+}{4}=\frac{\pi^2 r_+^3}{2}$,
$f(R,Q,r_+) \equiv 1 - \frac{r_+^2
+ \frac{\mu Q^2}{r_+^2}}{R^2} + \frac{\mu Q^2}{R^4}$,
$\mu=\frac{4}{3\pi}$, and $r_+=r_+(T,R,Q)$.
To help in the analysis, we plot in Fig.~\ref{fig:freeenergyBH}
$F_{\rm bh}$ as a function of the temperature
parameter $RT$,
for fixed electric charge parameter $\frac{\mu Q^2}{R^4}$
in $d=5$.
Recall that in $d=5$, one has
the saddle electric charge parameter value
$\frac{\mu Q_s^2}{R^4} = \frac{(68 - 27 \sqrt{6})^2}{250}
= 0.014$, the last equality being approximate.
%
%%%%%%%%%%%%%%%%%%%%%%%%%%%%%%%%%%%%%%%%%%%%%%%%%%%%%%%%%%%%%%%%%%%%%%%%%%
%%%%%%%%%%%%%%%%%%%%%%%%%%%%%%%%%%%%%%%%%%%%%%%%%%%%%%%%%%%%%%%%%%%%%%%%%%%%
%

A first set
of general and specific comments can be made directly from
Fig.~\ref{fig:freeenergyBH}, regarding the positivity of the 
free energy for each solution. 
For
relatively low
electric charge parameter
$0\leq\frac{\mu Q^2}{R^4}<\frac{\mu Q_s^2}{R^4}$,
where $\mu=\frac{4}{3\pi}$ in $d=5$,
the solution $\frac{r_{+1}}{R}$ 
has positive $F_{\rm bh}$
for all the temperatures in which the solution exists.
The same happens for the solution $\frac{r_{+2}}{R}$, 
but this solution is of no interest here
since it is unstable.
The solution $\frac{r_{+3}}{R}$ 
has a temperature $T_{F_{\rm bh}=0}$ depending on the
electric charge
at which the free energy becomes zero, and
so  $\frac{r_{+3}}{R}$  can have $F_{\rm bh}$ positive or negative.
In the figure, this range of
the electric charge parameter is represented by
the case $\frac{\mu Q^2}{R^4}=0.001$.
We see that for $\frac{\mu Q^2}{R^4}=0.001$,
one has for the   $\frac{r_{+3}}{R}$  solution that
$T_{F_{\rm bh}=0}=0.367$ approximately.
For the saddle charge $\frac{\mu Q^2}{R^4}=\frac{\mu Q_s^2}{R^4}$,
with $\frac{\mu Q_s^2}{R^4} = 0.014$ approximately,
the solution $\frac{r_{+1}}{R}$ is positive, the point
$\frac{r_{+1}}{R}=
\frac{r_{+2}}{R}=
\frac{r_{+3}}{R}
$
is positive, and the
solution $\frac{r_{+3}}{R}$   
has a temperature $T_{F_{\rm bh}=0}=0.37$
at which the free energy becomes zero.
For
relatively high
electric charge parameter
$\frac{\mu Q_s^2}{R^4}<\frac{\mu Q^2}{R^4}<1$,
the only solution is $\frac{r_{+4}}{R}$
and it 
has a temperature $T_{F_{\rm bh}=0}$ depending on the
electric charge. So $F_{\rm bh}$
of the black hole $\frac{r_{+4}}{R}$
can be positive or negative.
In the figure, this range of
$\frac{\mu Q^2}{R^4}$
is represented by
the case $\frac{\mu Q^2}{R^4}=0.1$.
We see that for $\frac{\mu Q^2}{R^4}=0.1$,
one has that the
solution $\frac{r_{+4}}{R}$
has
$T_{F_{\rm bh}=0}=0.387$ approximately.
Quite generally, one can calculate $T_{F_{\rm bh}=0}$
by solving $F_{\rm bh}=0$, with 
$F_{\rm bh}$ given in Eq.~\eqref{eq:freeennergybh5d}
for either the solution 
$\frac{r_{+3}}{R}$
or $\frac{r_{+4}}{R}$.
One obtains a quartic equation for the mass $2\mu m = r_+^2
+ \frac{\mu Q^2}{r_{+}^2}$, with here $\mu=\frac{3}{4\pi}$,
as a function of the electric charge, 
see Appendix~\ref{app:BHzerofreeenergy}. 
For temperatures lower than 
$T_{F_{\rm bh}=0}$, the solutions have positive free energy and for 
temperatures higher
than $T_{F_{\rm bh}=0}$, the solutions have negative free energy.

A second set
of general and specific comments can be made directly from
Fig.~\ref{fig:freeenergyBH}, regarding the favorability 
between black hole solutions.
For a range of low
electric charge parameter
$0\leq\frac{\mu Q^2}{R^4}< \frac{\mu Q_s^2}{R^4}$, the
solutions $\frac{r_{+1}}{R}$ and $\frac{r_{+3}}{R}$ 
have the same free energy 
at a specific temperature $T_f$, i.e., the phase
favorability temperature which
depends on $\frac{\mu Q^2}{R^4}$. 
For temperatures lower than $T_f$, the solution
$\frac{r_{+1}}{R}$ either has lower free energy than $\frac{r_{+3}}{R}$
or it is the only existing solution,
and so $\frac{r_{+1}}{R}$ is more favorable. 
For a temperature equal to $T_f$, the solutions $\frac{r_{+1}}{R}$ and
$\frac{r_{+3}}{R}$ have the same free energy and they
are equally favorable, meaning they coexist equally.
For temperatures higher than
$T_f$, the solution $\frac{r_{+3}}{R}$ either has lower free energy than 
$\frac{r_{+1}}{R}$ or it is the only existing solution, and so 
$\frac{r_{+3}}{R}$ is more favorable.
This is represented
for  $\frac{\mu Q^2}{R^4}= 0.001$ in the figure. We see that in this
case, the favorability temperature is $RT_f=0.347$ approximately. 
Also, for $RT<0.32$, there is only the
$\frac{r_{+1}}{R}$ solution, whereas for $RT>0.50$ there is only the
$\frac{r_{+3}}{R}$ solution. The solution $\frac{r_{+2}}{R}$ is
unstable and does not enter in this analysis, however it is plotted in
the figure to show a continuity of the free energy on the three solutions.
For saddle charge
$\frac{\mu Q^2}{R^4}=\frac{\mu Q_s^2}{R^4}
= 0.014$, the latter equality being approximate,
which is shown in the figure, 
the temperature $T_f = 0.30$, approximately, 
is the temperature at which
$\frac{r_{+1}}{R}= \frac{r_{+2}}{R}= \frac{r_{+3}}{R}$, and all have
the same free energy, i.e., $\frac{r_{+1}}{R}$ and $\frac{r_{+3}}{R}$
coexist. For temperatures lower than $T_f$, the solution
$\frac{r_{+1}}{R}$ is the only existing solution.  For temperatures
higher than $T_f$, the solution $\frac{r_{+3}}{R}$ is the only existing
solution.
For the higher values of the electric charge parameter, i.e., for
$\frac{\mu Q_s^2}{R^4}<\frac{\mu Q^2}{R^4}<1$,
there is only one black hole solution
$\frac{r_{+4}}{R}$ that is stable, and so it is favorable. This is 
represented in the figure by the case $\frac{\mu Q^2}{R^4} = 0.1$.
We can now consider phase transitions between the two
stable black hole solutions.
One has a first
order phase transition from $r_{+1}$ to $r_{+3}$, for the electric
charge parameter in the range $0 < \frac{\mu Q^2}{R^{4}} <\frac{\mu
Q^2_s}{R^{4}}$.  Moreover, in the limit of the electric charge
parameter given by the value
$\frac{\mu Q^2}{R^{4}}=\frac{\mu Q^2_s}{R^{4}}$, this first
order phase transition becomes a second order phase transition. This
can be seen from Fig.~\ref{fig:freeenergyBH}, since the intersection
point represents a first order phase transition, and at the limit of
the critical charge, this point represents a second order phase
transition.

\begin{figure}[h]
\centering
\includegraphics[width=\linewidth]{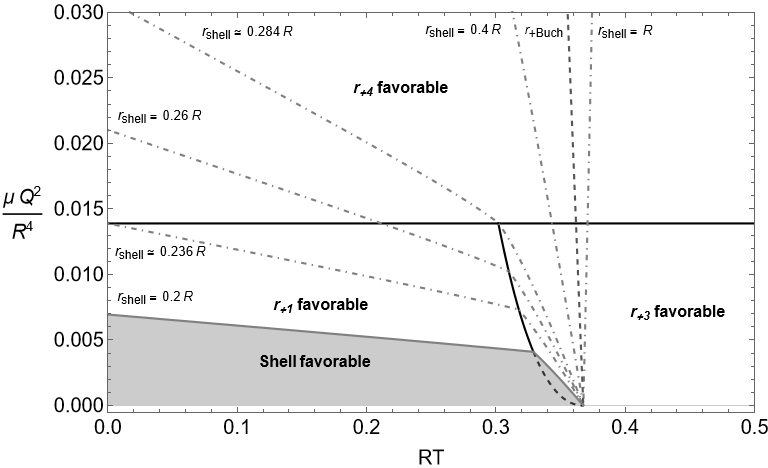}
\caption{\label{fig:favorable}
Favorable states of the 
canonical ensemble of
an electrically 
charged black hole inside a cavity 
in $d=5$
in an electric charge  $Q$ times
temperature $T$, more precisely,
$\frac{\mu Q^2}{R^4}\times RT$ plot.
It is displayed
the region where 
the black hole
$r_{+1}$ is a
favorable phase,
the region
where the black hole
$r_{+3}$ is a favorable
phase, and the region
where the black hole
$r_{+4}$ is a favorable phase.
The delimiters of the favorable 
regions of the black hole solutions are the black lines, 
including the 
dashed line.
It is also incorporated the
solution for a nongravitating 
electrically charged shell
as a simulator for hot flat space.
The electrically charged shell with 
$\frac{r_{\rm shell}}{R} = 0$ is never favored.
The electrically charged shell with 
$\frac{r_{\rm shell}}{R} = 0.2$ is favored in 
the region in 
gray, this case is given as an example.
The upper delimiter of the region of favorability 
of electrically charged shells with
$\frac{r_{\rm shell}}{R} = 0.236$
approximately, 
$\frac{r_{\rm shell}}{R} = 0.26$, 
$\frac{r_{\rm shell}}{R} = 0.284$
approximately, $\frac{r_{\rm shell}}{R} = 0.4$ and 
$\frac{r_{\rm shell}}{R} = 1$, which better simulates
hot flat space, are given by
the dot-dashed lines. The Buchdahl condition
line, i.e., 
$r_{+\rm Buch}$, above which there is presumably
collapse is given by a thick black dash line.
See text for details.}
\end{figure}

We now compare, in $d=5$, the black hole phases discussed just above
with hot flat space phase which we have emulated by a
nonself-gravitating shell, see Fig.~\ref{fig:favorable}.
%
%
%
%%%%%%%%%%%%%%%%%%%%%%%%%%%%%%%%%%%%%%%%%%%%%%%%%%%%%%
%%%%%%%%%%%%%%%%%%%%%%%%%%%%%%%%%%%%%%%%%%%%%%%%%%%%%%
%
%
%
The favorable states for each electric
charge and temperature,
and for various values of the shell radius 
can be seen in the figure. 
The free energy of the shell for the 
case $d=5$ is
\begin{align}
F_{\rm shell} = \frac{Q^2}{2r_{\rm shell}^2}
\left(1-\frac{r_{\rm shell}^2}{R^2} \right)
\,,
\label{eq:freeennergyshell5d}
\end{align}
where $r_{\rm shell}$ is
the radius of the shell, see Eq.~\eqref{eq:freeennergyshell}.
So the shell free energy $F_{\rm shell}$ has a dependence on
electric charge $Q$, on $r_{\rm shell}$, and on $R$, but
as a function of the
temperature $T$, the free energy is a constant.
Due to the term $\frac{Q^2}{r_{\rm shell}^2}$, 
the free energy becomes divergent for 
a very small shell and fixed electric charge.
Therefore, the region of 
favorability for the very small shell
lies in very small values of the
electric charge $Q$.
There are the cases of intermediate shell radius
that are represented in the figure, namely the cases 
$\frac{r_{\rm shell}}{R}={0.2, 0.236, 0.26, 0.284,0.4}$, 
with $0.236$ and $0.284$ being approximate values.
The more interesting limiting 
case is when the electric
charge is near or at the boundary of the cavity,
$\frac{r_{\rm shell}}{R}=1$. The 
free energy of the shell in this limit is
zero.
The black hole solution is favored compared to the shell when $F_{\rm
bh} <F_{\rm shell}$, while both the black hole and the shell coexist
equally when $F_{\rm bh} =F_{\rm shell}$, and the black hole is not
favored compared to the shell when $F_{\rm bh} > F_{\rm shell}$. The
gray dashed curves in the figure represent the condition $F_{\rm bh}
=F_{\rm shell}$ for each shell radius, delimiting the regions where
the black hole is favorable, for higher temperature, and where the
shell is favorable, for lower temperature.  When the radius of the
shell is at the cavity radius, $\frac{r_{\rm shell}}{R}=1$, the free
energy of the shell becomes zero, emulating hot flat space with free
energy
$F_{\rm shell}=F_{\rm hfs}=0$. This is the case of hot flat space
with electric charge at the boundary.
Again, the black hole is not favored compared to hot flat space when 
$F_{\rm bh} >0$, while both the black hole and hot flat space 
coexist equally when 
$F_{\rm bh} =0$, and
the black hole is favored compared to hot flat space when 
$F_{\rm bh} <0$. The gray dashed curve $r_{\rm shell} = R$ 
in the figure corresponds to the boundary of the regions of 
favorability $F_{\rm bh} =0$, and for higher temperature, the 
black hole is favorable, while for lower temperature, hot flat space 
is favorable.
If for some reason the system is in an
unfavored phase, then a first order phase transition occurs
to a favored phase.

\begin{figure}[h]
\centering 
\includegraphics[width=\linewidth]{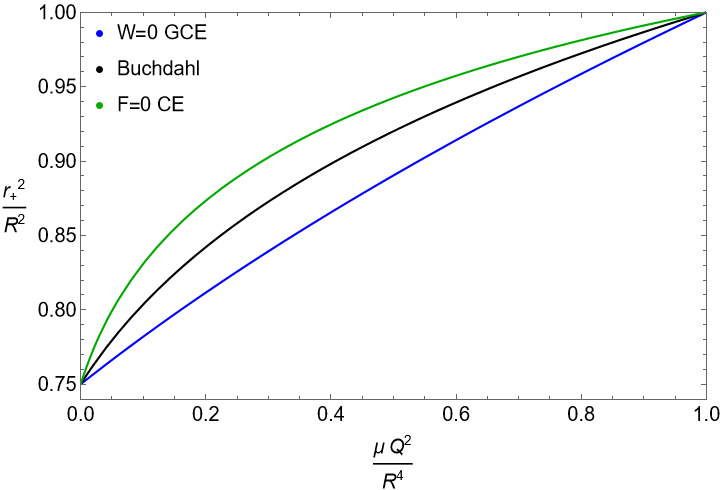}
\caption{\label{fig:Buchdahl}
Ratio $\frac{r_+^2}{R^2}$ in terms of
the electric charge parameter
$\frac{\mu Q^2}{R^4}$, $\mu=\frac{4}{3\pi}$,
for $d=5$ for three different cases: 
given by the condition 
$F_{\rm bh}=0$ in the canonical ensemble in green,
representing the stable solution $\frac{r_{+3}}{R}$; 
given by the condition $W_{\rm bh} =0$ 
in the grand canonical ensemble in blue,
representing the only stable solution; 
and given by generalized Buchdahl condition in black.}
\end{figure}

The problem of the thermodynamic phases is more involved
as we mentioned already.  When there is no electric charge, 
one has Schwarzschild space in $d=5$. It was found in
\cite{Andre:2020,Andre:2021} that, in the canonical ensemble 
of Schwarzschild space in $d=5$, the
condition $F_{\rm bh}=0$ corresponds to a value for $\frac{r_+}{R}$ that is
equal to the generalized Buchdahl bound radius \cite{Wright:2015},
which is the value $\left(\frac{r_+}{R}\right)_{\rm Buch}$ for
gravitational collapse of a self-gravitating system of energy $E$ and
radius $R$. Since we are maintaining $R$ fixed, we write
$\left(\frac{r_+}{R}\right)_{\rm Buch}\equiv \frac{r_{+\rm Buch}}{R}$, 
and in $d=5$, one has $\frac{r_{+\rm
Buch}}{R}=\frac{\sqrt3}{2}=0.86$, the latter equality being
approximate. Since for $Q=0$, the free energy of hot flat space is
zero, $F_{\rm hfs}=0$, meaning that there is no further coexistence with
hot flat space as soon as the black hole phase is favored, 
because the system tends to collapse.
For nonzero electric charge
parameter
$\frac{\mu Q^2}{R^4}$
there is no coincidence.
To compare the free energies,
we consider the case in which the shell
has radius equal to the cavity radius,
$\frac{r_{\rm shell}}{R}=1$, and so $F_{\rm shell}=0$,
meaning that the shell is a surrogate for hot flat space, i.e.,
$F_{\rm shell}=F_{\rm hfs}=0$,
indeed it is hot flat space
with electric charge at the boundary.
For nonzero 
$\frac{\mu Q^2}{R^4}$, we find that in the
canonical ensemble $F_{\rm bh}=0$ results in a $\frac{r_+}{R}$ value, both
for $\frac{r_{+3}}{R}$ and $\frac{r_{+4}}{R}$, 
that is higher than the generalized Buchdahl bound,
which is the value of $\frac{r_+}{R}$ for gravitational collapse
of a self-gravitating system of energy $E$, electric charge $Q$, and
radius $R$, see Fig.~\ref{fig:Buchdahl}.  
For an electric charge parameter lower or equal than the saddle
value
$\frac{\mu Q_s^2}{R^4}$, there is a temperature $RT$ at which
the solution $\frac{r_{+3}}{R}$ can assume the 
value of the Buchdahl bound, corresponding to a positive free energy
lower than the free energy of $\frac{r_{+1}}{R}$. 
For a system with this $RT$ or
higher, the system must suffer gravitational collapse into a black hole
with the corresponding $\frac{r_{+3}}{R}$. For an electric charge
higher than the saddle
value $y_s$, there is again a temperature $RT$ at which 
$\frac{r_{+4}}{R}$ assumes the Buchdahl bound,
with positive value of $F_{\rm bh}$. For a system with this $RT$ or
higher, then the system must collapse 
gravitationally into a black hole with the
corresponding $\frac{r_{+4}}{R}$.
%%%%%%%%%%%%%%%%%%%%%%%%%%%%%%%%%%%%%%%%%%%%%%%%%
%%%%%%%%%%%%%%%%%%%%%%%%%%%%%%%%%%%%%%%%%%%%%%%%%
Interesting to note that the picture in 
the grand canonical ensemble is different. It was found
\cite{Fernandes:2023} that the equation $W_{\rm bh}=0$, with $W_{\rm
bh}$ denoting the grand potential, results in a $\frac{r_+}{R}$ value 
for the single stable black hole, that is lower than the generalized
Buchdahl bound. One has thermodynamically that when 
the system has $W_{\rm bh}=0$
the black hole phase and hot flat space phase coexist, for $W_{\rm
bh}<0$ the black hole phase dominates, and for a certain definite
negative value of $W_{\rm bh}$ the value of $\frac{r_+}{R}$ of the
system is the same as the value of the Buchdahl bound. For larger 
temperatures, therefore the system must collapse 
gravitationally. The only phase of the system is the black hole phase and 
so there is no more coexistence, see Fig.~\ref{fig:Buchdahl} and
Appendix~\ref{app:BHzerofreeenergy}.  We admit we have not done a
thermodynamic treatment of gravitational collapse, but the arguments
given in this paragraph are plausible enough to assure us that once
there is sufficient thermodynamic energy inside the cavity, collapse
to a black hole sets in inevitably.

%%%%%%%%%%%%%%%%%%%%%%%%%%%%%%%%%%%%%%%%%%%%%%%%%%%%%%%%%%%%%%%%%
\section{Infinite cavity radius: The Davies limit and the Rindler
limit}
\label{sec:infiniteradiuscavity}
%%%%%%%%%%%%%%%%%%%%%%%%%%%%%%%%%%%%%%%%%%%%%%%%%%%%%%%%%%%%%%%%%

\subsection{Ensemble solutions in the $R\to +\infty$ limit:
Davies and Rindler}
\label{sec:infiniteradiuscavityR}

%\subsubsection{Ensemble solutions in the $R\to +\infty$ limit}

\begin{figure}[h]
    \centering
    \includegraphics[scale=0.40]{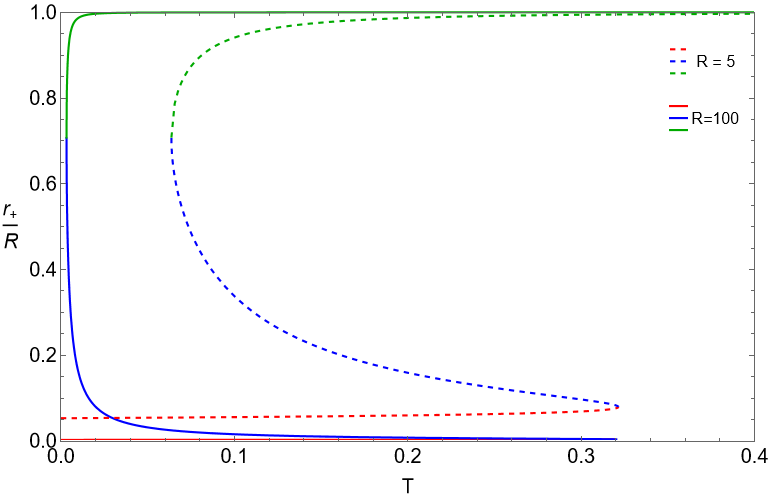}
    \caption{Plot of the solutions $r_{+1}$ in 
    red, $r_{+2}$ in blue and $r_{+3}$ in green 
    of the canonical ensemble in $d=5$ 
    as $\frac{r_+}{R}$ as a function of $T$ in Planck units, 
    for $\mu Q^2 = 0.005$, $\mu=\frac{4}{3\pi}$,
    and for two values of $R$, $R=5$ in dashed lines, and $R=100$ 
    in filled lines. One can see the emergence of the $r_{+1}$ and 
    $r_{+2}$ solution
    limits corresponding to the Davies limit
    as they get closer to the $\frac{r_+}{R} = 0$ axis, 
    and the $r_{+3}$ solution limit corresponding to the Rindler limit 
    as it gets closer to the $\frac{r_+}{R}=1$ 
    axis.
    \label{fig:rindlerdaviesind5}}
\end{figure}

We now analyze the infinite cavity radius limit, and discuss each 
solution that arises from this limit. As it turns, by
performing $R\to +\infty$ limit while keeping $T$ fixed and $Q$ fixed,
three different solutions are found. One observes from
Sec.~\ref{sec:ZeroLoopSolutions}, that there are three solutions for
$r_+(R,T,Q)$ if $\frac{\mu Q^2}{R^{2d-6}} < \frac{\mu Q_s^2}{R^{2d-6}}$.
By performing the
$R\to +\infty$ limit, the term $\frac{\mu Q^2}{R^{2d-6}}$ approaches
zero, and so the
solutions of the ensemble
in this limit
should correspond
to these three solutions under the
$R\to +\infty$ limit.
 For the  smallest and intermediate solutions, the
limit $R\to+\infty$ must be performed by fixing $T$ and $Q$, while
doing $\frac{r_+}{R}\to 0$.
For the largest solution, the limit $R\to
+\infty$ must be performed by fixing $T$ and $Q$, while doing
$\frac{r_+}{R}\to 1$.
The  smallest and intermediate solutions
correspond to Davies thermodynamic
solutions, while
the largest solution limit corresponds to
the Rindler solution.
These solution limits
occur for any $d$, in particular for $d=4$ and $d=5$
that we have been analyzing in more detail. 
In Fig.~\ref{fig:rindlerdaviesind5},
the behavior of the three solutions in $d=5$
can be seen
for a charge $\mu Q^2 = 0.005$, $\mu=\frac{4}{3\pi}$,
for two different $R$, $R=5$
and $R=100$, where the latter $R$
gives an idea of the $R\to\infty$ limit.
In this limit the scale $R$ is lost,
the scales set by the electric charge $Q$
and temperature $T$ at infinity are 
now the only two scales of the canonical ensemble.
We now comment briefly on
each solution.

The Davies solution corresponds to the smallest and intermediate
solution limits of the canonical ensemble when taking $R\to + \infty$,
with fixed $T$ and $Q$.  Thus, these are the solutions of the
electrically charged black hole in the canonical ensemble with
reservoir at infinity. This can be seen directly from the expression
of the temperature in Eq.~\eqref{eq:beta1}.  Since for these solutions
the behavior is $\frac{r_+}{R}\to 0$, one can maintain $r_+$ finite
during the limit $R\to +\infty$, thus obtaining the temperature
formula $T= \frac{d-3}{4\pi} \frac{r_+^{d-3} -\frac{\mu
Q^2}{r_+^{d-3}}}{r_+^{d-2}}$, which is obeyed by
the smallest and intermediate solutions.
This is precisely the Hawking temperature for
the electrically charged black hole.
From Fig.~\ref{fig:rindlerdaviesind5}
we see that the two
solutions tend to the axis $\frac{r_+}{R} = 0$ and  seem to get 
overlapped, which is due to the vertical axis being $\frac{r_+}{R}$. If one
regularizes the solutions through multiplying by $R$, one obtains 
the two solutions in $d$ dimensions,
which for $d=4$
are the Davies thermodynamic solutions.
Moreover, one can see that 
the solutions do not exist for all temperatures. This is because the 
two solutions only exist up to a critical temperature, the 
generalized Davies temperature,
after which there are no solutions. In the case represented  in 
Fig.~\ref{fig:rindlerdaviesind5}
which is $d=5$, the generalized Davies temperature, i.e.,
the temperature when
$R\to\infty$,
has the expression
$T_s = \frac{4}{10\pi \left(\sqrt{5 \mu Q^2} \right)^{\frac{1}{2}}}$,
and so
for $\mu Q^2 = 0.005$ as in the
figure it yields $T_s = 0.320$, with the last equality 
being approximate.

The Rindler solution is
the largest solution limit
that can be obtained from the ensemble
by keeping $T$ and $Q$ fixed, 
while doing $R\to +\infty$ and $r_+ \to R$ in Eq.~\eqref{eq:beta1}.
 In 
Fig.~\ref{fig:rindlerdaviesind5}, this solution
is the one that tends to
$\frac{r_+}{R}=1$.
The temperature dependence on the charge goes with 
$\frac{\mu Q^2}{R^{d-3}r_+^{d-3}}$, 
therefore such dependence in the limit $r_+ \to R$ and $R\to +\infty$
disappears. This happens
because the horizon radius of the black
hole tends to infinity and 
any contributions given by the charge become negligible. The 
expression for the temperature is now
the temperature of an
electrically uncharged black hole 
$T = \frac{d-3}{4\pi r_+ \sqrt{1 - \frac{r_+^{d-3}}{R^{d-3}}}}$.
Imposing that 
$T$ is fixed and finite leads to the condition that 
$r_+ \sqrt{1 - \frac{r_+^{d-3}}{R^{d-3}}}$ must 
tend to some constant when $R\to+\infty$ and 
$r_+ \to R$. One can show that in this limit 
the event horizon of
the black hole reduces to the Rindler horizon
and the cavity boundary is
accelerated  to yield the Unruh temperature $T$
set by the reservoir.

We now analyze in full detail the smallest and intermediate
solution limits 
arising from $R\to +\infty$, i.e., the Davies
solution. These are relevant since the formalism in 
this limit yields the Davies' thermodynamic theory of black holes for $d=4$.
We also
analyze in full detail the largest solution limit 
arising from $R\to +\infty$, i.e., the Rindler
solution.

%%%%%%%%%%%%%%%%%%%%%%%%%%%%%%%%%%%%%%%%%%%%%%%%%%%%%%%%%%%%%%%%%%%%%%
\subsection{Infinite cavity radius and Davies' thermodynamic theory
of black holes: Canonical ensemble,  thermodynamics, and stability of
electrically charged black hole solutions in the $R\to +\infty$ limit}
\label{sec:infiniteradiuscavityD}
%%%%%%%%%%%%%%%%%%%%%%%%%%%%%%%%%%%%%%%%%%%%%%%%%%%%%%%%%%%%%%%%%%%%%%

\subsubsection{The action for the canonical ensemble
in the $R\to +\infty$ limit}

The limit of infinite cavity radius for the small and intermediate 
solutions yields that
the canonical ensemble is
essentially defined by the temperature $T$ and the electric
charge $Q$ at infinity. It is this $R\to +\infty$ limit
that in four dimensions
gives Davies results \cite{Davies:1977}. This means that
Davies' thermodynamic theory
of black holes,
in this case of electrically charged black holes, can be
seen within the canonical ensemble formalism. Here we
have results for $d$ dimensions
in the $R\to +\infty$ limit, $d=4$ being a particular
case.

In the limit of infinite radius, the analysis above needs to be 
taken with care, since the quantities above
depend on the scale
given by the cavity radius $R$. To proceed with this limit, one must 
start from the reduced action in Eq.~\eqref{eq:actioninrp} and perform
the $R\to +\infty$
limit to obtain
\begin{align}
I_* = \frac{\beta}{\mu}\left(\frac{r_+^{d-3}}{2} 
+ \frac{\mu Q^2}{2r_+^{d-3}}\right)
-\frac{\Omega_{d-2} r_+^{d-2}}{4}\,.
\label{eq:actioninfinity}
\end{align}
The extrema of the action occurs when
\begin{align}
\beta=\iota(r_+)\,,\quad\quad
\iota(r_+)\equiv \frac{4\pi}{(d-3)}\frac{r_+^{d-2}}{r_+^{d-3}
-\frac{\mu Q^2}{r_+^{d-3}}}\,.
\label{eq:Hawkbeta}
\end{align}
This is the inverse Hawking temperature of the Reissner-Nordstr\"om 
black hole measured at infinity, i.e., performing the limit of 
infinite radius into Eq.~\eqref{eq:beta1}.

\subsubsection{Solutions and stability of the canonical ensemble
in the $R\to +\infty$ limit}

To find the solutions of this canonical ensemble, we must invert
Eq.~\eqref{eq:Hawkbeta} to get $r_+(\beta,Q)$, i.e., $r_+(T,Q)$.  This
can be done by solving the following equation
\begin{align}
\left( \frac{(d-3)}{4\pi T}\right)(r_+^{2d-6}-\mu Q^2) 
- r_+^{2d-5}= 0\,,
\end{align} 
which generally is not solvable analytically
for generic $d$, although
one can perform some qualitative analysis.
The function $\iota(r_+)$
in Eq.~\eqref{eq:Hawkbeta} 
has a minimum at 
$r_{+s1}^{d-3}= \sqrt{(2d-5)\mu}\,Q$,
which is a saddle point of the 
action for the black hole
and which we write as 
\begin{align}
r_{+s1}= \left(\sqrt{(2d-5)\mu}\,Q\right)^\frac{1}{d-3}
\label{r+davies}\,.
\end{align} 
This saddle point of the 
action of the black hole has the temperature 
$
T_{s1}^{-1}=  
\frac{2\pi}{(d-3)^2}(2d-5)
(\sqrt{(2d-5)\mu}Q)^{\frac{1}{d-3}}$, i.e., 
\begin{align}
T_{s1}=
\frac{(d-3)^2}
{2\pi (2d-5)
(\sqrt{(2d-5)\mu}Q)^{\frac{1}{d-3}}}
\label{Tdavies}\,.
\end{align} 
In $d=4$, this $T_{s1}$ is the Davies temperature, 
and so Eq.~\eqref{Tdavies} is the generalization of 
Davies temperature for higher dimensions.

By inspection,
one finds that for temperatures $T \leq T_{s1}$
there are two black holes, and for
$T > T_{s1}$
there are no black hole  solutions.
Indeed, for  temperatures in the
interval $0 <T \leq T_{s1}$,
there are two
solutions, the solution
$r_{+1}(T,Q)$ and the solution $r_{+2}(T,Q)$.
The solution
$r_{+1}(T,Q)$ is bounded in the interval 
$(\mu Q^2)^{\frac{1}{2d-6}}<r_{+1}(T,Q)\leq r_{+s1}$, where 
$r_{+1}(T\rightarrow 0,Q) = (\mu Q^2)^{\frac{1}{2d-6}}={r_{+}}_e$,
${r_{+}}_e$ being the radius of the extremal black hole, 
and 
$r_{+1}(T_{s1},Q) = r_{+s1}$. Moreover,
$r_{+1}(T,Q)$ is an increasing monotonic
function in $T$.
The solution $r_{+2}(T,Q)$ is bounded 
from below, i.e., $r_{+2}(T,Q) > r_{+s1}$, where 
$r_{+2}(T_{s1},Q) = r_{+s1}$, and
is unbounded from above, since 
at $T \rightarrow 0$, the 
solution $r_{+2}$ tends to infinity. Moreover,
$r_{+2}(T,Q)$ is a decreasing monotonic
function in $T$.
We note that the action given in Eq.~\eqref{eq:actioninfinity}
with $r_+$ holding for $r_{+1}(T,Q)$ or 
$r_{+2}(T,Q)$ is the action in zero loop approximation
that has been found in
\cite{fernandeslemosdavies:2024} directly from the
Gibbons-Hawking approach, rather than from York's approach
for a given $R$ with subsequently taking the $R\to\infty$ limit,
as we have been doing here.

Regarding stability, a solution is 
stable if $\frac{\partial \iota(r_+)}{\partial r_+}<0$,
as we have seen in the case of finite cavity.
This gives
\begin{align}
r_+\leq r_{+s1}\,,
\label{stabilityRinfinity}
\end{align}
with $ r_{+s1}$ given in Eq.~\eqref{r+davies}.
This means
that the solution is stable if the radius $r_+$ increases as the 
temperature increases. Therefore, the solution $r_{+1}$ is stable 
since it has this monotonic behavior, while the solution $r_{+2}$ 
is unstable since it has an opposite monotonic behavior.

\subsubsection{Thermodynamics in the $R\to +\infty$ limit}

\centerline{\small\it (i) Entropy, electric potential, and energy}
\vskip 0.1cm

\noindent
With the solutions of the canonical ensemble found in the limit of 
infinite radius of the cavity,  $R\to +\infty$,
one can 
find $I_0$, i.e., the action in the zero loop approximation 
given in Eq.~\eqref{eq:actioninfinity}
evaluated at the extrema of Eq.~\eqref{eq:Hawkbeta}.
The thermodynamics for the system follows 
through the correspondence $F = T I_0$,  where $F$ again is the 
Helmholtz free energy of the system and thus it can be written for this 
case as
\begin{align}
F = \frac{1}{\mu}\left(\frac{r_+^{d-3}}{2} 
+ \frac{\mu Q^2}{2r_+^{d-3}}\right)
-\frac{T \Omega_{d-2} r_+^{d-2}}{4}\,,
\label{eq:freeenergyinfinity}
\end{align}
where $r_+$ can be $r_{+1}(T,Q)$ or
$r_{+2}(T,Q)$.
Using the same calculation method from 
Sec.~\ref{sec:thermoquantities}, we obtain the entropy
as 
$S = \frac{\Omega_{d-2} r_+^{d-2}}{4}$, i.e.
\begin{align}
S = \frac14 A_+\,.
\end{align}
The thermodynamic pressure $p$ is zero,
\begin{align}
p=0\,.
\end{align}
The 
thermodynamic
electric potential is
\begin{align}
\phi = \frac{Q}{r_+^{d-3}}\,,
\end{align}
which is equal to the pure electric potential.
The energy, given by $E = F + TS$, can be written 
as 
$E = \frac{r_+^{d-3}}{2\mu} + \frac{Q^2}{2r_+^{d-3}}$.
But
the spacetime mass $m$ is given by
$m = \frac{r_+^{d-3}}{2\mu} + \frac{Q^2}{2r_+^{d-3}}$, see
also Appendix \ref{app:BHzerofreeenergy},  so
that the thermodynamic energy and the spacetime mass
are the same in the  $R\to +\infty$ limit, i.e., 
\begin{align}
E = m\,.
\label{eq:energyinfinity}
\end{align}
Thus, we can write the free energy
given in Eq.~\eqref{eq:freeenergyinfinity} as
\begin{align}
F=m-TS\,.
\label{eq:freeenergyinfinityagain}
\end{align}

We must note that the expressions for the entropy,
the pressure, the thermodynamic electric potential, 
and the energy are consistent with the limit of infinite 
radius to the respective expressions in 
Sec.~\ref{sec:thermoquantities}. Moreover, in this limit, 
the pressure $p$ vanishes, which is consistent with the 
absence of the variable $R$ in the action.

\vskip 0.4cm
\centerline{\small\it (ii) Smarr formula and the first law of black holes}
\vskip 0.15cm

\noindent
The energy in Eq.~\eqref{eq:energyinfinity} can be rewritten
in terms 
of the entropy and the charge as
$E = \frac{1}{2\mu}\left(\frac{4S}{\Omega_{d-2}}
\right)^{\frac{d-3}{d-2}} 
+ \frac{Q^2}{2}\left(\frac{4 S}{\Omega_{d-2}}
\right)^{\frac{3-d}{d-2}}$.
The energy function possesses the scaling property
$\nu^{\frac{d-3}{d-2}} E = E(\nu S, \nu^{\frac{d-3}{d-2}} Q)$, which allows
the use of the Euler relation theorem to have
$E= \frac{d-3}{d-2} TS + \phi Q$,
which is the formula obtained in Sec.~\ref{sec:thermoEuler} without 
the term $pA$. Indeed, the term $pA$ in the limit of infinite
reservoir radius 
has leading order $R^{-(d-3)}$, and so it vanishes. Since
from Eq.~\eqref{eq:energyinfinity}
$E=m$, we obtain 
\begin{align}
m = \frac{d-3}{d-2} TS + \phi Q\,,
\end{align}
which is the Smarr formula.

In this case the law
\begin{align}
dm =  TdS + \phi dQ\,,
\label{eq:fiirslawofbhm}
\end{align}
holds. This is exactly the first law of black hole mechanics.
This can be obtained from Eq.~\eqref{eq:firstlawoft}
in the $R\to\infty$ limit.
For $R$ finite, there is a first law of thermodynamics of the cavity
and does not correspond to the law of black hole mechanics.
For $R\to\infty$, the first law of black hole
thermodynamics and the first law of black hole mechanics
coincide into one same law, which is quite remarkable.
Moreover, in the electrically charged case, as opposed
to the Schwarzschild case, the thermodynamics of the canonical 
ensemble is valid,
since there is a region of the electric charge where the
system is thermodynamically stable.
It is from Eq.~\eqref{eq:fiirslawofbhm}
that Davies has started his  thermodynamic theory
of black holes for $d=4$. We have deduced it from the action
Eq.~\eqref{eq:actioninfinity}.

\vskip 0.4cm
\centerline{\small\it (iii)  Heat capacity and stability}
\vskip 0.1cm

\noindent
The thermodynamic
stability can be seen directly from applying the limit of 
infinite radius of the cavity in Eq.~\eqref{eq:Caq} and obtain the 
condition for the positivity of the heat capacity, which ensures 
that a solution is stable. The heat capacity in this limit is
\begin{align}
C_{Q}& = \frac{(d-2)\Omega_{d-2} r_+^{d-2}(r_+^{2d-6}-\mu Q^2)}
{4\left((2d-5)\mu Q^2 - r_{+}^{2d-6} \right)}\nonumber\\
&=
\hskip-0.1cm
\frac{S^3 E T}
{\frac{(d
\hskip-0.02cm
-
\hskip-0.03cm
3)\Omega_{\hskip-0.03cm d-2}^3}{4^5 \pi^2}
\hskip-0.15cm
\left[
\frac{
(\hskip-0.02cm
3d-8
\hskip-0.02cm
)
\mu^2 Q^4}
{\left(\frac{4S}{\Omega_{d-2}}\right)^{\hskip-0.06cm\frac{d-4}{d-2}}} 
\hskip-0.05cm
+
\hskip-0.05cm
(d
\hskip-0.02cm
-
\hskip-0.02cm
4)
\hskip-0.06cm
\left(\frac{4 S}{\Omega_{d-2}}\right)^{\frac{3d-8}{d-2}}
\right]
\hskip-0.16cm
-
\hskip-0.11cm
T^2
\hskip-0.06cm
S^3}\hskip-0.02cm,
\label{heatcapacityraw}
\end{align}
where
we have dropped the subscript $A$ in $C_{A,Q}$
since the evaluation is at infinity, and in the
second equality we wrote
the heat capacity in terms of the thermodynamic 
variables $S$, $E$, and $T$.
So there is stability if $C_Q\geq0$, i.e.,
$r_{+} \leq \left[(2d-5)\mu Q^2\right]^{\frac{1}{2d-6}}$,
which is Eq.~\eqref{stabilityRinfinity}
together with 
Eq.~\eqref{r+davies}.
This means that the solution $r_{+1}$ is thermodynamically
stable whereas the solution 
$r_{+2}$ is unstable. It must be noted also that $r_{+1}$ is an 
increasing monotonic function in $T$, which means the energy of the 
black hole increases of the temperature increases, as it is expected 
from a stable system. The opposite happens to the solution $r_{+2}$,
since it is a decreasing monotonic function in $T$ and so the energy
of the black hole decreases as temperature increases.

\subsubsection{Favorable phases}

There are two stable phases.
The small black hole $r_{+1}$ and hot flat space
with electric charge at infinity.
Since the  black hole $r_{+1}$ has positive
free energy and 
 hot flat space
with electric charge at infinity has
zero free energy, and systems
with lower free energy are preferred,
whenever the system finds itself
in the black hole $r_{+1}$ solution
it tends to transition to the
hot flat space 
with electric charge at infinity phase.

\subsubsection{$d=4$: Analysis leading to 
Davies' thermodynamic theory of black holes and
Davies point}

The dimension $d=4$ is specially interesting since in
the $R\to\infty$ gives the
results of Davies' thermodynamic theory of black holes
\cite{Davies:1977}. In this
setting, the reservoir of temperature
$T$ and electric charge $Q$
is at infinity.

The reduced action in Eq.~\eqref{eq:actioninfinity} 
in $d=4$ gives
\begin{align}
I_* = \frac{\beta}{2}
\left(r_+
+ \frac{Q^2}{r_+}\right)
-\pi r_+^2
\,,
\label{eq:actioninfinityd=4}
\end{align}
where $\mu=1$
and $\Omega_2=4\pi$.
The stationary points in $d=4$ occur when
\begin{align}
\beta=\iota(r_+)\,,\quad\quad \iota(r_+)\equiv
\frac{4\pi r_+^2}{r_+
-\frac{Q^2}{ r_+}}\,,
\label{eq:Hawkbeta0d=4}
\end{align}
corresponding to the inverse Hawking temperature 
of a charged black hole in $d=4$.

We invert 
Eq.~\eqref{eq:Hawkbeta0d=4} to get the solutions 
$r_+(T,Q)$. This results in solving
\begin{align}
\left( \frac{1}{4\pi T}\right)(r_+^2- Q^2) 
- r_+^3= 0\,,
\label{r+d=infinted=4}
\end{align} 
although we do not present the solutions here. 
The minimum of function $\iota(r_+)$
in Eq.~\eqref{eq:Hawkbeta0d=4}
occurs at 
$r_{+s1}=\sqrt3\,Q$,
being a saddle point of the 
action of the black hole.
We write the horizon radius of the 
saddle point as 
\begin{align}
{r_+}_{\rm D}=\sqrt3\,Q\,,
\label{r+daviesd=4}
\end{align} 
as in $d=4$ it gives the Davies horizon radius.
Since $r_+=m+\sqrt{m^2-Q^2}$, this means $m=\frac{2}{\sqrt3} Q$
at the saddle point,
a result that can be found in \cite{Davies:1977}. 
The temperature corresponding to the saddle point
is Eq.~\eqref{Tdavies} in $d=4$, or explicitly
\begin{align}
T_{\rm D}=
\frac{1}
{6\sqrt{3}\pi Q}\,,
\label{Tdavies4d}
\end{align} 
which is the Davies temperature, and it
is a result that can be extracted from \cite{Davies:1977}.

We present a summary of the behavior of the solutions for 
$d=4$.
For $0<T \leq T_{\rm D}$,
there are two
solutions, the solution
$r_{+1}(T,Q)$ and the solution $r_{+2}(T,Q)$.
The solution
$r_{+1}(T,Q)$ increases monotonically with $T$ and 
lies in the interval 
$ {r_{+}}_e
<r_{+1}(T,Q)\leq {r_+}_{\rm D}$,
where 
$r_{+1}(T\rightarrow 0,Q) = {r_{+}}_e=Q$
and 
$r_{+1}(T_{\rm D},Q) = {r_+}_{\rm D}= \sqrt3\, Q
$.
The solution $r_{+2}(T,Q)$ decreases monotonically with $T$ and
lies in the interval ${r_+}_{\rm D}<r_{+2}(T,Q) <\infty$, where 
$r_{+2}(T_{\rm D},Q) = {r_+}_{\rm D}= \sqrt3\, Q$.
For $T_{\rm D}<T $, there are no black hole solutions. 
Regarding stability, a solution is
stable if $\frac{\partial \iota(r_+)}{\partial r_+}\leq0$, i.e.
\begin{align}
r_+\leq {r_+}_{\rm D}\,.
\label{stabilityRinfinityd=4}
\end{align}
With ${r_+}_{\rm D}$ given in Eq.~\eqref{r+daviesd=4},
Eq.~\eqref{stabilityRinfinityd=4} can be turned in to the 
region in the electric charge
$\frac1{\sqrt3}r_+\leq Q\leq r_+$,
the latter term being simply the restriction to nonextremal case.
From Eq.~\eqref{stabilityRinfinityd=4},
we have that the solution $r_{+1}$ is stable 
while the solution $r_{+2}$ 
is unstable.

We summarize now the results for thermodynamics in $d=4$.
The free energy of the system is $F = T I_0$, coming from the 
zero loop approximation of the path integral.
From Eq.~\eqref{eq:actioninfinityd=4}, the free energy is
\begin{align}
F =
 \frac12
\left(r_+
+ \frac{Q^2}{r_+}\right)
-T\,\pi r_+^2
\,.
\label{eq:Fd=4}
\end{align}
From the derivatives of the free energy, 
we obtain the entropy 
$S = \pi r_+^2$, i.e., $S = \frac14 A_+$,
the thermodynamic pressure $p=0$ since there is no area dependence, 
the electric potential $\phi = \frac{Q}{r_+}$,
and the energy
$E = \frac12
\left(r_+
+ \frac{Q^2}{r_+}\right)$, from $E = F + TS$. 
Considering that this is the expression for
the spacetime mass $m$, we have
$E = m$. The free energy of
Eq.~\eqref{eq:Fd=4} is then $F=m-TS$.

The Smarr formula for $d=4$ is 
\begin{align}
m = \frac12\, TS + \phi Q\,.
\end{align}
Indeed, the first law of black hole mechanics 
$dm = TdS + \phi dQ$ coincides with the 
first law of thermodynamics, see above. 
The 
first law of black hole mechanics
is the expression from which Davies 
\cite{Davies:1977} started his analysis.
We have started our analysis from
the action Eq.~\eqref{eq:actioninfinityd=4}
and actually derived the first law from first principles.
Moreover, the system is stable thermodynamically in a 
range of values of the electric charge. 
On the other hand, 
the electrically charged case in the grand canonical
ensemble with the reservoir at infinity
is unstable. Gibbons and Hawking
through the action and the path integral
approach \cite{Gibbons:1977}
noticed this instability problem but
did not venture into the electric canonical ensemble to cure it.

The heat capacity of Eq.~\eqref{heatcapacityraw}
is for $d=4$ given by
\begin{align}
C_Q =
\frac
{2\pi r_+^2 \left(1-\frac{Q^2}{r_+^2}\right)} 
{3\frac{Q^2}{r_+^2}-1}=
\frac{S^3 E T}{\frac{\pi Q^2}{4} -
T^2
S^3}\,,
\label{heatcapacityraw4d}
\end{align}
where in the
second equality we wrote
the heat capacity in terms of the thermodynamic 
variables $S$, $E$, and $T$.
The system is thermodynamically stable if
$Q\geq\frac1{\sqrt3}r_+$, i.e.,
$\frac1{\sqrt3}r_+\leq Q\leq r_+$,
the latter term being the condition for nonextremal case.
The system is thermodynamically unstable if 
$0\leq Q<\frac1{\sqrt3}r_+$.
This is the same result as given in 
Eq.~\eqref{stabilityRinfinityd=4}
together with 
Eq.~\eqref{r+daviesd=4}. The heat capacity
$C_Q$ is infinitely positive 
at the point
$Q=\frac1{\sqrt3}r_+$
if one approaches it from 
higher $Q$, 
the heat capacity $C_Q$
is infinitely negative
if one approaches the point
$Q=\frac1{\sqrt3}r_+$ from lower $Q$. 
The heat capacity goes to zero at the extremal case 
$Q=r_+$. 
Precisely at the point $Q=\frac1{\sqrt3}r_+$,
this behavior of the heat capacity was 
found in \cite{Davies:1977}, and it was 
classified as being similar to a second order phase transition.
However, this
point  is a turning point rather than a
second order phase transition. This turning point
indicates the ratio of scales at which one has stability.
Indeed, when analyzing the heat capacity in terms of the 
temperature and electric charge, one has two distinctive 
curves, one for each solution, diverging at this point. 
But the unstable solution cannot be considered as a phase, 
due to its instability. The system will always remain in the 
stable configuration.
Note that the formula for $C_Q$ in 
the second line of Eq.~\eqref{heatcapacityraw4d}
is the same formula found in \cite{Davies:1977} by 
performing in Eq.~\eqref{heatcapacityraw4d}
the redefinitions 
$S \rightarrow 8\pi S$,
$T \rightarrow \frac{1}{8\pi}T$ and 
$\frac{C_Q}{8\pi} \rightarrow C_Q$.

\subsubsection{$d=5$: Analysis}

The dimension $d=5$ is a
typical higher dimension that we have been analyzing. 
We present here the summary for this specific case
in the $R\to +\infty$ limit.

The reduced action in Eq.~\eqref{eq:actioninfinity} 
in $d=5$ can be written simply as
\begin{align}
I_* = \frac{\beta}{2}
\left(\frac{3\pi r_+^2}{4} 
+ \frac{Q^2}{r_+^2}\right)
-\frac{\pi^2 r_+^3}{2}\,,
\label{eq:actioninfinityd=5}
\end{align}
where we have used $\mu=\frac{4}{3\pi}$
and $\Omega_3=2\pi^2$.
The stationary points are described by
\begin{align}
\beta=\iota(r_+)\,,
\quad\quad
\iota(r_+)\equiv
\frac{2\pi r_+^3}{r_+^2
-\frac{4 Q^2}{3\pi r_+^2}}.
\label{eq:Hawkbeta0d=5}
\end{align}
again corresponding to the inverse Hawking 
temperature of a charged black hole in $d=5$.

\begin{figure}[h]
\centering
\includegraphics[width=\linewidth]{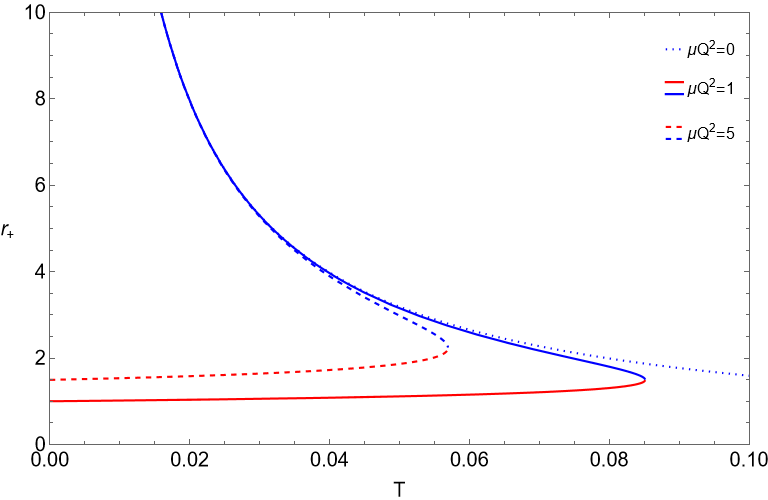}
\caption{Plot of the two solutions $r_{+1}(T,Q)$, in red, 
and $r_{+2}(T,Q)$, in blue, of the charged black hole in the 
canonical ensemble for infinite cavity radius, for two values of the 
charge, $\mu Q^2 = 1$ in filled lines, and $\mu Q^2 = 5$ in dashed 
lines, $\mu=\frac{4}{3\pi}$, in $d=5$.
\label{fig:rtinfd5}}
\end{figure}

The solutions are found by inverting
Eq.~\eqref{eq:Hawkbeta0d=5} to get $r_+(\beta,Q)$, i.e.,
$r_+(T,Q)$. 
This is the same as solving 
\begin{align}
\left( \frac{1}{2\pi T}\right)(r_+^4-\frac{4}{3\pi} Q^2) 
- r_+^5= 0\,,
\label{r+s=infinted=5}
\end{align} 
which cannot be done analytically. However, it can be 
analyzed qualitatively or solved numerically,
see Fig.~\ref{fig:rtinfd5} for this case  
of five dimensions. The function $\iota(r_+)$
in Eq.~\eqref{eq:Hawkbeta0d=5}
possesses a minimum at 
\begin{align}
    r_{+s1}= \left(\sqrt{\frac{20}{3\pi}}\,\,Q\right)^\frac{1}{2}\,,
    \label{r+sdaviesd=5}
    \end{align} 
which corresponds to a saddle point of the 
action of the black hole.
This generalizes the Davies radius to $d=5$.
The temperature at this saddle point is
$
T_{s1}^{-1}=  
\frac{10\pi}{4}
(\sqrt{\frac{20}{3\pi}}\,Q)^{\frac12}$, i.e., 
\begin{align}
T_{s1}=
\frac{4}
{
10\pi
\left(\sqrt{\frac{20}{3\pi}}\,Q\right)^{\frac12}
}
\label{Tdaviesd=5}\,.
\end{align} 
This generalizes the Davies temperature for $d=5$.

We summarize the behavior of the solutions in $d=5$.
For temperatures $0<T \leq T_s$
there are two
solutions, the solution
$r_{+1}(T,Q)$ and the solution $r_{+2}(T,Q)$.
The solution
$r_{+1}(T,Q)$ increases monotonically with the 
temperature and
is bounded by 
${r_{+}}_e
<r_{+1}(T,Q)\leq
r_{+s}$,
where 
$r_{+1}(T\rightarrow 0,Q) 
= {r_{+}}_e=\left(\sqrt{\frac{4}{3\pi}}\,\,
Q\right)^\frac{1}{2}$
is the extremal black hole, and 
$r_{+1}(T_{s1},Q) = r_{+s}=\left(\sqrt{\frac{20}{3\pi}}\,\,
Q\right)^\frac{1}{2}$.
The solution $r_{+2}(T,Q)$ decreases monotonically 
with the temperature and assumes values in the interval
$r_{+s1}<r_{+2}(T,Q)<\infty$, where 
$r_{+2}(T_s,Q) = r_{+s}$.
See Fig.~\ref{fig:rtinfd5}
for the plots of $r_{+1}$ and $r_{+2}$.
Regarding stability, a stable solution 
obeys $\frac{\partial \iota(r_+)}{\partial r_+}\leq0$.
This condition becomes
\begin{align}
r_+\leq r_{+s1}\,.
\label{stabilityRinfinityd=5}
\end{align}
With $r_{+s1}$ given in Eq.~\eqref{r+sdaviesd=5},
Eq.~\eqref{stabilityRinfinityd=5} can be transformed to
$\left(\frac{3\pi}{20}\right)^{\frac12}
r_{+}^2 \leq Q\leq\left(\frac{3\pi}{4}\right)^{\frac12}
r_{+}^2$, the latter term being 
the restriction to the nonextremal case.
From Eq.~\eqref{stabilityRinfinityd=5},
we obtain that $r_{+1}$ is stable and that 
$r_{+2}$ is unstable.
%%%%%%%%%%%%%%%%%%%%%%%%%%%%%%%%%%%%%%%%%%%%%%%%%%%%%%%%%%%%%%%%%%%%%%
%%%%%%%%%%%%%%%%%%%%%%%%%%%%%%%%%%%%%%%%%%%%%%%%%%%%%%%%%%%%%%%%%%%%%%
The plots
in Fig.~\ref{fig:rtinfd5} show the discussion above, 
namely the stable branch $r_{+1}$ and the unstable 
branch $r_{+2}$.
It is also seen clearly
that the plot of Fig.~\ref{fig:rtinfd5}
is the limiting $R\to\infty$ case
of Fig.~\ref{fig:rt5}. From Fig.~\ref{fig:rt5}
one finds that when $R\to\infty$,
the solution $r_{+3}$ disappears, leaving
$r_{+1}$ and $r_{+2}$, with
$r_{+1}$ and $r_{+2}$ meeting at a maximum temperature.
Also, from Fig.~\ref{fig:rt5}
we see that the $r_{+2}$ and $r_{+3}$ branches
meet at a minimum temperature, and these branches are
the ones that appears in the zero charge case of York,
here slightly modified due to the existence of
an electric charge $Q$.
More specifically,  comparing Fig.~\ref{fig:rtinfd5}
with  Fig.~\ref{fig:rt5}, one notes
that
the red and blue lines of Fig.~\ref{fig:rtinfd5}
are the stable and
unstable black holes of Davies, here in $d=5$,
and the red and blue lines 
of Fig.~\ref{fig:rt5} are precisely these branches
of black holes for finite reservoir radius $R$.
The blue and green branches in 
 Fig.~\ref{fig:rt5} correspond to
 York black holes. Thus, Fig.~\ref{fig:rt5}
 is a unified plot of
 York and Davies black holes.
Note further 
from Fig.~\ref{fig:rtinfd5}, that for the
electric charge going to zero,
the branch that survives in Fig.~\ref{fig:rtinfd5}
is
the blue branch, which corresponds to the unstable black 
hole $r_{+2}$,
and the solution goes up to the point characterized
by $T=\infty$ and $r_+=0$. This branch corresponds to
the original unstable Hawking black hole, the black hole
also found in the Gibbons-Hawking 
path integral approach.

We present the summary of 
the results for the thermodynamics in $d=5$.
The free energy can be obtained 
from the zero loop approximation of the 
path integral as $F = T I_0$.
From Eq.~\eqref{eq:actioninfinityd=5}, 
the free energy takes the form
\begin{align}
F = \frac{1}{2}
\left(\frac{3\pi r_+^2}{4} 
+ \frac{Q^2}{r_+^2}\right)
-T\frac{\pi^2 r_+^3}{2}\,.
\label{eq:Fd=5}
\end{align}
From its derivatives,
we obtain the entropy
as $S = \frac14 A_+$,
$A_+ = 2\pi^2 r_+^3$, the thermodynamic
pressure as $p=0$,
the 
thermodynamic
electric potential as $\phi = \frac{Q}{r_+^2}$,
and
the energy, given by $E = F - TS$, 
as 
$E = \frac{3\pi r_+^2}{8} + \frac{Q^2}{2r_+^2}$.
Note that this is exactly the expression for
the spacetime mass $m$, so the mean energy is
$E=m$.
The free energy of
Eq.~\eqref{eq:Fd=5} becomes $F=m-TS$.

The Smarr formula in $d=5$ takes the form
\begin{align}
m = \frac23\, TS + \phi Q\,.
\end{align}
Also, one has that the law
$dm =  TdS + \phi dQ$
holds. And so the first law of black hole mechanics 
coincides with the first law of thermodynamics.
Also, the system is stable thermodynamically in 
a small region of the charge, so this correspondence 
is valid.

The heat capacity of Eq.~\eqref{heatcapacityraw}
is now in $d=5$ given by
\begin{align}
C_Q =&
\frac
{3\pi^2  r_+^3 
\left(1-\frac{4}{3\pi}\frac{Q^2}{r_+^4}\right)} 
{2\left(\frac{20}{3\pi}\frac{Q^2}{r_+^4}-1\right)}\nonumber\\
 =&
\frac{S^3 E T}{\frac{7\pi^2}{36}Q^4
\left(\frac{2S}{\pi^2}\right)^{-\frac{1}{3}} 
+ \frac{\pi^4}{4^3}\left(\frac{2S}{\pi^2}\right)^{\frac{7}{3}} -
T^2
S^3}\,,
\label{heatcapacityraw5d}
\end{align}
where in the
second equality is in terms of the thermodynamic 
variables $S$, $E$, and $T$.
One has instability if 
$0\leq Q<\left(\frac{3\pi}{20}\right)^{\frac12}
r_{+}^2$, with $Q$ meaning its absolute modulus.
One has  thermodynamic stability if
$\left(\frac{3\pi}{20}\right)^{\frac12}
r_{+}^2\leq Q\leq\left(\frac{3\pi}{4}\right)^{\frac12}
r_{+}^2$, the latter term being the condition for 
the nonextremal case, and
this can also be derived from 
Eq.~\eqref{stabilityRinfinityd=5}
together with 
Eq.~\eqref{r+sdaviesd=5}. The heat capacity
$C_Q$ is infinitely positive 
at the point
$Q=\left(\frac{3\pi}{20}\right)^{\frac12}
r_{+}^2$, if this point is approached from 
higher $Q$, 
the heat capacity $C_Q$
is infinitely negative, if the point is approached
from lower $Q$. 
%%%%%%%%%%%%%%%%%%%%%%%%%%%%%%%%%%%%%%%%%%%%%%%%%%%%%%%
\begin{figure}[h]
\centering
\includegraphics[width=\linewidth]{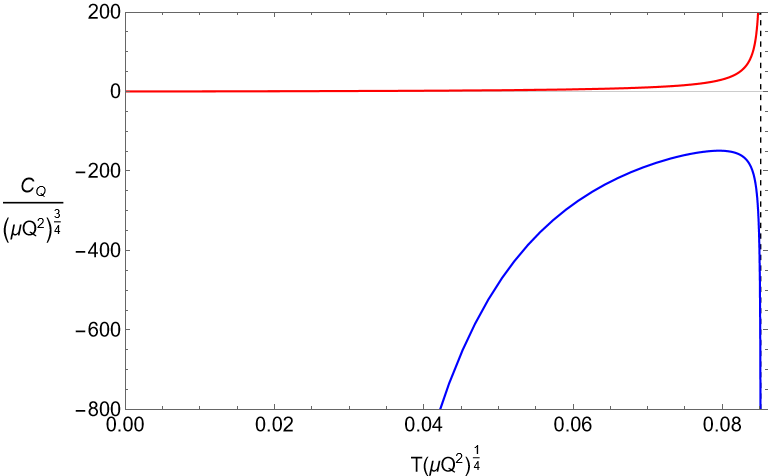}
\caption{The heat capacity $C_Q$ in $(\mu Q^2)^{\frac{3}{4}}$ units,
$\frac{C_Q}{(\mu Q^2)^{\frac{3}{4}}}$, 
is given as a function of the temperature $T (\mu Q^2)^{\frac{1}{4}}$
in $d=5$. In red, the heat capacity of $r_{+1}$ is represented, while 
in blue, the heat capacity of $r_{+2}$ is shown. 
There is a turning point at $T (\mu Q^2)^{\frac{1}{4}} =
\frac{4}{10\pi 5^{\frac{1}{4}}}$.}
\label{CAq5dRinfty}
\end{figure}
%%%%%%%%%%%%%%%%%%%%%%%%%%%%%%%%%%%%%%%%%%%%%%%%%%%%%%%
This is a turning point of the solutions, indicating 
the condition for stability. This is properly 
seen when analyzing the heat capacity with fixed 
temperature and electric charge, see Fig.~\ref{CAq5dRinfty}.
Indeed, the heat capacity is described by two curves, one for 
each solution $r_{+1}$ and $r_{+2}$, being positive for 
$r_{+1}$ and negative for $r_{+2}$. The system cannot be 
sustained in the solution $r_{+2}$ since it is unstable and 
so it can only be at the stable solution $r_{+1}$.

%%%%%%%%%%%%%%%%%%%%%%%%%%%%%%%%%%%%%%%%%%%%%%%%%%%%%%%%%%%%%%%%%%%%%%
\subsection{Infinite cavity radius and the Rindler limit: Cavity
boundary at the Unruh temperature}
%%%%%%%%%%%%%%%%%%%%%%%%%%%%%%%%%%%%%%%%%%%%%%%%%%%%%%%%%%%%%%%%%%%%%%

The largest solution limit can be obtained by keeping $T$ and $Q$ fixed, 
while doing $R\to +\infty$ and $r_+ \to R$ in Eq.~\eqref{eq:beta1}. 
The temperature dependence on the charge goes with 
$\frac{\mu Q^2}{R^{2d-6}}$, 
therefore such dependence in the limit $r_+ \to R$ and $R\to +\infty$
disappears. Intuitively, the black hole becomes very large such that 
any contributions from the charge become negligible. Then, the 
expression for the temperature reduces to the noncharged case, 
$T = (d-3)
(4\pi r_+)^{-1} (1 - \frac{r_+^{d-3}}{R^{d-3}} )^{-\frac12}$, 
but we still need to apply the limit. The requirement that 
$T$ is fixed and so finite leads to the condition that 
$r_+ \sqrt{1 - \frac{r_+^{d-3}}{R^{d-3}}}$ must 
tend to some constant under the limit of $R\to+\infty$ and 
$r_+ \to R$. Still, it seems unclear a priori
what the system in this limit describes.

In order to understand the limit, one can first consider the 
Euclidean Schwarzschild metric
$ds^2 = R^2\frac{4 r_+^2}{R^2(d-3)^2}
(1 - \frac{r_+^{d-3}}{r^{d-3}})\,d\tau^2 + 
(1 - \frac{r_+^{d-3}}{r^{d-3}})^{-1}R^2 \,d(\frac{r}{R})^2 
+ R^2(\frac{r^2}{R^2}) d\Omega^2_{d-2}$,
where we introduced the normalization by $R$ in the line element, 
with $0\leq\tau<2\pi$ and $r_+<r \leq R$. First, we need to 
consider $r_+\to R$ in the limit of infinite cavity and only then perform
$R\to \infty$. Therefore, we must consider the near horizon 
expansion of the metric. The normalized proper radial length is given 
by
$\epsilon(r) = \frac{1}{R}
\int_{r_+}^r (1 - \frac{r_+^{d-3}}{\rho^{d-3}})^{-\frac12}
d\rho= \frac{2}{d-3} 
(\frac{R}{r_+})^{\frac{d-5}{2}}\sqrt{(\frac{r}{R})^{d-3}
- (\frac{r_+}{R})^{d-3}}$,
valid 
at the near horizon, spanning 
the interval
$0<\epsilon< \epsilon(R)$. One can 
therefore rewrite the Schwarzschild metric in this limit
as
$ds^2 = (R^2 \epsilon^2 + \mathcal{O}(\epsilon^4))d\tau^2 + R^2 d\epsilon^2 
+ (R^2 + \mathcal{O}(\epsilon^2))d\Omega^2$.
Notice however that as $r_+\to R$, the total normalized radial proper 
length $\epsilon(R)$ tends to zero. It is now that we perform the limit 
$R\to +\infty$ but such that $R\epsilon(R)$ tends to a constant,
which we write as $\bar R$, ${\bar R}\equiv R\epsilon(R)$.
Thus, we have a new proper length
$\bar r$, defined as
\begin{align}
{\bar r}\equiv R\epsilon(r)\,,\quad\quad 0<{\bar r}<{\bar R}\,.
\label{barr}
\end{align}
The metric becomes in this limit 
\begin{align}
ds^2 = {\bar r}^2 d\tau^2 + d{\bar r}^2 + R^2 d\Omega^2\,,
\label{rindlermetric}
\end{align}
i.e.,
it becomes
the two-dimensional Euclideanized Rindler metric times 
a $(d-2)$-sphere with infinite radius. The metric on the $(d-2)$-sphere 
can be normalized by choosing a specific point on the sphere and performing 
the expansion around such point, obtaining $R^2 d\Omega^2 = \sum_{i=1}^{d-2} 
(dx^i)^2$, where $x^i$ are the new coordinates. The metric then reduces to the 
$d$ dimensional Euclideanized Rindler space. The system can now be interpreted 
as follows. The event horizon of the black hole reduces to the Rindler horizon 
at ${\bar r} =0$, while the cavity boundary is located at ${\bar R}$
and it is being accelerated. The proper acceleration 
of the cavity is precisely $\frac{1}{{\bar R}}$ and the 
temperature measured at the boundary of
the cavity is $T = \frac{1}{2\pi {\bar R}}$.

We now analyze what happens to the thermodynamic quantities in this 
Rindler solution limit. First, the temperature in Eq.~\eqref{eq:beta1} 
is finite and equals to $T = \frac{1}{2\pi {\bar R}}$.
Since $T$ is fixed by the ensemble
this gives the solution for the cavity boundary,
namely 
\begin{align}
{\bar R} = \frac{1}{2\pi T}\,.
\label{RrindlerfromT}
\end{align}
To be in equilibrium with the temperature $T$ of the reservoir,
the boundary itself ${\bar R}$ has to have a
Rindler acceleration that matches its Unruh temperature.
The free energy in Eq.~\eqref{eq:freeennergy} 
diverges negatively, $F\to-\infty$.
It diverges as $F= \frac{R^{d-3}}{\mu} 
- \frac{\Omega_{d-2}}{8\pi {\bar R}}R^{d-2}$,
and is negative since the 
power $R^{d-2}$ is always larger than $R^{d-3}$ for 
$R\to + \infty$.
This divergence is due to the fact that the area is
divergent. Thus, it is better to work
with a specific free energy, ${\bar F}$,
a free energy per unit area, defined as
 ${\bar F}\equiv \frac{F}{\Omega_{d-2}R^{d-2}}$.
Then,
\begin{align}
{\bar F}=-\frac{1}{8\pi {\bar R}}\,,
\label{Frindler}
\end{align}
so it is negative.
From 
Eq.~\eqref{eq:entropy}, the entropy also diverges,
$S\to\infty$, it diverges as 
$S = \frac{\Omega_{d-2}R^{d-2}}{4}$. 
Defining a specific entropy
${\bar S}\equiv \frac{S}{\Omega_{d-2}R^{d-2}}$
\begin{align}
{\bar S}=\frac14\,,
\label{Srindler}
\end{align}
so it is a constant.
The thermodynamic pressure in 
Eq.~\eqref{eq:pressure} is  finite, which we write as
\begin{align}
{\bar p}= \frac{1}{8\pi {\bar R}}\,, 
\label{prindler}
\end{align}
so 
${\bar p}= \frac{T}{4}$.
The electric potential in Eq.~\eqref{eq:phi} is zero,
\begin{align}
{\bar \phi}=0\,. 
\label{phirindler}
\end{align}
The thermodynamic energy from Eq.~\eqref{eq:energy}
obeys $E\to\infty$,
it diverges as $E = \frac{R^{d-3}}{\mu}$ positively.
Defining a specific energy, ${\bar E}$,
as
 ${\bar E}\equiv \frac{E}{\Omega_{d-2}R^{d-2}}$,
 one obtains
\begin{align}
{\bar E}=0\,.
\label{Erindler}
\end{align}
The heat capacity in Eq.~\eqref{eq:Caq} goes
positively as
$C_{A} = \frac{(d-2)(d-3)\Omega_{d-2}}{2}R^{d-4}
{\bar R}^2$.
So 
$
C_{A}=4\pi{\bar R}^2\;\;{\rm for}\,d=4$ and
$C_{A}\to\infty \;\;{\rm for}\,d>4$, 
i.e., for $d=4$ is finite
and depends on the temperature as $C_{A}= \frac{1}{\pi T^2}$,
and for
$d>4$  diverges. Since $C_A$
is positive,
this solution can then be considered stable.
Defining a specific heat capacity, ${\bar C}_A$,
as
 ${\bar C}_A\equiv \frac{C_A}{\Omega_{d-2}R^{d-2}}$
 gives
\begin{align}
{\bar C}_A=0\,.
\label{CArindler}
\end{align}
Although this solution has divergent quantities,
when one resorts to specific quantities, as one should
since the system is infinite, one finds
finite quantities.

For the ensemble with infinite radius one can try to define what is
the most preferred phase thermodynamically. However, it seems that the
two limiting solutions have different character. In the Davies
solution there is still a net electrically charge $Q$ at infinity.  In
the Rindler solution the electric charge has disappeared from the
context, so it is in fact a zero electric charge solution. Although
the starting ensembles are the same, the final ensembles in the
infinite radius limit are different. From the free energies, given
that the stable black hole in Davies solution has positive free energy
and the Rindler one has infinite negative free energy, one would
conclude that the Rindler solution is the most preferred phase. But in
fact the two solutions belong to different ensembles and cannot be
compared. As we have mentioned, the Davies stable solution
tends to disperse to hot flat space with electric charge at infinity.

%%%%%%%%%%%%%%%%%%%%%%%%%%%%%%%%%%%%%%%%%%%%%%%%%%%%%%%%%%%%%%%%%%%%%%
\section{Conclusions}
\label{sec:Concl}
%%%%%%%%%%%%%%%%%%%%%%%%%%%%%%%%%%%%%%%%%%%%%%%%%%%%%%%%%%%%%%%%%%%%%%

We have analyzed the canonical ensemble of a Reissner-Nordstr\"om
black hole in a cavity for arbitrary dimensions. The construction of
the canonical ensemble was done through the computation of the
partition function in the Euclidean path integral approach. The
action is the usual Einstein-Hilbert-Maxwell action with the
Gibbons-Hawking-York boundary term and an additional Maxwell boundary
term so that the canonical ensemble is well defined, all
terms having been Euclideanized.  We assumed that
the heat reservoir has a spherical boundary at finite radius $R$,
where the temperature is fixed as the inverse of the Euclidean proper
time length at the boundary, and also the electric charge is fixed by
fixing the electric flux at the boundary. We then restricted to
spherically symmetric spaces and assume regularity boundary conditions
that avoid the presence of conical and curvature singularities.

The zero loop approximation was then performed by first imposing the
Hamiltonian and the Gauss constraints, obtaining a reduced action that
depends on the fixed inverse temperature $\beta$, electric charge $Q$,
and the radius of the boundary $R$, and also depends on the radius of
the event horizon $r_+$ as a variable that is integrated through the
path integral.  We then found the equation for the stationary points
of the reduced action which are the solutions $r_+[\beta,Q,R]$, and
the condition of stability of the solutions. The equation cannot be
solved analytically.

The existence of the solutions of the ensemble were
analyzed for arbitrary dimensions. For charges smaller than a saddle,
or critical,
electric charge, there are always three possible solutions where
the one with the smallest radius and the one with the largest radius are
stable, and the other with intermediate radius is unstable. The value
of the saddle charge and the value of the radii that bound these
solutions, which are saddle points of the reduced action, were found
analytically. For the saddle charge, the unstable solution reduces to
a point, having formally only two solutions which are stable.  For
charges larger than the saddle charge, there is only one solution, and
this solution is stable. This analysis was then applied to the four
and five dimensional cases. Regarding stability, the
solutions are stable if the radius of the event horizon increases as
the temperature increases. For this case, the condition is given in
terms of the saddle points of the reduced action.

The thermodynamics of the electrically charged black hole was obtained
using that the partition function is related to the Helmholtz free
energy of the system in the canonical ensemble. Through the zero loop
approximation, the free energy was obtained. The entropy, the
thermodynamic electric potential,
the thermodynamic pressure, and the
thermodynamic energy  were
retrieved through the derivatives of the free energy. More precisely,
the entropy is the Bekenstein-Hawking entropy, the pressure
has the same expression of the
pressure of a self-gravitating  charged
shell with radius $R$, and the
thermodynamic electric potential is given by the usual expression. The
mean thermodynamic energy, which can be identified with a quasilocal
energy, was calculated through the definition of
free energy. Regarding thermodynamic stability, the configurations are
stable if the heat capacity with constant charge and area is
positive. The integrated first law, i.e., the Euler formula,
and the Gibbs-Duhem relation were
also found.

We analyzed the favorable states in the canonical ensemble.  A
favorable state is a stable state of the ensemble that has the lowest
value of the free energy. In some sense, transitions can occur between
phases. Here, for an electric charge lower than the critical charge,
there are two stable black hole solutions that are in competition,
with an existing first order phase transition between them. For the
critical charge, this first order phase transition becomes a second
order phase transition.  For a charge larger than the critical charge,
there is only one stable black hole solution. In the uncharged case,
there is a stable solution and hot flat space.  Pure hot flat space
does not seem to be a solution of the canonical ensemble since the
charge is fixed. Instead, we compare the stable solutions with a
nonself-gravitating charged sphere.  This covers two limits, the case
where we have flat space with a charge at the center, which is not a
solution and is never favorable, and another case where the charge
resides near the cavity or at the cavity. In this last case, it would
act as a hot flat space with electric charge at the boundary and the
corresponding free energy vanishes.
Considering this latter case There is a first order phase transition
between the largest black hole and hot flat space with electric charge
at the boundary.  The black hole solutions and the charged shell model
have been compared in a phase diagram.

In this work of the canonical ensemble of a Reissner-Nordstr\"om
black hole in a cavity for four and higher dimensions
there are several main achievements which can be stated:

First, the construction of the canonical ensemble and the thermodynamic
analysis of all generic $d$ dimensions in a unified way was 
done. Moreover,
significant cases were presented in all the detail, namely,
the dimension $d=4$ as the most important case, and the dimension
$d=5$ as a typical higher dimensional case.

Second, in the analysis of the specific heat $C_{A,Q}$ in terms of the
temperature and the electric charge, we found the existence of a
second order phase transition between the two stable solutions for a
critical electric charge parameter
$\frac{\mu Q_s}{R^{2d-6}}$ in arbitrary
dimensions.  For lower electric charge $\frac{\mu Q}{R^{2d-6}}$, we found two
turning points indicating the stability of the solutions, where the
heat capacity diverges and is double valued.  For higher charge
$\frac{\mu Q}{R^{2d-6}}$, we found that the heat capacity is always
positive.

Third, since in the canonical ensemble one can have two stable
black hole solutions, an analysis of the free energy has
enabled us to pick the black hole solution that is most favored
according to the temperature
and electric charge of the ensemble and find the possible
first order phase transitions.  Moreover, a comparison
with the free energy of hot flat space, emulated by an electric shell
at the boundary, has revealed the thermodynamic phase that is favored.
We have also argued that the Buchdahl bound is important in this
context, and the free energies for which the bound is superseded were
found, for higher free energies gravitational collapse sets in.

Fourth, the Davies thermodynamic theory of black holes has been shown
to follow from the electric charged canonical ensemble in the infinite
large reservoir limit when $d=4$.  The two ensemble solutions of lower
radii maintain, in this limit, their black hole character. One, with
smallest radius, is the stable one, and the other with intermediate
radius is unstable.  These two solutions meet at a saddle point.  The
thermodynamic quantities were found and in particular, the heat
capacity at constant area and charge was found.  In $d=4$, the
expression of the heat capacity reduces to the expression found by
Davies.  We have started from the action and the path integral
approach for a reservoir at infinity and showed that the formalism
gives the first law of black hole mechanics which, of course, is also
the first law of thermodynamics for black holes.  Davies, in the $d=4$
formulation of the theory, started directly from the first law of
black hole mechanics.  These results, reached through different means,
point toward the equivalence between black hole mechanics and black
hole thermodynamics through the canonical ensemble.

Fifth, the limit of infinite radius of the
boundary of the cavity, has revealed a surprise solution.
Indeed, the largest black hole solution of the ensemble, changes
character in this limit. The black hole solution
turns into a Rindler solution with the ensemble fixed temperature
being the Unruh temperature of the now accelerated boundary.

Sixth and last, the York path integral procedure, which was originally
applied to Schwarzschild black holes, has been followed throughout
this work for Reissner-Nordstr\"om black holes. We have shown that the
black hole solutions found represent the unification of York electrically
uncharged black holes and Davies electric charged black holes, in a
remarkable way. Indeed,
the two York type solutions, one larger and
stable, one smaller and unstable,
do appear, and the two Davies type solutions, the smaller and
unstable, and the even smaller and stable also do appear, in a
remarkable way.  York and Davies results follow from two different
limits of our work. York results follow from taking the zero electric
charge limit. Davies results follow from taking the infinite cavity
radius limit, i.e., by putting the heat reservoir at infinity.  This
latter case can also be seen to stem from York's generic
reduced action approach with the
boundary at infinity, which in turn yields the Gibbons-Hawking path
integral formulation to black hole thermodynamics.  The
Gibbons-Hawking approach was originally applied to electrically
uncharged black holes and it was found that there was an unstable
black hole solution, the Hawking black hole, and thus no consistent
thermodynamics. It was also applied to an electrically charged black
hole in the grand canonical ensemble, and it was found a solution that
was unstable.  Had it been applied directly to electrically charged
black holes in the canonical ensemble,
one would have found that thermodynamic stable solutions exist
to vindicate the approach. We have filled this gap here.

What does remain to be understood? Here the interest has been in the
thermodynamic interaction of a black hole in a cavity with a boundary
of finite size and fixed temperature, as well as in the interaction of
the gravitational field with the electromagnetic field in such a
system.  The formalism by its very distinctive features, i.e., its
Euclidean character, applies only to the outside of a black hole event
horizon.  The black hole interior and its singularity are not
considered in the analysis.
Thus, the question about the nature of the singularity remains.  
It is expected that the
singularity is 
described by a Planck scale object, however
intricate the description might be. 
A canonical formalism for micro black holes, say of the order
of ten Planck radii, seems valid, after all Hawking
radiation, a tamed radiation at most of the scales,
if left by itself, slowly peels the singularity away.
If that radiation interacts harmoniously with the boundary
of a cavity, a thermodynamic procedure might be valid
and show how the black hole horizon and the singularity
fuse into one single describable object.

%%%%%%%%%%%%%%%%%%%%%%%%%%%%%%%%%%%%%%%%%%%%%%%%%%%%%%%%%%%%%%%%%%%%%%
\section*{Acknowledgements}
%%%%%%%%%%%%%%%%%%%%%%%%%%%%%%%%%%%%%%%%%%%%%%%%%%%%%%%%%%%%%%%%%%%%%%
We thank conversations with Francisco J. Gandum and
Oleg B. Zaslavskii. We acknowledge
financial support from Funda\c c\~ao para a Ci\^encia e
Tecnologia - FCT through the project~No.~UIDB/00099/2020 and
project~No.~UIDP/00099/2020. T.V.F acknowledges a grant from FCT 
no. RD0970 and the support from
the project 2024.04456.CERN.

%\vfill

\appendix

%%%%%%%%%%%%%%%%%%%%%%%%%%%%%%%%%%%%%%%%%%%%%%%%%%%%%%%%%%%%%%%%%%%%%%
\section{The Euclidean action for the canonical ensemble,
boundary conditions, Ricci scalar, Euler characteristic,
and the action with boundary conditions}
\label{bcRicciandEuler}
%%%%%%%%%%%%%%%%%%%%%%%%%%%%%%%%%%%%%%%%%%%%%%%%%%%%%%%%%%%%%%%%%%%%%%

In this appendix, we derive the conditions that were set 
in Sec.~\ref{sec:Canonical1} to find the reduced action
from the general Euclidean action.
Some of the equations appearing in that section are
repeated here for the sake of completeness and self-containment.

The system consisting of an electrically charged black hole 
inside a cavity in $d$ dimensions has an Euclidean action 
\begin{align}
    &I = - \frac{1}{16\pi}\int_M R \sqrt{g}d^d x
- \frac{1}{8\pi} \int_{\partial M} (K-K_0)\sqrt{\gamma} d^{d-1}x
     \nonumber\\
&    + \frac{(d-3)}{4\Omega_{d-2}}\int_M F_{ab}F^{ab}\sqrt{g}d^dx
\nonumber\\
& + \frac{(d-3)}{\Omega_{d-2}}\int_{\partial M}F^{ab}A_{a}n_b 
    \sqrt{\gamma}d^{d-1}x\,.
    \label{appeq:action1}
\end{align}
$R$ is the Ricci scalar of the space,
$g$ is the determinant of the metric $g_{ab}$, 
the extrinsic curvature of the boundary of the cavity 
is $K_{ab}$, $K$ is its trace, 
$K_0$ denominates the trace of the
extrinsic curvature of the boundary of the cavity 
embedded in flat Euclidean space, 
$\gamma$ is the determinant of the induced 
metric $\gamma_{\alpha\beta}$ on the boundary of the cavity, 
$\Omega_{d-2}$ is the surface area of a
$d-2$  unit sphere and appears 
for practical purposes,
$F_{ab} = \partial_a A_b - \partial_b A_a$ is the Maxwell
tensor, $A_a$ is the
electromagnetic vector potential, and
$n_b$ is the outward unit normal vector to the boundary
of the cavity. The indices $a,b$ label the spacetime 
indices running from $0$ to $d-1$, and $\alpha, \beta$ 
are indices on the boundary running from $0$ to $d-2$.
To prescribe 
the canonical 
ensemble, one has to 
set a boundary term related to the Maxwell tensor 
\cite{Braden:1990}. 
This term fixes the electric flux given by the 
integral of the Maxwell tensor on a $(d-2)$-surface,
i.e., it fixes
the electric charge. If instead, the potential vector 
is fixed, one is in the presence of the grand canonical
ensemble,
see \cite{Fernandes:2023} for this case.
Note that Eq.~\eqref{appeq:action1} corresponds to
Eq.~\eqref{eq:action1} in the main text.
It
is useful to rewrite the Maxwell boundary term in the action
Eq.~\eqref{appeq:action1}. Using the 
divergence theorem and that $\nabla_b(F^{ab}A_a) = - \frac{1}{2}
F_{ab}F^{ab} + \nabla_bF^{ab} A_a$, one transforms the boundary 
Maxwell term into a bulk term, obtaining the action
\begin{align}
    &I = - \frac{1}{16\pi}\int_M R \sqrt{g}d^d x
- \frac{1}{8\pi} \int_{\partial M} (K-K_0)\sqrt{\gamma} d^{d-1}x
\nonumber \\
 &   - \frac{(d-3)}{4\Omega_{d-2}}\int_M F_{ab}F^{ab}\sqrt{g}d^dx
\nonumber \\
    & + \frac{(d-3)}{\Omega_{d-2}}
    \int_M  A_a\nabla_b F^{ab} \sqrt{g}d^dx
    \,.
\label{appeq:action2}
\end{align}

Now, we develop
the line element.
We want to treat spherically symmetric Euclidean spaces,
so that the Euclidean path integral is
to be performed along metrics 
which have spherical symmetry.
The space is then 
given by the warped product $\mathbb{R}^2\times \mathbb{S}^{d-2}$
with $\mathbb{R}^2$ being the Euclidean two-space, 
$\mathbb{S}^{d-2}$ being a 
$(d-2)$-sphere with radius $r$, and 
$r^2$ being
the warping function.
The line element $ds^2$  of such a space is given by
\begin{align}
ds^2 = b^2(y) d\tau^2 + \alpha^2(y) dy^2 + r^2(y)d\Omega_{d-2}^2\,,
\label{appeq:Euclideanmetric}
\end{align} 
where
$\tau$ is the periodic Euclidean time with 
range $0\leq\tau<2\pi$,
and is in fact an angular coordinate, 
 $y$ is a spatial radial coordinate with range $0\leq y\leq1$, 
$b(y)$ and $\alpha(y)$ 
are functions of $y$,
the radius of the $(d-2)$-sphere is $r(y)$,
and $d\Omega_{d-2}^2$ is the line element of the unit 
$(d-2)$-sphere with total area 
$\Omega_{d-2} = \frac{2 \pi^{\frac{d-1}{2}}}{\Gamma(\frac{d-1}{2})}$, 
where $\Gamma$ is the gamma function. 
Since $0\leq y\leq1$,
it is clear that the boundaries to this space
given by the line element
of Eq.~\eqref{appeq:Euclideanmetric}  are at $y=0$ and $y=1$.
The functions 
$b(y)$, $\alpha(y)$, and $r(y)$ are
to be integrated in the path integral.

Given the line element Eq.~\eqref{appeq:Euclideanmetric},
we can
develop the action of Eq.~\eqref{appeq:action2}
with the considered terms involved.
The Ricci tensor $R_{ab}$
and its
contraction Ricci scalar $R$
which depend on second derivatives of the metric 
$g_{ab}$, form the Einstein tensor 
 $G_{ab}$, with
$G_{ab}=R_{ab}-\frac12 g_{ab}R$.
Then, the Ricci scalar for the metric in
Eq.~\eqref{appeq:Euclideanmetric} is 
given by
\begin{align}
    -\frac{1}{16\pi}R = 
    \frac{1}{8\pi \alpha b r^{d-2}}
    \left( \frac{r^{d-2}b'}{\alpha}\right)'
    + \frac{1}{8\pi}{G^{\tau}}_{\tau}\,,
    \label{appeq:ricciscalar0} 
\end{align}
where ${G^{\tau}}_{\tau}$ is the time-time component of the 
Einstein tensor and is given by 
\begin{align}
    {G^{\tau}}_{\tau} = \frac{(d-2)}{2 r' r^{d-2}}\left[r^{d-3}
\left(\frac{r'^2}{\alpha^2} -1\right) \right]'\,.
\label{appeq:Einsteintensor}
\end{align}
The Gibbons-Hawking-York 
boundary term is given by
\begin{align}
&-\frac{1}{8\pi}\left(K-K_0\right)_{y=1} 
=  \left(\frac{(d-2)}{8\pi r}\left(1 -
\frac{r'}{\alpha}\right)\right)_{y=1} \nonumber\\
&- \left(\frac{1}{8\pi b r^{d-2} }
\left(\frac{r^{d-2}b'}{\alpha}\right)\right)_{y=1}\,,
\label{appeq:Gibbonshawkingyork}
\end{align}
where it was used that the extrinsic curvature of a constant $y$ 
hypersurface is $\mathbf{K} = \frac{bb'}{\alpha}d\tau 
+ \frac{rr'}{\alpha}d\Omega_{d-2}^2$
and that $\mathbf{K}_0=rd\Omega_{d-2}^2$
is the extrinsic curvature of the hypersurface embedded in flat 
space. With respect to the bulk terms depending on the Maxwell field,
one has
\begin{align}
    -\frac{(d-3)}{4\Omega_{d-2}}F_{ab}F^{ab} = 
    -\frac{(d-3)}{2\Omega_{d-2}} 
    \frac{{A'_{\tau}}^2}{\alpha^2 b^2}\,,
\label{appeq:Maxwelldinamicterm}
\end{align}
where $F_{y\tau} = A'_\tau$ was used, and
\begin{align}
\frac{(d-3)}{\Omega_{d-2}}\nabla_b F^{ab}A_{a}
= -\frac{(d-3)}{\Omega_{d-2} \alpha b r^{d-2}}
\left(\frac{r^{d-2}A'_{\tau}}{b\alpha}\right)'
A_{\tau}\,,
\label{appeq:MaxwellEq}
\end{align}
where $\nabla_a F^{\tau a} = 
-\frac{1}{\alpha b r^{d-2}}
\left(\frac{r^{d-2}A'_\tau}{\alpha b}\right)'$
was used.

We now study the boundary conditions.
We study first the boundary conditions for the geometry
at
 $y=0$ and 
 at $y=1$, and afterward
the boundary conditions for the Maxwell field
at
 $y=0$ and 
 at $y=1$.

The boundary conditions for the geometry at
 $y=0$ have several important features.
 We also comment on the connection of these 
to the Euler characteristic.
We assume that the hypersurface $y=0$ corresponds 
to the bifurcate two-surface 
of the event horizon of the 
electrically charged black hole, so we must impose the conditions
\begin{align}
& b(0) = 0\,, \label{appeq:condb0}\\
& r(0) = r_+\label{appeq:condrp}\,,
\end{align}
where $r_+$ is the horizon radius. 
The conditions
given in Eqs.~\eqref{appeq:condb0} and~\eqref{appeq:condrp}  
impose that the $y=0$ hypersurface corresponds 
to $\{y=0\}\times\mathbb{S}^{d-2}$, 
i.e., a point times a $(d-2)$-sphere.
The $y=0$ point in the $(\tau,y)$ sector
coincides with the central point 
of the $\mathbb{R}^2$ plane in polar coordinates, 
since $\tau$ is an angular coordinate
and $y$ is a radial coordinate. 
The $y=0$ hypersurface can be seen as 
the limit $y\rightarrow 0$ of   $y={\rm constant}$ hypersurfaces,
with these latter having  an
$\mathbb{S}^1 \times \mathbb{S}^{d-2}$ topology. 
For the metric to be smooth as $y$ goes to zero, these
$y={\rm constant}$ hypersurfaces $\mathbb{S}^1 \times \mathbb{S}^{d-2}$
must go smoothly to $\{y=0\}\times\mathbb{S}^{d-2}$.
Analytically, one 
can expand the line element given in
Eq.~\eqref{appeq:Euclideanmetric}
around $y=0$ with the boundary 
conditions set
in Eqs.~\eqref{appeq:condb0}-\eqref{appeq:condrp}.
This yields 
\begin{align}
&ds^2 = \left[\left(\frac{b'}{\alpha}\right)^2_{\hskip-0.1cm y=0}
\varepsilon^2 
+ \left(\frac{b'}{\alpha^2}
\left(\frac{b'}{\alpha}
\right)'\right)_{\hskip-0.1cm y=0}\,\varepsilon^3 
+ \mathcal{O}(\varepsilon^4)\right]d\tau^2 
\notag \\
&+ d\varepsilon^2  + \left[r_+ + \left( r'\alpha\right)_{y=0}\,
\varepsilon 
+ \mathcal{O}(\varepsilon^2)\right]^2d\Omega_{d-2}^2\,,
\label{appeq:expansionmetric}
\end{align} 
where $b'$ is defined as $b' = \frac{db}{dy}$, 
$r'$ is defined as $r' = \frac{dr}{dy}$,
$\left(\frac{b'}{\alpha}\right)'$ is defined as 
$\left(\frac{b'}{\alpha}\right)'= 
\frac{d}{dy}\left(\frac{1}{\alpha}\frac{db}{dy}\right)$,
$\left(\frac{b'}{\alpha}\right)_{\hskip-0.1cm y=0}$ means 
$\left(\frac{b'}{\alpha}\right)$ evaluated at $y=0$, $\varepsilon$ is 
defined as
$\varepsilon = \int^\delta_0 \alpha dy$ for small $\delta$
and small  $\varepsilon$, assuming that the integral is well-defined, as
it should be if the metric is smooth. 
The term $\left(\frac{b'}{\alpha}\right)_{\hskip-0.1cm y=0}$
may be absorbed into a redefinition 
of $\tau$ with the caveat that the period of $\tau$ becomes 
$2\pi \left(\frac{b'}{\alpha}\right)_{\hskip-0.1cm y=0}$.
This means there is a deficit angle 
and so a conical singularity. Therefore, for smoothness of 
the metric, 
we impose a third condition 
\begin{align}
\left(\frac{b'}{\alpha}\right)(0)= 1\,,
\label{appeq:condbprime0}
\end{align}
where
$\left(\frac{b'}{\alpha}\right)(0)
\equiv\left(\frac{b'}{\alpha}\right)_{\hskip-0.1cm y=0}$.
With Eq.~\eqref{appeq:condbprime0} considered, one can compute 
the Ricci scalar of the metric in Eq.~\eqref{appeq:expansionmetric}
and obtain the  problematic terms at $y=0$. One finds
\begin{align}
\hskip -0.2cm
R = -\frac{2(d-2)}
{\varepsilon r_+} \left(\frac{r'}{\alpha}\right)_{\hskip-0.1cm
y=0}
- 
\frac{2}
{\varepsilon} \left(\frac{1}{\alpha}
\left(\frac{b'}{\alpha}\right)'\right)_{\hskip-0.1cm y=0}
+ \mathcal{O}(1)\,,
\label{appeq:Riccicondbprime0}
\end{align}
where we have used Eq.~\eqref{appeq:condbprime0}.
For the curvature invariant $R$ to be 
well-defined and so for the space to be smooth, one must impose 
a fourth and a fifth condition, namely,
\begin{align}
    & \left(\frac{r'}{\alpha}\right)(0) = 0\,,
    \label{appeq:condalphar0}\\
    & \left(\frac{1}{\alpha}\left(\frac{b'}{\alpha}\right)'\right)
    \hskip-0.1cm (0) = 0\,,
    \label{appeq:condb2prime0}
\end{align}
with $\left(\frac{r'}{\alpha}\right)(0)
\equiv\left(\frac{r'}{\alpha}\right)_{\hskip-0.1cm y=0}$
and
$ \left(\frac{1}{\alpha}\left(\frac{b'}{\alpha}\right)'\right)
\hskip-0.1cm (0)
\equiv
\left(\frac{1}{\alpha}\left(\frac{b'}{\alpha}\right)'\right)
_{\hskip-0.1cm y=0}$.
In even dimensions, 
the condition
given in Eq.~\eqref{appeq:condalphar0} is 
equivalent to requiring that the Euclidean space considered
has an Euler characteristic $\chi = 2$ by the Chern-Gauss-Bonnet 
formula. For odd dimensions, the Euler characteristic vanishes and 
so this requirement is not satisfactory. Nevertheless, the 
requirement that the Ricci scalar is well-defined suffices. 
One can also see that this condition is equivalent to requiring 
that the event horizon
of the black hole is a null hypersurface, if one performs a 
Wick transformation. The condition Eq.~\eqref{appeq:condb2prime0}
means that $b$ does not have a dependence in $\varepsilon^3$, but, 
for some coordinate $y$, it may indicate that 
if
$\left(\frac{b'}{\alpha}\right)'_{\hskip-0.1cm y=0}$
is nonzero and finite, then 
$\alpha|_{y=0}$
must diverge. Indeed, this is satisfied by the 
Reissner-Nordstr\"om line element with coordinate choice $y=r$ 
found by solving Einstein's
equations, as it is the case in this setting.
We note that condition 
Eq.~\eqref{appeq:condb2prime0} is not referred 
in~\cite{York:1986,Braden:1990,Fernandes:2023}. 
The boundary conditions for the geometry at
 $y=1$ are now given.
Here, we impose the condition
\begin{align}
    &  b(1)=\frac {\beta}{2\pi}\,,
    \label{appeq:betacond}\\
    & r(1) = R\,,
    \label{appeq:r1}
\end{align}
where,
$\beta$ is the inverse temperature of the cavity.
The condition 
Eq.~\eqref{appeq:betacond} is usually written as
$\beta = 2\pi b(1)$.
This condition,  Eq.~\eqref{appeq:betacond},
comes from the definition of the path integral 
as stated in Sec.~\ref{sec:partition}, 
and it 
imposes 
that the total Euclidean proper time of the boundary
of the cavity is fixed 
and it is equal to the inverse temperature of the cavity. 
The condition given in  Eq.~\eqref{appeq:r1}
states that 
the hypersurface $y=1$ corresponds to the boundary of the cavity 
with radius $r(1) = R$. 

The boundary conditions for the Maxwell field
are now given.
Due to spherical symmetry and admitting the nonexistence of 
magnetic monopoles, the only nonvanishing components of the 
Maxwell tensor $F_{ab}$ are $F_{y\tau} = -F_{\tau y}$. Moreover,
we choose a gauge where the only nonvanishing component of the 
vector potential is $A_{\tau}(y)$. Therefore, the Maxwell tensor 
$F_{ab}$ is only described by $F_{y\tau} = A'_\tau$.
Therefore, the boundary condition for the Maxwell field
at $y=0$ is given by the requirement that
\begin{align}
    A_{\tau}(0) = 0\,.\label{appeq:Atau0}
\end{align}
At $y=1$, we fix the electric charge by fixing the electric 
flux given by $\int_{\substack{y=1 \\ \tau = c}} F^{ab}dS_{ab} =
2 i\Omega_{d-2} Q$,
where $c$ is a constant, $Q$ is the charge of the black
hole and $dS_{ab}$ is the surface 
element of the $y=1$ and $\tau = 0$ surface,
i.e.,
\begin{align}
    & \int_{\substack{y=1 \\ \tau = c}} F^{\tau y}dS_{\tau y} =
    i\Omega_{d-2} Q\,.\label{appeq:Fcond}
\end{align}

Putting together all these conditions with the action 
Eq.~\eqref{appeq:action1}, or
Eq.~\eqref{appeq:action2},
the partition function is given only 
in terms of the radius of the cavity $R$, the inverse temperature 
$\beta$ and the charge $Q$, which are fixed quantities of the system.
Given the boundary conditions just found
one can use them in
Eqs.~\eqref{appeq:ricciscalar0}-\eqref{appeq:MaxwellEq}
to find the final form of the action
 Eq.~\eqref{appeq:action1}, or Eq.~\eqref{appeq:action2}.
We observe that the first term integrated over $y$ of 
Eq.~\eqref{appeq:ricciscalar0} yields 
$\frac{1}{8\pi}\left(\frac{r^{d-2}b'}{\alpha}\right)_{y=1} 
- \frac{1}{8\pi}\left(\frac{r^{d-2}b'}{\alpha}\right)_{y=0}$, i.e., 
a boundary term at $y=0$ and a boundary term at $y=1$. The 
boundary term at $y=1$ cancels with the last term 
in Eq.~\eqref{appeq:Gibbonshawkingyork}, therefore the only surviving
boundary term of the Ricci scalar is
$- \frac{1}{8\pi}\left(\frac{r^{d-2}b'}{\alpha}\right)_{y=0}$. 
Moreover, by using the boundary condition 
Eq.~\eqref{appeq:condbprime0}, this 
term becomes $- \frac{1}{8\pi}\left(\frac{r^{d-2}b'}{\alpha}\right)_{y=0} 
= -\frac{1}{8\pi}r^{d-2}_+$. One can proceed with the integrations 
at the cavity, since the integrands do not depend on time or on 
the angles, and one obtains the full action 
\begin{align}
    &I[\beta,Q,R;b,\alpha,r,A_\tau] = \frac{\beta R^{d-3}}{\mu}
    \left(1 - \left(\frac{r'}{\alpha}\right)(1)\right) \notag \\
    &- \frac{\Omega_{d-2}}{4}r_+^{d-2} -\frac{(d-3)}{\Omega_{d-2}}
    \int_M \left(\frac{r^{d-2}A'_\tau}{b\alpha}\right)'
    A_\tau d\tau dy d\Omega_{d-2} \notag\\
    &  + 
    \int_M \frac{\alpha b r^{d-2}}{8\pi}
    \left({G^\tau}_\tau 
    - \frac{4\pi(d-3)}{\Omega_{d-2}}\frac{{A'_\tau}^2}
    {\alpha^2 b^2}\right)d\tau dy d\Omega_{d-2}\,,
    \label{appeq:actionwithcond}
\end{align}
where it was used that the time length at the cavity is given by 
$\beta = 2\pi b(1)$ and that 
$r(1)=R$,
see Eqs.~\eqref{appeq:betacond}
and \eqref{appeq:r1}. We have then 
the action as a functional of $b$, $\alpha$, $r$ and $A_\tau$
to be integrated in all paths, in the path integral.
This is the action displayed in Eq.~\eqref{eq:actionwithcond}.
When one integrates the action in 
$b$, $\alpha$, $r$ and $A_\tau$ in the path integral
one indeed obtains a partition function that
depends on the radius of the cavity $R$, the inverse temperature 
$\beta$ and the charge $Q$, the fixed quantities in the ensemble.

%%%%%%%%%%%%%%%%%%%%%%%%%%%%%%%%%%%%%%%%%%%%%%%%%%%%%%%%%%%%%%%%%%%%%%
\section{Calculation of the radii where the free energies of the
electrically charged black hole are zero: Results for different
ensembles and the generalized Buchdahl radius in $d$ dimensions
\label{app:BHzerofreeenergy}}
%%%%%%%%%%%%%%%%%%%%%%%%%%%%%%%%%%%%%%%%%%%%%%%%%%%%%%%%%%%%%%%%%%%%%%

\subsection{The electric uncharged case: Canonical ensemble radius
and the generalized Buchdahl radius in $d$ dimensions}

We want to analyze a thermodynamic energy or
mass to radius ratio
for the $d$-dimensional
canonical ensemble, namely, the  energy or
mass for which the black hole free
energy is zero, $F=0$.  We want to compare this
mass to the Buchdahl bound mass
in $d$ dimensions.

In the canonical ensemble of an uncharged spherically symmetric black
hole in $d$ dimensions~\cite{Andre:2021}, which is described by the
Euclidean Schwarzschild-\hskip-0.05cm{}Tangherlini black hole space,
the canonical ensemble
is realized with a fixed temperature 
at the boundary of the cavity.
There are two black hole solutions, where the one with the largest mass is
stable and the one with the least mass is unstable.
Here we are interested in the large stable black hole.
The free energy of the ensemble also has a critical point
 at zero horizon radius, which is a
minimum, the hot flat space case.  Therefore,
one can analyze which are the favorable states in comparing the free
energies of the zero horizon radius, i.e., hot flat space, and the
stable black hole solution. The free energy of hot flat space is
zero.  The black hole solution also has zero free energy for a given
horizon radius, which is thus an important thermodynamic radius.  The
larger the temperature of the ensemble, the larger this radius, and
the lower the corresponding free energy.  Thus, one can argue that a
stable black hole is favored to hot flat space when the free energy of
the black hole is lower than the zero, which is the free energy of hot
flat space.
The radius of the black hole horizon
that yields zero free energy, i.e., $F=0$, is
$\left(\frac{r_+}{R}\right)_{F=0}=
\left(\frac{4(d-2)}{(d-1)^2}\right)^{\frac{1}{d-3}}$.
In terms of the spacetime mass $m$ this is
\begin{align}
\left(\frac{\mu m}{R^{d-3}}\right)_{F=0}= \left(
\frac{2(d-2)}{(d-1)^2}\right)\,.
    \label{appeq:thermomassF=0}
\end{align}

The Buchdahl bound radius
marks the maximum compactness of a
spherically symmetric star before
spacetime turns singular.  The Buchdahl bound for a star
or matter configuration of
gravitational
radius $r_+$ and radius $R$ is \cite{Wright:2015}
$\left(\frac{r_+}{R}\right)_{\rm Buch}=
\left(\frac{4(d-2)}{(d-1)^2}\right)^{\frac{1}{d-3}}$, 
which in terms of the spacetime
mass $m$ and radius $R$ is
\begin{align}
\left(\frac{\mu m}{R^{d-3}}\right)_{\rm Buch}
= \left(
\frac{2(d-2)}{(d-1)^2}\right)\,.
    \label{appeq:buchmass}
\end{align}
It is a structural bound coming from
mechanics. Self-gravitating matter for which the
mass, or the energy, content within a radius $R$ is above the bound,
in principle collapses to a black hole.

We see that both masses, or radii, although
conceptually different, have the same
expression, indeed,
$\left(\frac{\mu m}{R^{d-3}}\right)_{F=0}=
\left(\frac{\mu m}{R^{d-3}}\right)_{\rm Buch}$.
Therefore, one can argue that as soon as the
black hole phase is
thermodynamically 
favorable over the hot flat space, it is actually
the only phase that exists,
the energy within the reservoir collapses to form a
black hole. This could indicate that there is a link
between black hole thermodynamics and matter mechanics.

\subsection{The electric charged case: Canonical and
grand canonical ensembles radii and the
generalized Buchdahl radius in $d$ dimensions}

We now
want to analyze a thermodynamic energy or mass to radius ratio
for  two ensembles, one is
the
$d$-dimensional canonical ensemble
with electric charge that is being treated here, and
the other is the grand canonical
ensemble that we treated before,  for which the black hole free
energies are zero, i.e., $F=0$, and $W=0$, respectively.
We want to compare these two energy or mass to radius ratio
to the generalized Buchdahl bound, i.e., the Buchdahl
bound in the electric charged
case in $d$ dimensions, also
called the Buchdahl-Andr\'easson-Wright
bound, see \cite{Wright:2015}.

In the canonical ensemble of a charged black hole inside a 
cavity in $d$
dimensions, the construction has been described throughout 
the paper. The canonical ensemble
in this case is realized with a fixed temperature 
and fixed electric charge at the boundary of the cavity. One has in 
this case two stable black hole solutions for a charge below 
a saddle, or critical,
charge $Q_s$, and one stable black hole solution for 
a charge larger than $Q_s$. In this case,
it can be shown that the stable solution with the
largest mass for every charge can have a negative free energy, 
if the black hole has a larger mass than the one that solves this 
equation
$
a \left(\frac{\mu m}{R^{d-3}}\right)^4 
\hskip -0.15cm
+
\hskip -0.05cm
b \left(\frac{\mu m}{R^{d-3}}\right)^3
\hskip -0.15cm
+
\hskip -0.05cm
c \left(\frac{\mu m}{R^{d-3}}\right)^2
\hskip -0.15cm
+
\hskip -0.05cm
d \left(\frac{\mu m}{R^{d-3}}\right)
\hskip -0.05cm
+
\hskip -0.05cm
e
\hskip -0.05cm
=
\hskip -0.05cm
0
$,
where
$a
\hskip -0.05cm
=
\hskip -0.05cm
\left(
\hskip -0.05cm
\left(\frac{d-3}{d-2}
\hskip -0.05cm
\right)^2
\hskip -0.05cm
-
\hskip -0.05cm
4
\hskip -0.05cm
\right)^2
\hskip -0.05cm
$,
%%%
$b
\hskip -0.05cm
=
\hskip -0.05cm
-4
\hskip -0.05cm
\left(
\hskip -0.05cm
4
\hskip -0.05cm
+
\hskip -0.05cm
8 y
\hskip -0.05cm
-
\hskip -0.05cm
\left(
\hskip -0.05cm
\frac{d-3}{d-2}
\hskip -0.05cm
\right)^2
\hskip -0.05cm
(
\hskip -0.05cm
3
\hskip -0.05cm
+
\hskip -0.05cm
2y
\hskip -0.05cm
)
\hskip -0.05cm
\right)$,
$c = -2\left(\frac{d-3}{d-2}\right)^4 y 
- 2 \left(\frac{d-3}{d-2}\right)^2 (y^2 - y +2)
+ 4+24(y +6y^2)$,
$ d = -4y\left((1+y)(1+2y)+\left(\frac{d-3}{d-2}\right)^2
(3+2y) \right)$,
$ e = \left(\frac{d-3}{d-2}\right)^4 y^2 + y^2 (1+y)^2 
+ 2 \left(\frac{d-3}{d-2}\right)^2 y (1+y)(2+y)
$,  with $y$ being the electric charge parameter
given by 
$y\equiv\frac{\mu Q^2}{R^{2d-6}}$,
as before. We see that the equation
is a quartic equation in $\frac{\mu m}{R^{d-3}}$.
The solution can be written
formally as
\begin{align}
\left(\frac{\mu m}{R^{d-3}}\right)_{F=0} =
g\left(d,\frac{\mu Q^2}{R^{2d-6}}\right)\,,
\label{quartic}
\end{align}
for some calculable function $g\left(d,\frac{\mu Q^2}{R^{2d-6}}\right)$.
In the case $Q=0$ one gets
$\left(\frac{\mu m}{R^{d-3}}\right)_{F=0}= \left(
\frac{2(d-2)}{(d-1)^2}\right)^{\frac{1}{d-3}}$ as required, see
Eq.~\eqref{appeq:thermomassF=0}.
The largest stable black hole 
with this mass has a zero
Helmholtz free energy, $F=0$.
Contrasting to the canonical ensemble of the electrically
uncharged black hole discussed above, 
the free energy 
in the electrically charged
case does not include the zero horizon radius 
case. The minimum possible horizon radius is the extremal black hole 
point ${r_{+}}_e = (\mu Q^2)^{\frac{1}{2d-6}}$, yielding a free energy 
$F_{{r_{+}}_e} = \frac{Q}{\sqrt{\mu}}$. 
To emulate hot flat space, we used an electrically charged
nonself-gravitating shell. 
We have then compared the black hole 
configuration with the electrically charged 
shell with no self-gravity
at the boundary of the cavity, having 
then hot flat space inside the cavity
with the electric charge near the boundary. 
This configuration would require
one to look into the matter sector which
we have not done
here. It is unclear if a
transition can occur between 
hot flat space with electric
charges near the cavity and the stable 
black holes. Nevertheless, we still regard the thermodynamic radius 
of zero free energy in the canonical ensemble as an important quantity.

In the grand canonical ensemble of a charged
Reissner-Nordstr\"om  black hole inside a 
cavity for $d$ dimensions, the
construction and its thermodynamics were described
in \cite{Fernandes:2023}.
The grand canonical ensemble is realized with a fixed temperature 
and fixed electric potential
at the boundary of the cavity.
In this ensemble,  the
partition function in the zero loop approximation is given in terms of
the grand potential, or
Gibbs free energy, $W=E - TS - Q\phi$, where $E$ is the mean energy,
$T$ is the temperature, $S$ is the entropy, $Q$ is the mean charge and
$\phi$ is the electric potential. The grand potential yields $W[r_+,Q]
= \frac{R^{d-3}}{\mu}\left(1 - \sqrt{f}\right)- Q \phi - T\frac{\Omega_{d-2}
r_+^{d-2}}{4}$, with $f = \left(1 -
\frac{r_+^{d-3}}{R^{d-3}}\right) \left(1 - \frac{\mu Q^2}{(r_+
R)^{d-3}} \right)$, and the equilibrium equations that yield the black
hole solutions are $\frac1T = \frac{4\pi}{(d-3)}
\frac{r^{d-2}_+}{r^{2d-6}_+ - \mu Q^2} \sqrt{f}$ and $\phi =
\frac{Q}{\sqrt{f}}\left(\frac{1}{R^{d-3}}
-\frac{1}{r_+^{d-3}}\right)$, where the convention
for the electromagnetic coupling and electric charge was chosen so
that $Q\rightarrow \sqrt{(d-3)\Omega_{d-2}} Q$ and $\phi \rightarrow
(\sqrt{(d-3)\Omega_{d-2}})^{-1}\phi$ in the expressions in
~\cite{Fernandes:2023}.
One has in 
this case up to two solutions, depending on the fixed quantities 
$T$ and $\phi$, with only one being stable.
The grand canonical
free energy of the ensemble also has a critical
point at zero horizon radius, which is a minimum, it is the
hot flat space case.
The stable black hole solution also has zero free energy for
a given horizon radius, which is thus an important thermodynamic
radius. The larger the temperature of the
ensemble, the larger this radius, and the lower the corresponding
free energy. Thus, one can argue that a stable
black hole is favored to hot flat space when the free energy
of the black hole is lower than the zero, which is
the free energy of hot flat space. The radius of the black
hole horizon that yields zero
grand potential energy, i.e., $W = 0$
is complicated to find, but 
the corresponding mass has a simple expression given by
\begin{align}
 \hskip-0.1cm
 \left(\frac{\mu m}{R^{d-3}}\right)_{W=0}\hskip-0.15cm
=&\frac{-4(d-2)^2}{(d-1)^2(d-3)^2}\hskip-0.05cm
+ \hskip-0.05cm\frac{2(d-2)((d-2)^2 + 1)}{(d-1)^2 (d-3)^2}
\notag \\
&
\times
\sqrt{1 + \frac{(d-1)^2 (d-3)^2}{4(d-2)^2}\frac{\mu Q^2}{R^{2d-6}}}\,.
\label{eq:thermodynamicradiusgrandcanonical}
\end{align}
Since hot flat space 
is described here by the grand potential $W[r_+,Q]$,
a possible  transition can 
occur from the charged hot flat space to the stable black hole for 
temperatures corresponding to stable black holes with higher mass 
than Eq.~\eqref{eq:thermodynamicradiusgrandcanonical}. 
In the case $Q=0$, one has that
$W=F$, so one
gets
$\left(\frac{\mu m}{R^{d-3}}\right)_{W=0}
=\left(\frac{\mu m}{R^{d-3}}\right)_{F=0}= \left(
\frac{2(d-2)}{(d-1)^2}\right)^{\frac{1}{d-3}}$ as required, see
Eq.~\eqref{appeq:thermomassF=0}.

The 
Buchdahl bound was originally given
for the electrically uncharged case and in
$d=4$.
For electrically charged matter in $d$  
dimensions  one has
the generalized Buchdahl bound
that is given by \cite{Wright:2015}
\vfill
\begin{align}
\left(\frac{\mu m}{R^{d-3}}\right)_{\rm Buch}
& = \frac{d-2}{(d-1)^2} +
\frac{1}{d-1}\frac{\mu Q^2}{R^{2d-6}}
\notag \\
&+ \frac{d-2}{(d-1)^2}\sqrt{1 + (d-1)(d-3)
\frac{\mu Q^2}{R^{2d-6}}}\,.
\end{align}
In the no charge case, $Q=0$, one
gets 
$\frac{\mu m}{R^{d-3}}= \left(
\frac{2(d-2)}{(d-1)^2}\right)^{\frac{1}{d-3}}$,
as required.

We see that the three mass to radius ratios, are
conceptually different, and now in the electrically
charged case,
have generically different expressions, indeed, 
$\left(\frac{\mu m}{R^{d-3}}\right)_{F=0}$, 
$\left(\frac{\mu m}{R^{d-3}}\right)_{W=0}$, and
$\left(\frac{\mu m}{R^{d-3}}\right)_{\rm Buch}$
are not equal.
One has 
$\left(\frac{\mu m}{R^{d-3}}\right)_{F=0}\geq
\left(\frac{\mu m}{R^{d-3}}\right)_{\rm Buch}\geq
\left(\frac{\mu m}{R^{d-3}}\right)_{W=0}$. This is 
an interesting result.
In the canonical ensemble,
the thermodynamic energy content within the cavity
when the black hole phase starts to be favorable, i.e.,
when $F=0$,
is higher than the 
Buchdahl bound, and so
even before the black hole is thermodynamically  favored,
collapse should occur, i.e., 
as soon as a black hole forms
there is no possibility of a thermodynamic
phase transition
to hot flat space, indeed the black hole
has been formed dynamically. 
In the grand canonical ensemble,
the energy content within the cavity
when the black hole phase starts to be favorable, i.e., 
when $W=0$, is less
than the 
Buchdahl bound, and so there
should be no collapse at this stage, indeed, collapse should only occur
when the energy content is increased
above the bound. In the grand canonical ensemble
this occurs only for some negative $W$.
Both
thermodynamic mass to radius
ratios are equal to the generalized Buchdahl bound
bound
when the electric 
charge is put to zero, and all the three are also
equal at the extremal point.
The plots given in Fig.~\ref{fig:Buchdahl}
for $d=5$
help in the understanding of this behavior.
These results present a counter example to the
possible link between the black hole thermodynamics and stability 
of spherically symmetric matter. The uncharged case seems to be 
a coincidence.

\vfill

%\vskip 10cm
%\vfill

\end{document}